\documentclass[a4paper]{JHEP3}
\usepackage{amsmath}
\usepackage{amsfonts}
\usepackage{bbm}
\newcommand{\ben}{\begin{eqnarray}}
\newcommand{\een}{\end{eqnarray}}
\newcommand{\bee}{\begin{equation}}
\newcommand{\eee}{\end{equation}}
\newcommand{\nn}{\nonumber}
\newcommand{\hth}{\hat{\theta}}

\newcommand{\p}{\partial}

\newcommand{\cx}{\mathcal}
\newcommand{\ti}{\tilde}
\newcommand{\Ga}{\Gamma}
\newcommand{\bb}[1][\overline]{1}

\newcommand{\la}{\lambda}
\newcommand{\zh}{\hat{z}}

\newcommand{\nus}{\nu^\star}

\newcommand{\nub}{{\bar{\nu}^\star\!}}

\newcommand{\noi}{\noindent}
\newcommand{\al}{\alpha}
\newcommand{\IP}{\mathbb P}
\newcommand{\IZ}{\mathbb Z}
\newcommand{\IC}{\mathbb C}
\newcommand{\be}{\beta}
\newcommand{\om}{\omega}
\newcommand{\Om}{\Omega}

\newcommand{\ga}{\gamma}
\newcommand{\lb}{{\underline{l}}}

\newcommand{\Si}{\Sigma}

\newcommand{\x}{cl{x}}

\newcommand{\mcs}{\mathcal{M}_{CS}}
\newcommand{\cxH}{\mathcal{D}}
\newcommand{\tha}{\vartheta}
\def\x{}
\def\Lbulk{\cx L^{bulk}}\def\Lbdry{\cx L^{bdry}}\def\th{\theta}

\def\ux#1{\underline{#1}}
\def\hth{{\hat{\theta}}}\def\taun{\pi}
\def\SP{{\cx W}}  
\def\sp{W}  
\def\pmk{\kappa}
\def\newpag{\ \\[2mm]}
\title{Type II/F-theory Superpotentials with Several Deformations and $\cx N=1$ Mirror Symmetry}
\author{Murad Alim$^1$,  Michael Hecht$^2$, Hans Jockers$^{3,4}$, Peter Mayr$^2$, Adrian Mertens$^2$ and Masoud Soroush$^2$ 
\\
\\ $^1$Jefferson Physical Laboratory, Harvard University, Cambridge, MA, 02138, USA
\\ $^2$Arnold Sommerfeld Center for Theoretical Physics, LMU, Theresienstr. 37, D-80333 Munich, Germany
\\ $^3$Department of Physics, Stanford University, Stanford, CA 94305-4060, USA
\\ $^4$Kavli Institute for Theoretical Physics, Santa Barbara, CA 93106, USA}

\abstract{We present a detailed study of D-brane superpotentials depending on several open and closed-string deformations. The relative cohomology group associated with the brane defines a generalized hypergeometric GKZ system which determines the off-shell superpotential and its analytic properties under deformation. Explicit expressions for the $\cx N=1$ superpotential for families of type II/F-theory compactifications are obtained for a list of multi-parameter examples. Using the Hodge theoretic approach to open-string mirror symmetry, we obtain new predictions for integral disc invariants in the $A$ model instanton expansion. We study the behavior of the brane vacua under extremal transitions between different Calabi-Yau spaces and observe that the web of Calabi-Yau vacua remains connected for a particular class of branes.}
\preprint{LMU-ASC 74/10\\ SU-ITP-10/28\\ NSF-KITP-10-117}

\begin{document}
\newpage
\section{Introduction}
String backgrounds with mirror symmetry offer some of the rare occasions, where quantitative non-perturbative data on semi-realistic string theory compactifications can be obtained explicitly. An important example are $\cx N=1$ supersymmetric compactifications of type II strings on Calabi--Yau threefolds with branes, sometimes related to an F-theory compactification on a Calabi--Yau fourfold. The application of open-string mirror symmetry to this case has been pioneered in refs.~\cite{Vafa:1998yp,Kachru:2000ih,Kachru:2000an} and in particular in ref.~\cite{Aganagic:2000gs}, where the authors defined a large class of mirror pairs of brane geometries and obtained the first prediction for Ooguri--Vafa invariants \cite{Ooguri:1999bv} in non-compact geometries from a $B$-model computation. 

Motivated and guided by the results of ref.~\cite{Aganagic:2000gs}, a Hodge theoretic approach to the computation of D-brane superpotentials and open-string mirror symmetry was put forward in refs.~\cite{Lerche:2002ck,Lerche:2002yw}. It was argued, that the periods on the relative cohomology group associated with a $B$-type brane determine the superpotential as the solution of a GKZ generalized hypergeometric system and that the Hodge filtration defines the mirror map and the potentials for the $A$ model instantons, much as in the case of closed-string mirror symmetry treated in refs.~\cite{Candelas:1990rm,Candelas:1993dm,Batyrev:1994hm,Hosono:1993qy}. The Hodge theoretic framework applies also to compact geometries. In refs.~\cite{Walcher:2006rs,Morrison:2007bm,Pandharipande:2008} the first results on compact manifolds were obtained by computing the dependence of the superpotential on closed-string deformations for rigid branes from so-called normal functions. The case with open-string deformations has been solved in refs.~\cite{Jockers:2008pe,Alim:2009rf,Alim:2009bx} in the relative cohomology framework. Since the open-string degrees of freedom are frozen at a critical value in the first formalism, while the superpotential still depends on open-string deformations away from a critical value in the second formalism, we refer to the two cases as on- and off-shell approaches, respectively.

The off-shell superpotential for the $B$-type brane geometry on the threefold is often related to the GVW flux superpotential \cite{Gukov:1999ya} for an M/F-theory compactifications on a dual Calabi-Yau fourfold $X_4$ by an open-closed duality \cite{Mayr:2001xk,Alim:2009rf,Aganagic:2009jq}. The GVW superpotential on the fourfold $X_4$ can be computed from the integral fourfold periods by standard methods \cite{Greene:1993vm,Mayr:1996sh,Klemm:1996ts,Lerche:1997zb} and it agrees with the brane superpotential on the threefold at lowest order in $g_s$ \cite{Alim:2009rf,Alim:2009bx,Aganagic:2009jq,Grimm:2009ef,Li:2009dz}. The full F-theory superpotential computes $\cx O(g_s),\cx O(e^{-1/g_s})$ corrections to the superpotential of the local brane geometry \cite{Jockers:2009ti} and captures the superpotential of dual type II and heterotic compactifications on generalized Calabi-Yau manifolds \cite{Dasgupta:1999ss,Haack:2000di,Jockers:2009ti}.

The continuous brane deformations test off-shell directions of the superpotential in the open-string direction away from the critical point. Depending on the behavior of the superpotential near the critical locus this leads to two qualitatively different types of instanton expansions in the mirror $A$-model. Generically, the open-string deformations are obstructed classically and should be integrated out. Freezing the open-string parameters one obtains an instanton expansion of the critical superpotential in the closed-string moduli only, which leads to the modified disc invariants defined in refs.~\cite{Walcher:2006rs,Pandharipande:2008}. The other case is a critical locus with almost flat directions also in the open-string direction, where the $A$ model potential has an instanton expansion in closed- {\it and} open-string deformation parameters. This led to the first $B$ model predictions for genuine Ooguri--Vafa invariants in compact brane geometries in refs.~\cite{Alim:2009rf,Jockers:2009mn,Alim:2009bx}, generalizing the familiar large volume expansion of the closed-string mirror symmetry to the open-string sector.

For compact geometries, the predictions on off-shell superpotentials and invariants obtained from the generalized GKZ systems for relative cohomology pass some non-trivial consistency checks, but await for a verification by independent methods.\footnote{See ref.~\cite{Baumgartl:2010ad} for recent progress from matrix factorizations.} In this note we further test the Hodge theoretic approach in the more general situation of compact brane geometries with several deformations, near the critical points of the first, generic type. Explicit expressions for the $\cx N=1$ superpotential for brane compactifications on Calabi-Yau threefolds and related F-theory compactification on Calabi-Yau fourfolds are obtained for these examples. Particular emphasis is given to the relation of the off-shell approach of refs.~\cite{Alim:2009rf,Jockers:2009mn,Alim:2009bx} and the on-shell computations of refs.~\cite{Walcher:2006rs,Morrison:2007bm}.\footnote{See also refs.~\cite{Walcher:2007qp,Krefl:2008sj,Knapp:2008uw,Walcher:2009uj} for further examples and discussions.} Specifically the multi-parameter examples studied below lead, at the critical locus, to a class of Picard-Fuchs equations with complicated inhomogeneous pieces given by hypergeometric series with special properties. Applying the mirror map to these examples we obtain new predictions for integral disc invariants on the $A$ model side. 

The multi-parameter models allow us to study the fate of the domain walls under extremal transitions between closed-string compactifications on different manifolds, which are believed to connect the web of $\cx N=2$ vacua represented by different Calabi-Yau manifolds \cite{Candelas:1989ug,Greene:1995hu}. It is an interesting question to what extent the $\cx N=2$ web remains connected after adding D-branes. This was already studied in ref.~\cite{Walcher:2009uj} in one example. We find that for extremal transitions through points of enhanced non-Abelian gauge symmetries, the two vacuum branches stay connected for a particular set of domain walls and there is an interesting physical and group theoretic structure. If $G$ denotes the non-perturbative gauge group, the domain walls fall into representations of the Weyl group, with the disc invariants of the domain walls mapping to each other under the group action. At the locus of gauge symmetry enhancement, the domain wall tensions in non-trivial representations degenerate, which implies the existence of tensionless domain walls at this point.

The organization of this note is as follows. In sect.~\ref{sec:RelCoh} we outline the Hodge theoretic approach to the computation of type II and F-theory superpotentials and describe how the off-shell approach based on families of relative cohomology groups reduces at the critical points to the formalism of normal functions studied in refs.~\cite{Walcher:2006rs,Morrison:2007bm}. The crucial link is provided by a subset of the period integrals defined by the relative cohomology group. These determine the critical set as the vanishing locus of a certain period vector and induce an inhomogeneous term in the Picard-Fuchs equations upon restriction to the critical point. We describe the generalized GKZ type systems that annihilate the type~II/F-theory superpotential for brane geometries in toric hypersurfaces. In sect.~\ref{sec:ex} we turn to a detailed study of critical points of the massive type for a number of brane geometries with several parameters. We compute the type II/F-theory superpotential and disc invariants for these vacua and study the fate of the domain walls through extremal transitions to other Calabi--Yau manifolds. In sect.~\ref{sec:con} we present our conclusions. Finally in app.~\ref{sec:app} we collect some additional material, which supplements the analysis of the main text. Here we give a description of the studied threefolds and fourfolds for type II/F-theory compactifications as toric hypersurfaces. We also study local limits of the compact Calabi-Yau manifolds in the examples. For these local geometries, which can be associated to elliptic curves, we extract disc invariants. These invariants are related to a subset of disc invariants of the corresponding compact Calabi-Yau manifolds.

\section{Relative cohomology, generalized GKZ systems and superpotentials}\label{sec:RelCoh}
\subsection{Brane and flux superpotentials in type II and F-theory}
In this note we study the $\cx N=1$ superpotential $\cx W$ of $B$-type D-branes wrapped on even-dimensional cycles of a Calabi-Yau threefold $X$ and, by open-string mirror symmetry, the superpotential of the $A$ brane geometry related to it. The $B$ model compactification will also be related to an F-theory compactification on a dual Calabi-Yau fourfold $X_4$. In the Hodge theoretic approach of refs.~\cite{Lerche:2002ck,Lerche:2002yw,Jockers:2008pe,Alim:2009rf,Alim:2009bx}, the superpotential $\cx W$ of these theories is derived from the period integrals $\ux \Pi(z,\zh)$ on the relative cohomology groups defined by the branes, schematically 
\bee
\ux \Pi(z,\zh) = \cx W(C)-\cx W(C_*)=\cx W_{brane}(z,\zh)\,,
\eee
where $(z,\zh)$ are certain local coordinates on the open-closed deformation space $\cx M$ specified below,\footnote{The letters $z$ and $\zh$ are reserved for closed- and open-string deformations, respectively.} $C$ is a 2-cycle in $X$ wrapped by the D-brane and $C_*$ is a reference cycle in the same homology class, $[C_*]=[C]$. The above expression is equal to the tension of a domain wall interpolating between the configurations obtained by wrapping the D-brane either on $C$ or on $C_*$. The relative periods also capture the 3-form flux superpotential $W_{flux}=\int_X G\wedge \Om$ of refs.~\cite{Gukov:1999ya,Taylor:1999ii}, leading to a unified expression of the four-dimensional superpotential in terms of a general linear combination of all relative period integrals \cite{Lerche:2002ck,Lerche:2002yw}:
\bee\label{PiW}
\cx W_{\cx N=1}(z,\zh)=\sum \ux N_\Si \ux \Pi_\Si(z,\zh)=\cx W_{flux}(z)+\cx W_{brane}(z,\zh)\,,
\eee
The coefficients $\ux N _\Si$ are determined by the topological charges of the brane and flux background. Solving the vacuum condition $\frac{d}{d\zh}\cx W_{\cx N=1}=0$ in the open-string direction gives the {\it on-shell} (in the open-string direction) superpotential $W(z)$ as a function of the closed string moduli and the topological data $\ux N_\Si$.

To write the {\it off-shell} superpotential $\cx W(C)$ on a deformation space $\cx M$, one needs to specify extra data, in particular a concrete parametrization for the off-shell configurations. The off-shell deformation space for a brane on $C$ is generically infinite dimensional, with most of the deformations re\-pre\-senting heavy fields in space-time that should be integrated out. To define a suitable finite dimensional space $\cx M$ with obstruction potential $\cx W$ one therefore needs to choose an appropriate set of 'light' fields and integrate out infinitely many others, as is familiar in the effective action approach. The result at the critical locus is independent of the parametrization of the off-shell directions, but the off-shell values depend, in a well-defined way, on the parametrization.

Generically, there are many consistent choices for the set of light fields, corresponding to local coordinate patches of the off-shell deformation space of different dimension and range of validity. Each choice of  parametrization corresponds to a slightly different formulation as a relative cohomology problem. A preferred class of parametrizations favoured equally well by mathematics and physics arises from the following construction motivated by duality to M/F-theory. Embed $C$ into a 4-cycle $D$ and define $\cx M$ as the {\it unobstructed} deformation space of a holomorphic family $\cx D$ of such 4-cycles. Adding a D-brane charge on $C\subset H_2(D)$ induces a superpotential $\cx W(C)$ on $\cx M$ \cite{Witten:1997ep,Aganagic:2000gs}. Physicswise this can be viewed as perturbing the true moduli space $\cx M$ of an F-theory compactification with an unobstructed family of D-branes wrapped on the 4-cycles $D$ by adding a D-brane charge on a 2-cycle $C$ in $D$ \cite{Alim:2009bx,Aganagic:2009jq}. It was already observed in refs.~\cite{Lerche:2002ck,Lerche:2002yw,Jockers:2008pe,Alim:2009rf}, that this class of parametrizations is the one preferred by the topological open-closed string theory, as it leads to flat coordinates on the open-closed deformation space $\cx M$, which are in agreement with the expectations from the chiral ring in the topological string theory. Moreover the Hodge theoretic definition of the open-string mirror map obtained in this way yields consistent results for the $A$ model disc invariants, in agreement with localization computations in the $A$ model, if available. Mathematically, this class of parametrizations derives directly from the on-shell meaning of the superpotential as an Abel-Jacobi invariant measuring rational equivalence of the cycles $C$ and $C_*$, as explained in sect.~2.1 below. 

The perturbation idea becomes obvious in the framework of the dual M/F-theory compactification on a related fourfold $X_4$, which geometrizes the branes to flux \cite{Mayr:2001xk,Alim:2009rf,Aganagic:2009jq}. In this context, $\cx M$ maps to the {\it unobstructed} complex structure moduli space $\cx M_{CS}(X_4)$ of the fourfold $X_4$, which is the vacuum space of topological strings in the type IIA compactification on $X_4$, and open-closed mirror symmetry maps to closed-string mirror symmetry for fourfolds. Adding a 4-form flux $G$ induces the Gukov-Vafa-Witten superpotential \cite{Gukov:1999ya} on the moduli space $\cx M_{CS}(X_4)$, and this is the dual description of the off-shell deformation space $\cx M$ of the brane geometry and the obstruction superpotential $\cx W(C)$ on it. More precisely, the F-theory superpotential\footnote{See ref.~\cite{Aganagic:2009jq} for the discussion from the M-theory perspective.} on $X_4$ computes $g_s$ corrections to the superpotential $\cx W(C)$ as captured by the relation \cite{Alim:2009bx,Jockers:2009ti}
\bee\label{GVW}
\cx W_{GVW}(X_4)=\int_{X_4} G \wedge \Om^{(4,0)}=\sum_{\Si} N_\Si(G)\ \ux \Pi(z,\zh)+\cx O(g_s)+\cx O(e^{-1/g_s})\,,
\eee
where the leading term on the right hand side is the result  \eqref{PiW} for the $B$-type branes on the threefold with the linear combination of relative periods determined by the flux $G$ on the fourfold. We will only consider the leading term in $g_s$ in this paper, which can be computed from the integral periods of a certain non-compact limit $X_4^\sharp$ of $X_4$, related to the threefold $X$ by the open-closed duality \cite{Mayr:2001xk,Alim:2009rf,Aganagic:2009jq}. The details of the compactification $X_4$ of $X_4^\sharp$ affect only the higher terms in $g_s$ and can be computed similarly \cite{Jockers:2009ti}. More details and many examples on the computation of the fourfold superpotential from the geometric period integrals can be found in refs.~\cite{Mayr:1996sh,Klemm:1996ts,Lerche:1997zb}.

\subsection{\label{sec:22}Relative periods and domain wall tensions}
As alluded to above, a preferred parametrization adapted to topological string states and open-string mirror symmetry is to parametrize the off-shell deformations of the D-brane on a 2-cycle $C$ by the deformations of a holomorphic family of 4-cycles $\cx D$ that embed $C\in H_2(D)$. The relative periods capturing the superpotential for the brane on $C$ are obtained by restriction to the subspace $H_3(X,C)\subset H_3(X,D)$. Mathematically, this class of parametrizations derives directly from the concept of rational equivalence and the on-shell meaning of the superpotential as an Abel-Jacobi invariant, as will be discussed now.\footnote{For a related mathematical discussion, see ref.~\cite{Griffiths:1979}.}

To this end, consider a Calabi-Yau threefold $X_0$ together with an ample divisor $D_0$. We assume that $H^{1,0}(X_0)=H^{2,0}(X_0)=0$, such that the complex structure deformations of the pair $(X_0,D_0)$ are unobstructed. Then this pair $(X_0,D_0)$ extends to a family of Calabi-Yau threefolds together with a family of ample divisors $\pi: ({\cal X}, {\cal D}) \rightarrow \Delta$ fibered over the disc $\Delta$, which parametrizes a local patch of the combined moduli space $\cal M$ of the family obtained by deforming the central fiber $\pi^{-1}(0)=(X_0,D_0)$. $\cx M$ is a fibration $\hat{\cx M}\to \cx M \to \cx M_{CS}$, where the base $\cx M_{CS}$ corresponds to complex structure deformations $z$ of the family of Calabi-Yau threefolds $\cal X$, while the fiber $\hat{\cx M}$, parametrized by the coordinates $\hat z$, corresponds to the deformations of the family of divisors $\cal D$. In string theory, the former arise in the closed-string sector and the latter in the open-string sector.

Since the holomorphic three form $\Omega(z)$ of the Calabi-Yau threefold $X_z$ vanishes on the divisor $D_{(z,\hat z)}$, the three form $\Omega(z)$, which is an element of $H^3(X_z)$, lifts to an element $\underline{\Omega}(z,\hat z)$ of the relative cohomology group $H^3(X_z,D_{(z,\hat z)})$. We define the integral relative periods as\footnote{Objects defined in relative (co-)homology will be distinguished by an underline.}
\bee \label{relPer}
\underline{\Pi}(\ux \Ga;z,\hat z)\,=\, \int_{\underline{\Gamma}_{(z,\hat z)}} \underline{\Omega}(z,\hat z) \ ,
\eee  
where $\ux{\Gamma}_{(z,\zh)}$ is an integral relative cycle in $H_3(X_z,D_{(z,\zh)},\IZ)$ whose boundary $\p\underline{\Gamma}_{(z,\hat z)}$ is trivial as a class in $H_2(X,\IZ)$. For concreteness we often assume that the boundary is the difference $\p\underline{\Gamma}_{(z,\hat z)}=C_{(z,\zh)}^{+}-C_{(z,\zh)}^{-}$ of two 2-cycles in $H_2(D_{z,\zh})$ with $[C^{+}_{(z,\zh)}]=[C^{-}_{(z,\zh)}]$ in $H_2(X_z,\IZ)$.

Following the fundamental works \cite{Aganagic:2000gs,Witten:1997ep}, it was proposed in refs.~\cite{Lerche:2002ck,Lerche:2002yw,Jockers:2008pe} that the relative period \eqref{relPer} defines the off-shell tension $\cx T(z,\zh)$ of a physical D-brane wrapped on the chain $\underline{\Gamma}_{(z,\hat z)}$, that is $\cx T(z,\zh)=\ux\Pi(\ux \Ga;z,\zh)$. This D-brane represents a domain wall interpolating between the two configurations obtained by wrapping a D-brane on $C^+_{(z,\zh)}$ or on $C^-_{(z,\zh)}$ and its tension measures the difference of the value of the superpotentials for the two D-brane configurations
\bee\label{TWsplit}
\cx T(z,\zh)=\cx W(C^+_{(z,\zh)})-\cx W(C^-_{(z,\zh)})\ .
\eee
The vacuum condition in the open-string direction is $\frac{d}{d\zh}\cx W(C^\pm)|_{\zh=\zh_{crit}}=0$ and it holds if $C^\pm_z:=C^\pm_{(z,\zh_{\rm crit})}$ is a holomorphic curve \cite{Witten:1997ep}. Imposing this condition on both branes implies $\frac{d}{d\zh}\cx T(z,\zh)=0$ as well.

Mathematically speaking, the vacuum configurations hence lie within the so-called Noether-Lefshetz locus,  defined as \cite{Clemens}
\bee \label{critcon}
{\cal N}\,=\, \left\{(z,\hat z)\in \Delta \ \left| \  0\,\equiv\,\frac{d\underline{\Pi}(z,\hat z)}{d\hat z} \right. \right\} \  .
\eee
Equivalently the locus ${\cal N}$ can be specified by the vanishing condition
\bee \label{critcon2}
{\cal N}\,=\, \left\{(z,\hat z)\in \Delta \ \left| \ 0\,\equiv\,\vec\pi(z,\hat z; {\partial\underline{\Gamma}_{(z,\hat z)}}) 
\right.\right\} \ ,
\eee
for the period vector of the divisor $D_{(z,\hat z)}$
\bee \label{Defpi}
\vec\pi(z,\hat z; {\partial\underline{\Gamma}_{(z,\hat z)}})\,=\,\left( \int_{\partial\underline{\Gamma}_{(z,\hat z)}}\!\!\!\!\om^{(2,0)}_{\hat a}(z,\hat z)\right)\ , 
\quad \hat a = 1,\ldots,\dim H^{2,0}(D_{(z,\hat z)})  \ .
\eee  
Here $\om^{(2,0)}_{\hat a}(z,\hat z)$ is a basis of two forms for $H^{2,0}(D_{(z,\hat z)})$. Hence the critical locus of D-brane vacua is mapped to the subslice of complex structures on the surface $D_{(z,\zh)}$, where certain linear combinations of period vectors on the surface vanish. At such points in the complex structure the Picard lattice of the surface $D_{(z,\zh)}$ is enhanced due to the appearance of an additional integral $(1,1)$-form. 

At the Noether-Lefschetz locus $(z,\hat z_{\rm crit})\in\cal N$ there is an interesting connection between the relative periods and another mathematical quantity studied in refs.~\cite{Walcher:2006rs,Morrison:2007bm}. By the result of ref.~\cite{Clemens}, the relative period $\underline{\Pi}(z,\hat z)$ evaluated at the Noether-Lefschetz locus $(z,\hat z_{\rm crit})\in\cal N$ gives (modulo bulk periods) the Abel-Jacobi invariant associated to the normal function of the algebraic curve $\partial\underline{\Gamma}_{(z,\hat z_{\rm crit})}$:
\bee \label{DWcon}
\underline{\Pi}(z,\hat z_{\rm crit}) \,=\, \nu_{c_2^{\rm alg}(\partial\underline{\Gamma}_{(z,\hat z_{\rm crit})})} (z) \mod \quad (\text{bulk periods}) \ .
\eee  
Specifically, the Abel-Jacobi invariant is defined via the normal function $\nu_{c_2^{\rm alg}(\alpha)}(z)$ as 
\bee\label{DefAJ}
AJ\,:\ CH^2(X_z) \rightarrow  J^3(X_z) \simeq \frac{F^2H^3(X_z)^*}{H_3(X_z,\IZ)}\,;\qquad  \alpha \mapsto  \nu_{c_2^{\rm alg}(\alpha)}(z)\,,
\eee
where, in the concrete setting, the normal function is defined as the chain integral 
\bee\label{DefNF}
T(z) \,=\, \int_{\Gamma^\pm_z} \Omega(z)\,=\, \nu_{c_2^{\rm alg}(C^+_z - C^-_z)} (z) \mod \quad (\text{bulk periods}) \ .
\eee
Here $\partial\Gamma^\pm_z = C^+_z - C^-_z$, with $C^\pm_z$ the holomorphic curves at fixed $\zh=\zh_{\rm crit}$. The essential point is that (only)  at the critical locus, the above integral is well-defined in absolute cohomology, because the potentially dangerous boundary terms vanish by holomorphicity of the boundary $\partial\Gamma^\pm_z$ and the Hodge type of $\Om$. The normal functions \eqref{DefNF} have been introduced in refs.~\cite{Walcher:2006rs,Morrison:2007bm} to study the on-shell values of the superpotentials
$$
T(z)=W(C^+_z)-W(C^-_z)\ .
$$
By the above argument, these are the restrictions of the relative period integrals \eqref{relPer} to the critical locus $\cx N$.

There is also a partial inverse of this relation, which recovers the relative periods for the family of divisors starting from the normal functions. To this end, recall the meaning of rational equivalence and the Abel-Jacobi invariant. The second algebraic Chern class $c_2^{\rm alg}$ takes values in the second Chow group $CH^2(X_z)$, which consists of equivalence classes of algebraic cycles of co-dimension two modulo rational equivalence \cite{Hartshorne:1977}.\footnote{The second algebraic Chern class is a refined invariant of the topological second Chern class \cite{Hartshorne:1977}.} Two algebraic cycles $\alpha$ and $\beta$ of co-dimension two are rationally equivalent, if we can find a subvariety $V$ of co-dimension one, in which $\alpha$ and $\beta$ are rationally equivalent as co-dimension one cycles. This is the case if $\alpha$ and $\beta$ are given by two linearly equivalent divisors on $V$, that is $[\alpha-\beta]=0\in CH^1(V)$.\footnote{If the subvariety $V$ is not normal the cycles $\alpha$ and $\beta$ are rationally equivalent, if their Weil divisors $D_\alpha$ and $D_\beta$ are linearly equivalent in the normalization $\tilde V$ of $V$, namely $\alpha \sim \beta$ if $D_\alpha \sim D_\beta$ with $f:\tilde V \rightarrow V$ and $\alpha = f_*D_\alpha$ and $\beta=f_*D_\beta$.} Moreover, rational equivalence implies that the Abel-Jacobi invariant vanishes.

Starting from an algebraic cycle $\alpha$ of co-dimension two with $c_2^{\rm top}(\alpha)=0$ we can find a three chain $\Gamma^\alpha$ such that $\alpha = \partial\Gamma^\alpha$, and associate a normal function $\nu_{c_2^{\rm alg}(\alpha)}$ to it via the integral \eqref{DefNF}. By \eqref{DefAJ}, the normal function vanishes for algebraic two cycles $C^\pm_z$ that are rationally equivalent \cite{Morrison:2007bm}. On the contrary, if $C^+_z$ and $C^-_z$ are not rationally equivalent, we obtain an element in the relative cohomology of each family $\cx D(z,\zh)$ of divisors that contains the two holomorphic curves $C^\pm_z$ at a 'critical' value $\hat z=\hat z_{\rm crit}$. Indeed, since $C^+_z$ and $C^-_z$ are \emph{not} rationally equivalent, $C^+_z - C^-_z$ defines by Poincar\'e duality a non-trivial Element $\om\in {\rm Pic}(D_{(z,\hat z_{\rm crit})})\simeq H^{1,1}(D_{(z,\hat z_{\rm crit})}) \cap H^2(D_{(z,\hat z_{\rm crit})},\IZ)$. Since the algebraic cycle $\al$ is topologically trivial on $X_z$, the associated two form $\om$ is \emph{not} induced from the hypersurface $X_z$ and lifts to a relative three form $\underline\Theta_{(z,\hat z_{\rm crit})}$ by the relation 
$$
H^3(X_z,D_{(z,\hat z)}) \,\simeq\, {\rm coker}\left(i^*: H^2(X_z) \rightarrow H^2(D_{(z,\hat z)}) \right) \oplus 
{\rm ker}\left(i^*: H^3(X_z) \rightarrow H^3(D_{(z,\hat z)}) \right) \,,
$$
with $i: D_{(z,\hat z)} \hookrightarrow X_z$. By construction, the three-chain $\Gamma^\alpha\simeq\ux{\Gamma}_{(z,\zh_{\rm crit})}$ is a representative of the relative homology class in $H_3(X_z,D_{(z,\hat z_{\rm crit})})$  dual to $\underline\Theta_{(z,\hat z_{\rm crit})}$. Surjectivity of the boundary map of homology then asserts that the above construction assigns to each normal function a relative period on $H^3(X_z,D_{(z,\hat z)})$, which measures the superpotential of the off-shell deformation parametrized by the family $\cx D(z,\zh)$.

The relative (co-)homology groups $H_3(X_z,D_{(z,\hat z_{\rm crit})},\IZ)$ (and $H^3(X_z,D_{(z,\hat z_{\rm crit})},\IZ)$) are topological and do not depend on the open-closed deformation parameters, for a smooth family of the pair $({\cal X},{\cal D})$. As a consequence the relative three cycle $\Gamma_{(z,\hat z_{\rm crit})}$ (and three-form $\underline\Theta_{(z,\hat z_{\rm crit})}$) extends over the whole disc $(z,\hat z)\in \Delta$. Therefore we can define a relative three cycle $\Gamma_{(z,\hat z)}$ (and a relative three form $\underline\Theta_{(z,\hat z)}$) for all open parameters $\hat z$ and study the relative period integrals $\underline{\Pi}(z,\hat z)$ using the Mixed Hodge Variation on the family of relative cohomology groups over $\Delta$. The Gauss-Manin derivative on this local system provides a powerful framework to study the relative periods as solutions to a system of Picard-Fuchs equations and leads to a predictive proposal for off-shell mirror symmetry formulated in refs.~\cite{Lerche:2002ck,Lerche:2002yw,Jockers:2008pe,Alim:2009rf,Alim:2009bx}.

Using this connection between normal functions \eqref{DefNF}, that is to say domain walls between critical points $C^\pm_z$, and the off-shell tensions represented by the integral relative periods \eqref{relPer} ending on $C^\pm_{(z,\zh)}$, we may calculate the critical tensions as follows. First determine the possible critical points as the vanishing locus \eqref{critcon2} of the periods of the surface $D_{(z,\zh)}$. The critical domain wall tension is then given by the relative period associated with the vanishing period on the surface, evaluated at the critical point $z_{\rm crit}$
\bee
T(z)\,=\, \cx T(z,\hat z_{\rm crit})\,.
\eee
This determines the critical tension up to a possible addition of a bulk period $\Pi_{\rm Bulk}(z)$. 

The vanishing condition \eqref{critcon2}, classifying the critical points, can be studied very explicitly for off-shell deformations in a single open-string parameter $\hat z$, which is sufficient to determine the on-shell tensions. In this case the surface $D_{(z,\zh)}$ has geometric genus one and it is isogenic to a K3 surface \cite{Morrison:1887}, that is the integral Hodge structures of the surface $D_{(z,\zh)}$ can be mapped to the equivalent Hodge structure of its isogenic K3 surface. This has already been used in ref.~\cite{Alim:2009bx} and will simplify the discussion in some of the examples below.

One particular type of solutions to the vanishing condition arises at the discriminant locus of the isogenic K3 surface, where the period vector, associated with a geometrically vanishing cycle in the K3 surface, develops a zero. However, this type of solution is non-generic in the sense that it is often related to points in the deformation space with a domain wall with zero tension. The generic critical points arise instead from a zero of the period vector, which is a linear combination of volumes of geometric cycles in the K3 surface rather then the volume of an irreducible cycle. The typical example is a point where the volumes of two different cycles coincide, such that the period vector associated with the difference vanishes. At these particular symmetric points there is an 'accidental' global symmetry of the K3 lattice, exchanging the two cycles. 
More generally the generic critical points should be classified by special symmetric points in the K3 moduli studied in ref.~\cite{Nikulin:1999}.

\subsection{Generalized GKZ systems and Picard-Fuchs equations for type II/F-theory superpotentials}
As alluded to above, the flat Gauss-Manin connection on the relative cohomology bundle leads to a Picard-Fuchs type of differential operators for the relative periods, which provide an effective method to determine and to study the tensions $\cx T(z,\zh)$ \cite{Lerche:2002ck,Lerche:2002yw,Jockers:2008pe,Alim:2009rf,Alim:2009bx}.
These differential equations also reflect the duality of $B$-type branes on the threefold $X$ to an M/F-theory compactification on a fourfold $X_4$ determined by open-closed duality \cite{Mayr:2001xk,Alim:2009rf,Aganagic:2009jq}. Specifically, the set of differential operators for the relative periods on $X$ and for the fourfold periods on $X_4$ have the superpotential periods in eqs.~\eqref{PiW} and \eqref{GVW} as common solutions, and the superpotential can be equivalently computed on the threefold or on the fourfold. 

For concreteness, we assume that the holomorphic curves $C^\pm_z$ are contained in the intersection of the hypersurface $X:\ P=0$ with two hyperplanes $D_{1,2}$ defined in a certain ambient space. Choose coordinates such that the equation for $D_1$ does not depend on the closed-string moduli $z$, typically of the form\footnote{Note that the equation for $D_1$ is a priori defined in the ambient space. However, by restriction to the hypersurface $X$ we also identify $D_1$ with a divisor on the hypersurface $X$. For ease of notation we denote both the divisor of the ambient space and of the hypersurface with the same symbol $D_1$.}
$$
D_1:\ x_1^a + \eta\,  x_2^b=0\,,
$$ 
where $x_i$ are some homogeneous coordinates on the ambient space, $a,b$ some constants that depend on the details and $\eta$ a fixed constant, which is a phase factor in appropriate coordinates. This hyperplane can be deformed into a family $\cx D_1:\ x_1^a + \zh\,  x_2^b=0$ by replacing the constant $\eta$ by a complex parameter $\zh$. The relative 3-form $\ux \Om$ and the relative period integrals on the family of cohomology groups $H^3(X,D_1)$, satisfy a set of Picard-Fuchs equations \cite{Lerche:2002ck,Lerche:2002yw,Jockers:2008pe,Li:2009dz} 
$$
\cx L_a(\th,\hth)\, \ux\Om = \ux{d\om}^{(2,0)}\quad  \Rightarrow\quad   \cx L_a(\th,\hth)\ \cx T(z,\zh)=0\,,\qquad a=1,...,A\,,
$$
where $a$ is some label for the operators. The differential operators can be split into two pieces 
\bee\label{Lsplit}
\cx L_a(\th,\hth)=:\Lbulk _a-\Lbdry_a\, \hth\,,
\eee
where the bulk part $\Lbulk _a(\th)$ acts only on the closed-string moduli $z$ and the boundary part $\Lbdry_a(\th,\hth)\, \hth$ contains at least one derivative in the parameters $\zh$. Since the dependence on $\zh$ localizes on $D_1$, the derivatives $2\pi i\,\hth\,\cx T(z,\zh)$ are proportional to the periods \eqref{Defpi} on the surface $D_1$
\bee\label{surfper}
2\pi i\,\hth\, \cx T(z,\zh)= \pi(z,\zh)\ .
\eee
Rearranging eq.\eqref{Lsplit} and restricting to the critical point 
$\zh=\eta$ one obtains an inhomogeneous Picard-Fuchs equation
\bee\label{InhPF}
\Lbulk_a\,  T(z)=f_a(z) \,,
\eee
with $T(z)=\cx T(z,\eta)$ and 
\bee\label{inhtermsii}
2\pi i\, f_a(z)=\, \left.\Lbdry_a \, \pi(z,\zh)\right|_{\zh=\eta}\, .
\eee
In absolute cohomology the inhomogeneous term $f_a(z)$ is due to the fact that the bulk operators $\Lbulk_a$ satisfy 
\bee
\Lbulk_a\, \Om = d\beta\quad \Rightarrow \quad \Lbulk_a \int_{\Ga \in H_3(X,\IZ)} \Om = 0\ ,
\eee
where $d$ is the differential in the absolute setting. This is sufficient to annihilate the period integrals over cycles, as indicated on the right hand side of the above equation, but leads to boundary terms in the chain integral \eqref{DefNF}. In the absolute setting and based on Dwork-Griffiths reduction the inhomogeneous term $f_a(z)$ has been determined by a residue computation in ref.~\cite{Morrison:2007bm}. Here we see that the functions $f_a(z)$ are different derivatives of the surface period $\pi(z,\zh)$, restricted to the critical point. Hence, together with the bulk Picard-Fuchs operators, the surface period determine both the critical locus \eqref{critcon2} and the critical tension. 

\def\LD{\cx L^{\cx D}}
In the examples we find that the inhomogeneous terms $f_a(z)$ satisfy a hypergeometric differential equation as well:
\bee\label{Linh}
\cx L_a^{inh}\, f_a(z)=0\ .
\eee
The hypergeometric operators $\cx L_a^{inh}$ descend from the Picard-Fuchs operators $\LD$ of the surface, which annihilate the surface periods $\LD\pi(z,\zh)=0$.\footnote{For simplicity we suppress an index for distinguishing several Picard-Fuchs operators $\LD$.} Specifically, if $f_a(z)$ is non-zero, the operator $\cx L^{inh}_a$ can be defined as 
\bee
\cx L_a^{inh}=\left(\LD+[\Lbdry_a,\cx \LD]\, {\Lbdry_a}^{-1}\right)_{\zh=\eta}\ ,
\eee
where the operators on the right hand side are restricted to the critical point as indicated.

It follows from the above that the inhomogeneous terms $f_a(z)$ can be written as an infinite hypergeometric series in the closed-string moduli. However, on general grounds the $f_a(z)$ need to be well-defined over the open-closed moduli space, which simplifies {\it on-shell} to a finite cover of the complex structure moduli space $\cx M_{CS}(X)$ of the threefold \cite{Walcher:2009uj}. This implies that the hypergeometric series $f_a(z)$  can be written as rational functions in the closed string moduli and the roots of the extra equations defining the curves $C$.\footnote{We are grateful to Johannes Walcher for explaining to us this property of the inhomogeneous terms and for pointing out the results of ref.~\cite{AMS} on this issue.}

In the examples we observe that already the leading terms of the surface periods $\pi(z,\zh)$ become rational functions at the special symmetric points on the Noether-Lefshetz locus $\cx N$ in this sense. Hence there appears to be a connection between the enhancement of the Picard-lattice of the surface at these points, rationality of its periods and D-brane vacua. The rationality  property is preserved when acting with $\Lbdry$ in eq.~\eqref{inhtermsii} to obtain the inhomogeneous term $f_a$. In the examples we verify, that the contribution $f_a(C_{\al_\ell})$ of a particular boundary curve $C_{\al_\ell}$ to the inhomogeneous term can be written in closed form as follows.
\bee
f_a(C_{\al_\ell})\ =\ \frac{p_a(\psi,\al)}{q_a(\psi,\al)}|_{\al=\al_\ell(\psi)}\ = \ \frac{g_a(\psi,\al)}{\prod_i \Delta_l(C)^{\ga_i^a}}|_{\al=\al_\ell(\psi)}\, ,
\eee
where $p_a,q_a$ are polynomials in the variables $(\psi,\al)$. Here $\psi=\psi(z)$ is a short-hand for the fractional power of the closed string moduli $z$ appearing in the defining equation of the hypersurface $X$ and $\{\al_\ell\}$ are the roots of the extra equations defining the curves, with the root $\al_\ell$ corresponding to the component $C_{\al_\ell}$. Moreover, the zeros of the denominator appear only at the zeros of the components $\Delta_i(C)$ of the open-string discriminant, where different roots/curves coincide for special values of the moduli $\psi$. The exponents $\ga_i^a$ are some constants and $g_a(\psi,\al)$ some functions without singularities in the interior of the moduli space.

For Calabi-Yau hypersurfaces in toric varieties, the differential operators $\cx L_a$ can be derived from the GKZ type differential operators associated with the toric action on the ambient space \cite{Alim:2009bx,Alim:2009rf,Li:2009dz}. In particular, the holomorphic $(2,0)$ forms $\om^{(2,0)}$ on $D_1$ arise from the Lie derivatives of the holomorphic (3,0) form
$$
\om^{(2,0)}=i_{v_\hth} \Om|_{D_1}, 
$$
where $v_{\hth}$ is the vector field generating the toric $\IC^*$ action parametrized by $\zh$, e.g. $(x_1,x_2)\to (\la x_1,\la^{-1}x_2)$ in the above example. It is not hard to see that in the above situation, the differential operators for relative cohomology of ref.~\cite{Alim:2009bx} depending on the parameters $\zh$, reduce at $\zh=\eta$ to the type of differential operators derived in prop.~3.3. of ref.~\cite{Li:2009dz} in the absolute setting, i.e. without open-string deformations. Specifically, the derivative in the parameter $\hth$ becomes equivalent to the Lie derivatives in the direction of $x_1$ and $x_2$ at the critical point $\zh=\eta$ .

In the notation of refs.~\cite{Alim:2009rf,Alim:2009bx,Li:2009dz}, the GKZ system for the relative periods on $X$ (or equivalently the fourfold periods for F-theory compactification on $X_4$) are expressed in terms of {\it extended} charge vectors $l^a$ of the gauged linear sigma model (GLSM) associated with the brane geometry $(X,\cx D)$. In the final form these are given by\footnote{The same formula describes also the generalized hypergeometric operators of GKZ type for the closed-string compactification \cite{Batyrev:1994hm,Hosono:1993qy} and this will be used in the examples to determine the periods of the threefold $X$ and the surface $\cx D$ below. The distinction between the three different cases arises only from the different generators $l^a$, which encode the action of the gauge symmetry of the GLSM associated with the surface $\cx D$, the threefold $X$, the brane geometry $(X,\cx D)$ and the dual F-theory fourfold $X_4$, respectively, with the latter two cases having the identical generators in the decoupling limit of ref.~\cite{Alim:2009bx}.} 
\bee\label{gkz}
{\cx L}(l)=
\prod_{k=1}^{l_0}(\tha_0-k)\prod_{l_i>0}\prod_{k=0}^{l_i-1}(\tha_i-k)
-(-1)^{l_0} z_a \prod_{k=1}^{-l_0}(\tha_0-k)\prod_{l_i<0}\prod_{k=0}^{-l_i-1}(\tha_i-k)\,,
\eee
where $l$ is an arbitrary integral linear combination of the extended charge vectors $l^a$ and $\tha_i=a_i\frac{\p}{\p a_i}$ are logarithmic derivatives with respect to the parameters $a_i$ in the defining equations for $X_z$ and $D_{(z,\zh)}$. For details we refer to the examples in sect.~\ref{sec:ex} and to refs.~\cite{Alim:2009rf,Alim:2009bx,Aganagic:2009jq,Li:2009dz}. From the redundant parameters $a_i$ one may define torus invariant algebraic coordinates $z_a$ on the open-closed deformation space $\cx M$ by 
\bee\label{Defz}
z_a=(-)^{l^a_0} \prod_i a_i^{l^a_i}\ ,
\eee
where $l^a$, $a=1,...,\dim \cx M$ is a fixed choice of basis vectors. These describe the $h^{2,1}(X_z)$ complex structure moduli of $X_z$ and in addition the brane deformations $\zh$, providing coordinates on the fiber of $\hat{\cx M} \to \cx M$.
For appropriate choice of basis vectors $l^a$, solutions to the GKZ system can be written in term of the generating functions in these variables as
\begin{equation}\label{DefB}
B_{\{l^a\}}(z_a;\rho_a)=\sum_{n_1,...,n_N\in \IZ^+_0}\frac{\Gamma\left(1-\sum_a l_0^a(n_a+\rho_a)\right)}{\prod_{i>0}\Gamma\left(1+\sum_a l_i^a (n_a+\rho_a)\right)}\prod_a z_a^{n_a+\rho_a}\,.
\end{equation}
Under certain conditions discussed in refs.~\cite{Mayr:2001xk,Alim:2009bx,Aganagic:2009jq,Li:2009dz}, the extended GKZ system \eqref{gkz} for the relative periods for the brane compactification on the threefold $X$ can be associated also to the periods of the non-compact limit $X_4^\sharp$ of the dual fourfold $X_4$ for M/F-theory compactification. The solutions to this system then describe at the same time the relative period integrals $\ux\Pi$, which give rise to the leading term in eq.~\eqref{GVW}, and the periods of the non-compact 4-fold $X_4^\sharp$. A discussion of the quantum corrections in $g_s$, computed by the periods of the compact fourfold, can be found in ref.~\cite{Jockers:2009ti}.

\section{Examples}\label{sec:ex}
We proceed with the study of type II/F-theory superpotentials for a collection of examples of brane geometries on toric hypersurfaces with several open-closed string deformations. Combining the small Hodge variation associated with the surface periods \eqref{Defpi} and the GKZ system on the relative cohomology group \eqref{gkz} provides an efficient method to compute the integral relative period integrals and the mirror map for a large number of deformations. We obtain new enumerative predictions for the $A$ model expansion, consistent with the expectations, and study the behavior of the branes under extremal transitions between different topological manifolds through points with enhanced non-abelian gauge symmetries.

\newpag\subsection{Degree 12 hypersurface in $\IP_{1,2,2,3,4}$ \label{sec:X12234}}
The charge vectors of the GLSM for the $A$ model manifold are given by \cite{Hosono:1993qy}
\begin{center}
\bee\label{ls12234}\phantom{arrrgh}\eee\vskip-1.5cm
\begin{tabular}{cc|ccccccc}
& $\x_0$ & $\x_1$ & $\x_2$ & $\x_3$ & $\x_4$ & $\x_5$ & $\x_6$\\
\hline
$l^1$ & -6 & -1 & 1 & 1 & 0 & 2 & 3 \\
$l^2$ & 0 & 1 & 0 & 0 & 1 & 0 & -2 \\
\end{tabular} .
\end{center}
These vectors describe the relations between the vertices of a reflexive polyhedron described in app.~\ref{apptor}. Written in homogeneous coordinates of  $\IP_{1,2,2,3,4}$ the hypersurface constraint for the mirror manifold reads
\ben\label{W12234}
P &=&a_1x_1^{12}+a_2x_2^{6}+a_3x_3^{6}+a_4x_4^4+a_5x_5^{3}+a_0x_1 x_2 x_3 x_4 x_5 +a_6x_1^6x_4^2\\
&=& 
x_1^{12}+x_2^{6}+x_3^{6}+x_4^4+x_5^{3}+\psi\,x_1 x_2 x_3 x_4 x_5 +\phi\,x_1^6x_4^2\ .
\een
In the second equation the variables $x_i$ have been rescaled to display the dependence on the torus invariant parameters $\psi=z_1^{-1/6}z_2^{-1/4}$ and $\phi=z_2^{-1/2}\!$, with the $z_a$ given by \eqref{Defz}. On the mirror manifold, the Greene-Plesser orbifold group acts as $x_i\to \la_k^{g_{k,i}}\, x_i$ with weights\footnote{The other factors of the Greene-Plesser group give nothing new, using a homogeneous rescaling of the projective coordinates, e.g. for the factor generated by $g_3=(1,0,0,-1,0)$ with $\la_3^4=1$ one finds $g_3\sim g_1^3g_2^3$.}
\begin{equation}
\label{GP12234}
\IZ_6:\, g_1=(1,-1,0,0,0),\quad {\IZ_6}:\,g_2=(1,0,-1,0,0)\,,
\end{equation}
where we denote the generators by $\lambda_k$ with $\lambda_{1,2}^6=1$.
The closed-string periods near the large complex structure point can be generated by evaluating the functions
$
B_{\{l^a\}}(z_a;\rho_a)$ in \eqref{DefB}
and its derivatives with respect to $\rho_i$ at $\rho_1=\rho_2=0$ \cite{Hosono:1993qy}.

In this geometry we consider the set of curves defined by the equations
\ben
\label{C12234}
&& C_{\alpha,\pmk}=\{x_2=\eta x_3\;,\;x_4=\alpha x_1^3\;,\;x_5=\pmk \sqrt{\alpha\eta\psi} x_3x_1^2\}\,,\nn\\
&&\hskip1.5cm \eta^6=-1\,,\qquad \pmk^2=-1\,,\qquad\alpha^4+\phi\alpha^2+1=0\,.
\een
The labels $(\eta,\alpha,\pmk)$ are identified as $(\eta,\alpha,\pmk)\sim(\eta \lambda_1\lambda_2^{-1},\alpha \lambda_1^3\lambda_2^3,\pmk)$ under the orbifold group. In the following we choose to label each orbit of curves by $(\alpha,\pmk):= (e^{i\pi/6},\alpha,\pmk)$. Note that a rotation of $\eta$ corresponds to a change of sign for $\alpha$ in this notation, $(e^{3i\pi/6},\alpha,\pmk)=(-\alpha,\pmk)$. Instead of choosing a fixed $\eta$ we can also fix the sign of $\alpha$ and keep two choices for $\eta^3$.

To calculate the domain wall tensions and the superpotentials for the vacua  $C_{\alpha_1,\pmk}$ and $C_{\alpha_2,\pmk}$ we will study two families of divisors. The family $Q(\cx D_{1})=x_2^{6}+\zh x_3^{6}$ interpolates between vacua related by a sign flip of $\eta^3$ or of the root $\al$ of the quartic equation. The family $Q(\cx D_{2})=x_4^{4}+\zh x_1^{6}x_4^2$ interpolates between any two different roots $\al$.

\subsubsection*{\it First divisor}
We start with the analysis of the divisor 
\begin{equation}\label{D112234}
Q(\cx D_{1})=x_2^{6}+z_3 x_3^{6}\,.
\end{equation}
To obtain some geometrical understanding of the surface defined by the intersection $P=0=Q(\cx D_{1})$ we explicitly solve for $x_3=(-z_3)^{-1/6}x_2$ and rescale $x_2$ to find
\begin{equation}\label{D1W12234}
P_{\cx D_1}=x_1^{12}+x_2^6+x_4^4+x_5^3+\ti \psi x_1x_2^2x_4x_5+\phi x_1^6x_4^2\,.
\end{equation}
Here $\ti \psi = u_1^{-1/6}u_2^{-1/4}$, $\phi=u_2^{-1/2}$ are expressed in terms of the previous parameters as
\begin{equation}\label{D1M12234}
u_1=-\frac{z_1}{z_3}(1-z_3)^2,\qquad u_2=z_2\,.
\end{equation}
Changing coordinates to $\ti x_2=x_2^2$ displays the family $\cx D_1$ as a double cover of a family of toric K3 surfaces associated to a GLSM with charges
\begin{center}
$\cx D_1:\qquad $\begin{tabular}{cc|ccccccc}
& $\x_0$ & $\x_1$ & $\x_2$ & $\x_4$ & $\x_5$ & $\x_6$ & \\
\hline
$\ti l^1$ & -6 & -1 & 2 & 0 & 2 & 3 &\\
$\ti  l^2$ & 0 & 1  & 0 & 1 & 0 & -2 &\\ 
\end{tabular}
\end{center}

\vskip-1.5cm\bee\label{D1ls}\eee\vskip0.5cm

\noi and with the two algebraic K3 moduli \eqref{D1M12234}. The two covers are distinguished by a choice of sign for $x_2$.

The family of algebraic K3 manifolds obtained from \eqref{D1W12234} by the variable change $\ti x_2=x_2^2$ generically has four parameters with the two extra moduli multiplying the monomials $x_1^3x_4^3$ and $x_1^9x_4$. Since these terms are forbidden by the Greene-Plesser group of the Calabi-Yau threefold, the embedded surface is at a special symmetric point with the coefficients of these monomials set to zero. The periods on the K3 surface at this point can be computed from the GKZ system for the two parameter family, obtained from \eqref{gkz} with the charge vectors $\{\ti l\}$ in eq.~\eqref{D1ls}:
\ben
\cx L^{\cx D}_1&=&\ti\th_1 (2 \ti\th_1-1)\prod_{k=0}^2 (-3 \ti\th_1+2 \ti\th_2+k) -\frac{9}{2} u_1  (\ti\th_1-\ti\th_2)\hskip-6pt \prod_{k=1,2,4,5}(6 \ti\th_1+k)\,,\nn\\
\cx L^{\cx D}_2&=&\ti\th_2 (\ti\th_2-\ti\th_1)-u_2(2\ti\th_2-3\ti\th_1)(2\ti\th_2-3\ti\th_1+1)\,,
\een 
where $\ti \th_a=u_a\tfrac{d}{du_a}$.
Apart from the regular solutions this system has two extra solutions depending on fractional powers in the $u_i$:
\def\nnorm#1{\footnote{$\bigstar\bigstar$ {\bf #1}$\bigstar\bigstar$ }}
\begin{align}\label{D1P12234}
\taun_1(u_1,u_2)&=\frac{c_1}{2} \,B_{\{\ti l\}}(u_1,u_2;\tfrac{1}{2},0)
=\frac{4c_1}{\pi } \sqrt{u_1}\ _2F_1(-\tfrac{1}{4},-\tfrac{3}{4},\tfrac{1}{2},4u_2)+\cx O(u_1^{3/2})\,,\nn \\
\taun_2(u_1,u_2)&=\frac{c_2}{2} \,B_{\{\ti l\}}(u_1,u_2;\tfrac{1}{2},\tfrac{1}{2})
=\frac{12c_2}{\pi }\sqrt{u_1u_2}\ _2F_1(-\tfrac{1}{4},\tfrac{1}{4},\tfrac{3}{2},4u_2)+\cx O(u_1^{3/2})\,.
\end{align}
Here $c_a$ are some normalization constants not determined by the differential operators. Later they will be fixed to one by studying the geometric periods on the surface. 

As indicated, the exceptional solutions vanish at the critical point $u_1=0$ as  $\sim\sqrt{u_1}$, with the coefficient a hypergeometric series in the modulus $u_2=z_2$. These solutions arise as the specialization of the standard solutions of the four parameter family of K3 manifolds to the special symmetric point.\footnote{An explicit illustration of this fact is given in the case of the second family of divisors below.} Since $u_1=0$ is not at the discriminant locus of the K3 family for general $u_2$, there is no geometric vanishing cycle associated with the zero of $\taun_{1,2}$. Instead the zero at $u_1=0$ arises from the 'accidental' cancellation between the volumes of different classes at the symmetric point.\footnote{One parameter controlling the difference of these volumes is the direction of the off-shell modulus.} The periods \eqref{D1P12234} have the special property that their leading terms $\sim \sqrt{u_1}$ near the critical point $u_1=0$ can be written in closed form as 
\bee\label{k3rat12234}
\lim_{z_3\to 1}\frac{\taun_a(u_1,u_2)}{(1-z_3)}=\frac{4c_a}{\pi } \cdot \left.\frac{(i\al)(2\al^2-\phi)(\al^2+\phi)}{\psi^3}\right|_{\al=\al_{a,+}}\ ,
\eee
where
\begin{small}\bee\label{RD112234}
\al_{1,\pm}=\pm\sqrt{\frac{-\phi+\sqrt{\phi^2-4}}{2}}\,,\qquad
\al_{2,\pm}=\pm\sqrt{\frac{-\phi-\sqrt{\phi^2-4}}{2}}\,,
\eee\end{small}
denote the roots of the quartic equation $\al^4+\phi\al^2+1=0$ appearing in the definition \eqref{C12234}. Hence the leading part of the two K3 periods near  the symmetric point is proportional to a rational function in the coefficients of the defining equations for the curve, evaluated at the critical points.

We will first compute the domain wall tensions by integrating the periods $\taun_{1,2}$ of the surface $D_1$. Note that the K3 periods $\taun_a$ depend on $\xi = \sqrt{z_3}$ via their dependence on $u_1$ and the sign of the square root correlates with the sign of $\al$.
To obtain the off-shell tension, we integrate $\taun_a(\xi)$ as
\begin{equation}\label{Int12234}
\cx T_a^{(\pm)}(z_1,z_2,z_3)
=\frac{1}{2\pi i}\int_{\xi_0}^{\pm \sqrt{z_3}} \taun_a(\xi)\, \frac{d\,\xi}{\xi}\,,
\end{equation}
where $\xi_0$ denotes a fixed reference point. For example, the period $\taun_1$ integrates to 
\ben
&{\displaystyle \frac{4\pi i\,\cx T_1^{(\pm)}}{c_1}}=\int_{\xi_0}^{\pm\sqrt{z_3}} \sum_{n_1,n_2\geq 0} \frac{\Gamma\left(4+6 n_1\right)  \left(-\frac{z_1}{\xi^2}(1-\xi^2)^2\right)^{n_1+\frac{1}{2}}z_2^{n_2}}{\Gamma\left(2+2 n_1\right)^2\Gamma\left(1+ n_2\right)\Gamma\left(\frac{1}{2}- n_1+n_2\right)\Gamma\left(\frac{5}{2}+3 n_1-2n_2\right)} \frac{d\xi}{\xi}\nn\\
&=\ \left.\!\!\!\!\!\sum_{n_1,n_2\geq 0}\!\frac{\Gamma\left(4+6 n_1\right)(-z_1)^{n_1+\frac{1}{2}} z_2^{n_2}  \left(\xi^2-1\right)^{2n_1+2}
{}_2F_1\left(1,\frac{3}{2}+n_1,\frac{1}{2}-n_1,\xi^2\right) }
{\left(1+2n_1\right) \Gamma\left(2+2 n_1\right)^2\Gamma\left(1+ n_2\right)\Gamma\left(\frac{1}{2}- n_1+n_2\right)\Gamma\left(\frac{5}{2}+3 n_1-2n_2\right)\xi^{2n_1+1}} \right|_{\xi=\xi_0}^{\xi=\pm\sqrt{z_3}}\!\!
\een
where the contribution from the reference point $\xi_0$ can be set to zero by choosing $\xi_0=i$ as the lower bound. This will be used to split the result of  the integral for the domain wall tension into two contributions of the superpotentials from the endpoints as in eq.~\eqref{TWsplit}. This split is not obvious in general, and ambiguous with respect to adding rational multiples of bulk periods. In the example we can use the $\IZ_2$ symmetry acting on the curves to require that the superpotentials obey $\cx W_1^{(+)} = -\cx W_1^{(-)}$. With this convention and the particular choice of $\xi_0$ above, we obtain $\frac{1}{2\pi i}\int_{\xi_0}^{\pm \sqrt{z_3}} \taun_a(\xi)\, \frac{d\,\xi}{\xi}=\cx W_a^{(\pm)}$ or $\frac{1}{2\pi i}\int_{- \sqrt{z_3}}^{+ \sqrt{z_3}} \taun_a(\xi)\, \frac{d\,\xi}{\xi}=\cx W_a^{(+)}-\cx W_a^{(-)}=2\cx W_a^{(+)}$.

According to the discussion in sect.~\ref{sec:RelCoh}, the superpotentials $\cx W_a^{(\pm)}(z_1,z_2,z_3)$ restrict to the on-shell superpotentials $W_a^{(\pm)}(z_1,z_2)$ with vanishing derivative in the open-string direction $z_3$ at the critical point:
\bee\label{Trel}
W^{(\pm)}_a(z_1,z_2)=\left.\cx W^{(\pm)}_a\right|_{z_3=1}, \qquad 
\left.\xi\p_\xi \cx W^{(\pm)}_a(z_1,z_2,\xi^2)\right|_{z_3=1}=\pm \frac{1}{2\pi i}\, \left.\taun_a\right|_{u_1=0}=0\,.
\eee
For the above integrals one obtains
\begin{align}\label{D1T12234}
W_1^{(\pm)}&=\mp\frac{c_1}{8\pi}\!\!\! \sum_{n_1,n_2\geq 0}\frac{(-1)^{n_1+1}\,\Gamma \left(-n_1-\frac{1}{2}\right) \Gamma \left(6 n_1+4\right)\,z_1^{n_1+\frac{1}{2}} z_2^{n_2}}{\Gamma \left(n_1+\frac{3}{2}\right)
\Gamma \left(2 n_1+2\right) \Gamma \left(3 n_1-2 n_2+\frac{5}{2}\right) \Gamma \left(n_2+1\right) \Gamma \left(-n_1+n_2+\frac{1}{2}\right)}\,,\\
W_2^{(\pm)}&=\mp\frac{c_2}{8\pi}\!\!\! \sum_{n_1,n_2\geq 0}\frac{(-1)^{n_1+1}\,\Gamma \left(-n_1-\frac{1}{2}\right) \Gamma \left(6 n_1+4\right)\,z_1^{n_1+\frac{1}{2}} z_2^{n_2+\frac{1}{2}}}{\Gamma
\left(n_1+\frac{3}{2}\right) \Gamma \left(2 n_1+2\right) \Gamma \left(3 n_1-2 n_2+\frac{3}{2}\right) \Gamma \left(n_2+\frac{3}{2}\right)
\Gamma \left(-n_1+n_2+1\right)}\,.\nn
\end{align}
These functions can be expressed in terms of the bulk generating function as
\begin{align}\label{D1T1B12234}
W_1^{(\pm)}=\mp \frac{c_1}{8}\,B_{\{ l\}}\left(z_1,z_2;\tfrac{1}{2},0\right)\,,\qquad 
W_2^{(\pm)}=\mp \frac{c_2}{8}\,B_{\{ l\}}\left(z_1,z_2;\tfrac{1}{2},\tfrac{1}{2}\right)\,.
\end{align}

Complementary, the tensions $\cx T_a^{(\pm)}(z_1,z_2,z_3)$ and their on-shell restrictions $T_a^{(\pm)}(z_1,z_2)$ can be described as solutions to the large GKZ system for the relative cohomology problem derived in refs.~\cite{Alim:2009rf,Alim:2009bx,Li:2009dz}. For the family \eqref{D112234} the additional charge vector is 
\begin{center}
\begin{tabular}{cc|cccccc|ccc}
& $\x_0$ & $\x_1$ & $\x_2$ & $\x_3$ & $\x_4$ & $\x_5$ & $\x_6$ & $\x_7$ & $\x_8$\\
\hline
$l^3$ & 0 & 0 & 1 & -1 & 0 & 0 & 0 & -1 & 1
\end{tabular} .
\end{center}
Together with the charge vectors $l^1$ and $l^2$ for the Calabi-Yau hypersurface this defines the extended hypergeometric system of the form \eqref{gkz}, which can be associated with a dual fourfold $X_4$ for a M/F-theory compactification \cite{Alim:2009rf,Aganagic:2009jq,Jockers:2009ti}. For a description of $X_4$ as a toric hypersurface we refer to app.~\ref{apptor}. From the extended charge vectors one obtains after an appropriate factorization the system of differential operators\footnote{The first operator is obtained after a factorization similar to the one described in ref.~\cite{Hosono:1993qy} for the underlying threefold.} 
\begin{align}
\mathcal{L}_1 &= (\theta_1+\theta_3)(\theta_1-\theta_3)(3\theta_1-2\theta_2)-36z_1(6\theta_1+5)(6\theta_1+1)(\theta_2-\theta_1+2z_2(1+6\theta_1-2\theta_2))\,,\nn\\
\mathcal{L}_2 &= \theta_2(\theta_2-\theta_1)-z_2(3\theta_1-2\theta_2-1)(3\theta_1-2\theta_2)\,,\nn\\
\mathcal{L}_3 &= \theta_3(\theta_1+\theta_3)+z_3\theta_3(\theta_1-\theta_3)\,.
\end{align}
After a simple variable transformation $y=\ln(z_3)$, with the variable $y$ centered at the critical point, the solutions to this system describe the expansion of the periods on the relative homology $H^3(Z^*,\cx D_1)$ around the critical point. These include the off-shell tensions $\cx T^{(\pm)}_a(z_1,z_2,z_3)$ \eqref{Int12234}, which restrict to the functions \eqref{D1T1B12234}, and in addition the closed-string periods $\Pi(z_1,z_2)$. The integration from the geometric surface periods of the subsystem fixes the $z_3$-dependent piece. The GKZ system restricts the afore mentioned integration constant to a linear combination of the closed-string periods $\Pi(z_1,z_2)$. The rational coefficients appearing in this combination can be determined by a monodromy argument, as in ref.~\cite{Walcher:2006rs} and as exemplified for a non-compact limit of the Calabi-Yau threefold in sect.~\ref{sec:AJ}. 

Finally one may also characterize the critical tensions $T^{(\pm)}_a$, or, for the above reasons also the critical superpotentials $W^{(\pm)}_a$, as the solution to the inhomogeneous Picard-Fuchs equation \eqref{InhPF}, which makes contact to the normal function approach of \cite{Morrison:2007bm}. Due to
\bee
{\cx L}_1=\Lbulk_1(\th_1,\th_2)-(3\th_1-2\th_2)\th_3^2,\qquad 
{\cx L}_2=\Lbulk_2(\th_1,\th_2),\qquad 
\eee
we observe that only the first operator may acquire a non-zero inhomogeneous term at the critical point. This term is determined by the leading behavior of the surface periods $\taun_a$ in the limit $u_1\to 0$. Acting with $\Lbdry_1=(3\th_1-2\th_2)\th_3$ on the terms on the right hand side of  eqs.~\eqref{D1P12234} one obtains the inhomogeneous Picard-Fuchs equations 
\ben\label{IH12234a}
\begin{aligned}
\Lbulk_1 W_1^{(\pm)}&=\mp \frac{3 c_1}{2\pi^2}\, \sqrt{z_1}\ _2F_1(\tfrac{1}{4},-\tfrac{1}{4},\tfrac{1}{2},4z_2)
= f_1(\al_{1,\pm})\ ,\\[3mm]
\Lbulk_1 W_2^{(\pm)}\,&=\,\mp \frac{3 c_2}{2\pi^2}\, \sqrt{z_1z_2}\ _2F_1(\tfrac{3}{4},\tfrac{1}{4},\tfrac{3}{2},4z_2)
=  f_1(\al_{2,\pm}) \ ,
\end{aligned}
\een
while $\Lbulk_2\, W_a^{(\pm)}=0$. The roots \eqref{RD112234} of the quartic equation are identified with the label $(a,\pm)$ of the curves  in the right hand side of eq.~\eqref{IH12234a}. Indeed, as a consequence of eq.~\eqref{k3rat12234}, the inhomogeneous terms can again be written in closed form as $$\Lbulk_a W(\al) = f_a(z,\al)\,,$$ with $W(\al_{a,\pm})=W_a^{(\pm)}$ and the $f_a(z,\al)$ rational functions in the coefficients of the defining equation:
\ben\label{IH12234b}
f_1(z,\al)=\frac{3 c}{2\pi^2} \cdot \frac{i\, \phi \,\al(\al^2+\phi)}{\psi^3}  \,,\qquad f_2(z,\al)=0\,,
\een
for $c=c_1=c_2$. As is apparent from \eqref{IH12234a}, this function satisfies a hypergeometric equation $\cx L^{inh} f_1=0$. The hypergeometric operator is related to the surface operators by eq.~\eqref{Linh}. In the present case, the relevant operator arises from $\LD_2$, that is  $\cx L^{inh}=(\LD_2+[\Lbdry_1,\LD_2]{\Lbdry_1}^{-1})|_{\zh_{crit}}$, while $\LD_1$ becomes irrelevant. With
$$
\LD_2 |_{\zh_{crit}}=\th_2(\th_2-\tfrac{1}{2})-4z_2(\th_2-\tfrac{1}{4})(\th_2-\tfrac{3}{4})\,,\qquad
\Lbdry_1|_{\zh_{crit}}=i(\th_2-\tfrac{3}{4})\,,
$$
one obtains 
\bee\cx L^{inh}=\th_2(\th_2-\tfrac{1}{2})-4z_2(\th_2-\tfrac{1}{4})(\th_2+\tfrac{1}{4})\,.
\eee
In the above we have used that the relevant surface period is the solution to the Picard-Fuchs system $\{\LD_b\}$ with index $\tfrac{1}{2}$ in the variable $u_1$ to set $\ti\th_1=\tfrac{1}{2}$.

\subsubsection*{\it A-model expansion}
By mirror symmetry, these functions should have an integral instanton expansion when expressed in terms of the appropriate coordinates and taking into appropriately the contributions from multi-covers \cite{Ooguri:1999bv}. For the critical branes at fixed $\zh$, we use the modified multi-cover formulae of the type proposed in refs.~\cite{Walcher:2006rs,Pandharipande:2008,Walcher:2009uj}: 
\bee\label{MC2}
\frac{W^{(\pm)}_1(z(q))}{\omega_0(z(q))}=\frac{1}{(2\pi i)^2}\sum_{k\, odd}\sum_{d_1\, odd \atop d_2\geq 0} n^{(1,\pm)}_{d_1,d_2}\, \frac{q_1^{k d_1/2}\,q_2^{k d_2}}{k^2}\,,
\eee
\bee\label{MC22}
\frac{W^{(\pm)}_2(z(q))}{\omega_0(z(q))}=\frac{1}{(2\pi i)^2}\sum_{k\, odd}\sum_{d_1\, odd \atop d_2\,odd} n^{(2,\pm)}_{d_1,d_2}\, \frac{q_1^{k d_1/2}\,q_2^{k d_2/2}}{k^2}\,.
\eee
In this way one obtains the integer invariants in Tab.~\ref{Tab12234} for $c_a=1$. As can be guessed from these numbers, the  superpotentials for $a=1,2$ are in fact not independent, but related by a $\IZ_2$ symmetry. The family of Calabi-Yau hypersurfaces \eqref{W12234} develops a singularity at the discriminant locus $\Delta=1-4z_2=0$, which is mirror to a curve of $A_1$ singularities \cite{Katz:1996ht,Klemm:1996kv}. 
On the $B$ model side the $\IZ_2$ monodromy around the singular locus $\Delta=0$ exchanges the two sets of roots $\al_{1,\pm}$ and $\al_{2,\pm}$ in eq.~\eqref{RD112234}. Accordingly, the superpotentials $W_1^{(\pm)}$ and $W_2^{(\pm)}$ are also exchanged as can be seen from the structure of the inhomogeneous terms. On the level of periods this monodromy action yields 
\begin{equation}\label{Sy12234}
t_1\to t_1+3 t_2,\qquad t_2\to -t_2\ .
\end{equation}
As a result the invariants of $W_{2}$ are related to that of $W_1$ by the $\IZ_2$ quantum symmetry $q_1\to q_1q_2^3,\ q_2\to q_2^{-1}$ generated by \eqref{Sy12234}.\footnote{The $\IZ_2$ symmetry is also realized on the closed-string invariants, see the results of ref.~\cite{Hosono:1993qy}.}

\begin{table}
\begin{tiny}\hskip -1cm\vbox{\begin{center}
\begin{center}
$n^{(1,+)}_{d_1,d_2}$\\[2mm]
\begin{tabular}{|r|cccccc|}
\hline
$\!\!q_1^{1/2} \backslash q_2\!\!\!$&$0$&1&2&3&4&5\\[0.3mm]
\hline
1 & 16 & 48 & 0 & 0 & 0 & 0 \\
3&-432 & -480 & 38688 & 10800 & 0 & 0 \\
5&45440 & -78192 & 5472 & 92812032 & 146742768 & 26162880 \\
7&-7212912 & 25141920 & -165384288 & 61652832 & 327357559584 & 1094178697056 \\
9&1393829856 & -6895024080 & 49628432160 & -426927933792 & 261880092960 & 1383243224519472 \\
11&-302514737008 & 1905539945472 & -14487202588320 & 131586789107520 & -1448971951799232 & 1383991826496480 \\
13&70891369116256 & -538859226100800 & 4335978084777792 & -39691782337561536 & 440278250387930640 & -5799613460160838608 \\
15& -17542233743427360 & 155713098595732704 & -1328641212531217728 & 12308540119113753936 & -132576278776141577664 & 1710971659352271824160 \\
\hline
\end{tabular}\\[2mm]
$n^{(2,+)}_{d_1,d_2}$\\[2mm]

\begin{tabular}{|r|rcccccc|}
\hline
$q_1^{1/2} \backslash q_2^{1/2}$&$\!\!\!1$&3&5&7&9&11&13\\[0.3mm]
\hline
1& 48 & 16 & 0 & 0 & 0 & 0 & 0 \\
3& 0 & 10800 & 38688 & -480 & -432 & 0 & 0 \\
5& 0 & 82080 & 26162880 & 146742768 & 92812032 & 5472 & -78192 \\
7& 0 & -10780160 & 241323840 & 88380335472 & 702830702688 & 1094178697056 & 327357559584 \\
9& 0 & 1843890480 & -36172116480 & 932346639840 & 364829042312640 & 3751178206812144 & * \\
11& 0 & -369032481792 & 6979488962400 & -143329914498240 & 4246347124847520 & * & * \\
\hline
\end{tabular}
\end{center}\end{center}}\end{tiny}
\caption{\label{Tab12234} Disc invariants for the on-shell superpotentials $W_{a}^{(+)}$ of the threefold $\IP_{1,2,2,3,4}[12]$.}
\end{table}

\subsubsection*{\it Extremal transition and a non-compact limit}
The above results and the normalization obtained by integration from the subsystem can be verified by taking two different one parameter limits. At the singular locus $\Delta=0$, there is an extremal transition to the one parameter family mirror to a degree (6,4) complete intersection hypersurface in $\IP_{1,1,1,2,2,3}$. 
From eq.~\eqref{Sy12234} it follows that the transition takes place at $q_2=1$, predicting the relation 
\bee
\sum_{\ell=0}^{3k}n^{(a,+)}_{k,\ell} (\IP_{1,2,2,3,4}[12])=n_k(\IP_{1,1,1,2,2,3}[6,4])\ , \quad a=1,2 \ , 
\eee
where $(k,\ell)$ denote the degree in $q_1$ and $q_2$, respectively. The finiteness of the sum over $\ell$ follows from the symmetry \eqref{Sy12234}. From the left hand side of the above equation one gets 
\bee
n_k=64,\ 48\,576,\ 265\,772\,480,\ 2\,212\,892\,036\,032,\ 22\,597\,412\,764\,939\,776,\
\ldots 
\eee
for the first invariants of $\IP_{1,1,1,2,2,3}[6,4]$. This can be checked by a computation for the complete intersection manifold with the inhomogeneous Picard-Fuchs equation 
\bee
\cx L\ W(z)=\frac{4\sqrt{z}}{(2\pi i)^2}\ ,\qquad \cx L=\th^4 -48z(6\th+5)(6\th+1)(4\th+3)(4\th+1)\ .
\eee

Another interesting one modulus limit is obtained for $z_2\to 0$, where $X$ degenerates to the non-compact hypersurface 
\ben
X^\flat:\ &&y_1^2+y_2^3+y_3^6+y_4^6+y_5^{-6}+\hat\psi\,y_1y_2y_3y_4y_5=0,\qquad \hat\psi =\frac{\psi}{\sqrt{\phi}}=z_1^{-1/6}\,
\een
in weighted projective space $\IP^4_{3,2,1,1,-1}$, with the new variables $y_i$ related to the $x_i$ by 
$$
y_1=\phi^{1/2}x_4x_1^3,\ y_2=x_5,\ y_3=x_2,\ y_4=x_3,\ y_5= x_1^{-2}\ .
$$ 
The non-compact 3-fold $X^\flat$ is a local model for a certain type of singularity associated with the appearance of non-critical strings and has been studied in detail in ref.~\cite{Lerche:1996ni}. 

In this limit the curves $C_{\al_{2,\pm},\pmk}$ of eq.~\eqref{C12234} are pushed to the boundary of the local threefold geometry $X^\flat$ and the domain wall tension between $C_{\al_{2,+},\pmk}$ and $C_{\al_{2,-},\pmk}$ becomes independent of the modulus $z_1$, which is reflected by the fact that all the disc invariants of $W_2$ vanish in the limit $z_2\rightarrow 0$.  The curves $C_{\al_{1,\pm},\pmk}$ become
\bee \label{LimC12234}
C_{\varepsilon,\pmk}^\flat\,=\,
\left\{ y_3 = \eta\, y_4 \,,\  y_1 y_5^3 = \varepsilon \,, \ y_2y_5 = \pmk\,y_4\sqrt{\varepsilon\eta \hat\psi} \right\} \, , 
\quad \varepsilon=\pm i \, , \ \pmk =\pm i \, ,
\eee
where $\varepsilon=\pm i$ distinguishes between the two roots $\al_{1,+}$ and $\al_{1,-}$. In app.~\ref{sec:AJ} we show, that the 3-chain integral representing the domain wall tension in $X^\flat$ descends to an Abel-Jacobi map on a Riemann surface, which can be computed explicitly as an geometric integral. The invariants $n^{[6]}$ obtained for the superpotential in the non-compact geometry $X^\flat$ are reported in app.~\ref{sec:AJ} and they agree with the $q_2^0$ term of $T_1$, $n_{k,0}=n^{[6]}_k$.

\subsubsection*{\it A second family of divisors and symmetric K3s}
The same critical points can be embedded into a different family of divisors 
\begin{equation}\label{12234D2}
Q(\cx D_{2})=x_4^{4}+z_3z_2^{-1/2} x_1^{6}x_4^2\,.
\end{equation}
Our motivation to consider this second family in detail is two-fold. Firstly, the Hodge problem on the surface is equivalent to that of a two parameter family of K3 surfaces at a special point in the moduli, which can be studied explicitly without too many technicalities. We will explicitly show that the relevant zero of the period vector arises at an orbifold point of the K3, which has been interpreted as a point with a half-integral $B$-field for the closed-string compactification on the local geometry \cite{Aspinwall:1993xz}. Secondly, this family tests a different direction of the off-shell deformation space of the brane, leading to a different off-shell superpotential $\cx W$ for the deformation~\eqref{12234D2}. However, since the family contains the curves $C_{\alpha,\pmk}$ for  $z_3=-\alpha^2z_2^{1/2}$, the critical superpotential has to be the same as the one obtained for the family $\cx D_1$ in eq.~\eqref{D1T1B12234}. The agreement with the previous result and normalization gives an explicit illustration of the fact that different parametrizations of the off-shell directions, corresponding to a different choice of light fields represented by different relative cohomologies, fit together consistently near the critical locus.

As the critical point is determined by the vanishing condition \eqref{critcon2}, we again study the subsystem $P=Q(\cx D_{2})=0$ . Solving for $x_4$ and changing coordinates to $\ti x_1=x_1^4$, the surface can be described as a cover\footnote{The change from $x_1$ to $\ti x_1$ gives a fourfold cover acted on by a remaining $\IZ_2$ action generated by $g_1$ in \eqref{GP12234}.} of a mirror family of K3 hypersurfaces
$$
\ti x_1^3+x_2^6+x_3^6+x_5^3+\ti \psi \ti x_1x_2x_3x_5+\ti \phi (x_2x_3)^3=0\ .
$$
Here $\ti \psi^{-6}:=u=-\frac{z_1 z_2}{z_3^3}(z_2-z_3+z_3^2)^2$ and the parameter $\ti \phi$ is zero for the embedded surface. 
At $\ti \phi=0$, the GLSM for this family is defined by the charges 
\begin{center}
\begin{tabular}{cc|cccc}
& $\x_0$ & $\x_1$ & $\x_2$ & $\x_3$ & $\x_5$\\
\hline
$\ti l$ & -6 & 2 & 1 & 1 & 2 \\
\end{tabular} .
\end{center}
The GKZ system for this one modulus GLSM has an exceptional solution
\begin{equation}\label{D2tau12234}
\taun(u)=\frac{c}{2}\;B_{\{\ti l\}}(u;\tfrac{1}{2})=\frac{c}{2} \sum_{n=0}^{\infty}\frac{ \Gamma (4+6 n)}{ \Gamma (2+2n)^2 \Gamma
(\frac{3}{2}+n)^2}u^{n+\frac{1}{2}}\,,
\end{equation}
that vanishes at the critical point $u=0$. To get a better understanding of this solution and of the integral periods on the surface, one may describe $\taun$ as a regular solution of the two parameter family of K3 surfaces parametrized by $\ti \psi$ and $\ti \phi$, restricted to the symmetric point $\ti \phi=0$. The charges of the GLSM for the two parameter family of K3 manifolds are 
\begin{center}\vskip-0.2cm
\begin{tabular}{cc|ccccc}
&$\x_0$ & $\x_1$ & $\x_2$ & $\x_3$ & $\x_4$ & $\x_5$ \\
\hline
$\ti l^1$& -3 & 1 & 0 & 0 & 1 & 1 \\
$\ti l^2$& 0 & 0 & 1 & 1 & 0 & -2
\end{tabular} .
\end{center}
The two algebraic moduli of this family are $v_1=-\ti \phi \ti \psi^{-3}$ and $v_2=\ti \phi^{-2}$ and these are related to the single modulus of the embedded surface by $u=\ti \psi^{-6}=v_1^2v_2$. The principal discriminant locus for this family has the two components  
$$
\Delta=\Delta_0\cdot \Delta_1=(1+54v_1+729v_1^2-2916v_1^2v_2)\cdot(1-4v_2)\ .
$$

The periods near $\ti \phi=0$ can be computed in the phase of the two parameter GLSM with coordinates $u_1=v_1v_2^{1/2}$ and $u_2=v_2^{-1/2}$. The hypergeometric series
\begin{equation}\label{K32112Pera}
\tilde{\taun}(u_1,u_2)= \frac{c}{2\pi^2}\sum_{n=0}^{\infty}\sum_{p=0}^{\infty}\frac{ \Gamma (1+3 n) \Gamma
(\frac{1}{2}-n+p)^2}{  \Gamma (1+n)^2 \Gamma (2-n+2
p)}u_1^n u_2^{1+2p-n}\
\end{equation}
is a solution of the Picard-Fuchs equation that restricts to $\taun(\sqrt u)$ in the limit $u_2=0$. This series can be expressed with the help of a Barnes type integral as
\begin{align}
\tilde{\taun}(u_1,u_2)= &-\frac{c}{2\pi^2}\int_{\cx C_+} \sum_{n=0}^{\infty}\frac{ \Gamma (1+3 n) \Gamma
(\frac{1}{2}+s)^2 \Gamma(1+s)\Gamma(-s)(-1)^s}{  \Gamma (1+n)^2 \Gamma (2+n+2
s)} (u_1u_2)^n u_2^{1+2s} \\ &+  
\frac{c}{2\pi^2} \sum_{n=0}^{\infty}\sum_{p=1}^{\infty}\frac{ \Gamma (1+3 n) \Gamma
(\frac{1}{2}-p)^2}{  \Gamma (1+n)^2 \Gamma (2+n-2
p)}(u_1u_2)^n u_2^{1-2p}
\,,
\end{align}
where the contour $\cx C_+$ encloses the poles of the Gamma functions on the positive real line including zero. To relate the special solution $\ti \taun(u_1,u_2)$ to the integral periods on the K3, one may analytically continue it to large complex structure by closing the contour to the left and obtains 
\ben \label{K32112Per}
\tilde{\taun}(v_1,v_2)&=&\displaystyle \frac{c}{2\pi i} \sum_{n,p=0}^{\infty}\frac{\Gamma (1+3 n) v_1^n v_2^{p}\left(-i\pi +\ln(v_2)+2(\Psi(1+n-2p)-\Psi(1+p))\right)}{\Gamma (1+n)^2 \Gamma (1+n-2p)\Gamma(1+p)^2}\nn\\
&=& c\, \om_0\, \left(t_2^{K3}-\frac{1}{2}\right)\ .
\een
Here $\om_0=B_{\ti l}(v_a;0,0)$ is the fundamental integral period at large volume, and $t_2^{K3}=(2\pi i \om_0)^{-1}\p_{\rho_2} B_{\ti l}|_{\rho_a=0}$ is the integral period associated with the volume of another 2-cycle $C$, which is mirror to the base of the elliptic fibration defined by the GLSM of the $A$ model side. 

From the last expression it follows that the zero of the K3 period vector associated with the D-brane vacuum arises at the locus 
\bee
J^{K3}={\rm Im}\, t_2^{K3} = 0\ ,\qquad B^{K3}={\rm Re}\, t_2^{K3} =\tfrac{1}{2} \ ,
\eee
which, in the closed string compactification on this local K3 geometry, is interpreted as a 2-cycle of zero volume with a half-integral $B$-field. Indeed, in the limit $u=0=u_1$, eq.~\eqref{K32112Pera} becomes 
$$
\tilde{\taun}(u_1,u_2)|_{u_1=0}\,\sim\, \ln\left(\frac{1-2v_2-\sqrt{1-4 v_2}}{2v_2}\right)-i\pi\,,
$$
expanded around $v_2=\infty$. The first term on the right hand side is the period for the compact cycle of the $\IC^2/\IZ_2$-quotient singularity studied in ref.~\cite{Aspinwall:1993xz}, which is zero on the discriminant locus $\Delta_1=0$, but a constant at $v_2=\infty$. The zero associated with the critical point hence does not appear on the principal discriminant, but at an orbifold point with non-vanishing complex quantum volume. It has been argued in refs.~\cite{Walcher:2006rs,Morrison:2007bm}, that the $A$~model data associated with the critical points of the present type include $\IZ_2$-valued open-string degrees of freedom from the choice of a discrete gauge field on the $A$-brane. Here we see that to this discrete choice in the $A$ model there corresponds, at least formally, a half-integral valued $B$-field for the tension in the $B$-model geometry. It would be interesting to study this phenomenon and its $\IC^2/\IZ_n$ generalizations in more detail, and we hope to come back to this issue elsewhere.

As in the previous parametrization, the tensions can be computed from the integrals
$$
T_a=\frac{1}{2\pi i}\int_{*}^{\be_a} \taun(u(\xi))\frac{d\xi}{\xi}\ ,
$$
where
$\be_{1/2}=\pm i z_2^{1/4}\al_{1/2}$,
with $\al_{1/2}$ defined in eq.~\eqref{RD112234}. We again choose the reference point such that $W^{(+,\alpha)}=-W^{(-,\alpha)}$ and find 
\bee
W^{(\pm,\alpha_1)}=\mp\, \frac{c}{8} \cdot  \,B_{\{ l^1, l^2\}}\left(z_1,z_2;\tfrac{1}{2},0\right)\,,\ \ 
W^{(\pm,\alpha_2)}=\mp\, \frac{c}{8} \cdot  \,B_{\{ l^1, l^2\}}\left(z_1,z_2;\tfrac{1}{2},\tfrac{1}{2}\right)\,,
\eee
which is in agreement with \eqref{D1T1B12234} for $c=1$.

\newpag \subsection{Degree 14 hypersurface in $\IP_{1,2,2,2,7}$}
The charge vectors of the GLSM for the $A$ model manifold are given by \cite{Hosono:1993qy}
\begin{center}
\begin{tabular}{cc|ccccccc}
& $\x_0$ & $\x_1$ & $\x_2$ & $\x_3$ & $\x_4$ & $\x_5$ & $\x_6$\\
\hline
$l^1$ & -7 & -3 & 1 & 1 & 1 & 0 & 7 \\
$l^2$ & 0 & 1 & 0 & 0 & 0 & 1 & -2 \\
\end{tabular} .
\end{center}
The hypersurface constraint for the mirror manifold, written in homogeneous coordinates in $\IP_{1,2,2,2,7}$ as well, is 
\begin{equation}\label{W12227}
P = x_1^{14}+x_2^{7}+x_3^{7}+x_4^7+x_5^{2}-\psi\,x_1 x_2 x_3 x_4 x_5 +\phi\,x_1^7x_5\,,
\end{equation}
where $\psi=z_1^{-1/7}z_2^{-1/2}$  and $\phi=z_2^{-1/2}$. The orbifold group acts as $x_i\to \la_k^{g_k^i}x_i$ with $\lambda_k^7=1$ and weights
\begin{equation}
\IZ_7:\ g_1=(1,-1,0,0,0),\ \ \ \IZ_7:\ g_2=(1,0,-1,0,0),\ \ \IZ_7:\ g_3=(1,0,0,-1,0)\,.
\end{equation}
In this geometry we consider the set of curves
\ben\label{C12227}
&&C_{\alpha,\pm}=\{x_3=\eta x_4\,,\;x_5=\alpha x_1^7\;,\;x_2^3=\pm \sqrt{\alpha\eta\psi}\, x_4x_1^4\}\,,\nn\\&&\hskip1.5cm \eta^7=-1\,,\qquad \alpha^2+\phi\alpha+1=0\,,
\een
with the following identification under the orbifold group: $(\eta,\alpha,\pm)\sim(\eta \lambda_2\lambda_3^{-1},\alpha,\pm)$. By choosing representatives we can fix $\eta$ completely and label the orbits by $(\alpha,\pm)$. 

\subsubsection*{{\it First divisor}}
The family of divisors
\begin{equation}\label{D112227}
Q(\cx D_{1})=x_3^{7}+z_3 x_4^{7}
\end{equation}
contains the curves $C_{\al,\pm}$ for the critical value $z_3=1$. The periods on the family of surfaces is captured by the GLSM with charges
\begin{center}
\begin{tabular}{cc|cccccc}
& $\x_0$ & $\x_1$ & $\x_2$ & $\x_3$ & $\x_5$ & $\x_6$ \\
\hline
$\ti l^1$ & -7 & -3 & 1 & 2 & 0 & 7 \\
$\ti l^2$ & 0 & 1 & 0 & 0 & 1 & -2  \\ 
\end{tabular}.
\end{center}
with two algebraic moduli $u_1=-\frac{z_1}{z_3}(1-z_3)^2$ and $u_2=z_2$. The exceptional solutions 
\begin{align}\label{D1P12227} 
\taun_1&=\frac{c_1}{2} B_{\{\ti l\}}(u_1,u_2,\tfrac{1}{2},0)
=-\frac{c_1}{2\pi} \sqrt{u_1}\ _2F_1(-\tfrac{7}{4},-\tfrac{5}{4},-\tfrac{1}{2},4u_2)+\cx O(u_1^{3/2})\,,\nn \\
\taun_2&=\frac{c_2}{2} B_{\{\ti l\}}(u_1,u_2,\tfrac{1}{2},\tfrac{1}{2})
=\frac{35c_2}{2\pi}\sqrt{u_1}u_2^{3/2}\ _2F_1(-\tfrac{1}{4},\tfrac{1}{4},\tfrac{5}{2},4u_2)+\cx O(u_1^{3/2})\,,
\end{align}
vanish at the critical point $u_1=0$.
Note that these are series in $\sqrt{z_3}$ and the sign of the root distinguishes the two different holomorphic curves $C_{\al,+}$ and $C_{\al,-}$ in \eqref{C12227}.
The superpotentials obtained from integrals similar to \eqref{Int12234} are
\begin{align}\label{T112227}
W_1^{(\pm)}&=\pm\frac{c_1}{8}\sum_{n_i\geq 0} \frac{\Gamma \left(7 n_1+\frac{9}{2}\right)\,z_1^{n_1+\frac{1}{2}} z_2^{n_2}}{\Gamma \left(n_1+\frac{3}{2}\right)^3\Gamma \left(7 n_1-2 n_2+\frac{9}{2}\right) \Gamma \left(n_2+1\right) \Gamma \left(n_2-3 n_1-\frac{1}{2}\right)}\,,\nn \\
W_2^{(\pm)}&=\pm \frac{c_2}{8}\sum_{n_i\geq 0} \frac{\Gamma \left(7 n_1+\frac{9}{2}\right)\,z_1^{n_1+\frac{1}{2}} z_2^{n_2+\frac{1}{2}}}{\Gamma \left(n_1+\frac{3}{2}\right)^3 \Gamma \left(7 n_1-2 n_2+\frac{7}{2}\right) \Gamma \left(n_2+\frac{3}{2}\right) \Gamma \left(n_2-3 n_1\right)}\,.
\end{align}
They can be expressed in terms of the bulk generating function as
\begin{equation}\label{SP12227}
W_1^{(\pm)}=\pm \frac{c_1}{8}\,B_{\{ l^1, l^2\}}\left(z_1,z_2;\tfrac{1}{2},0\right)\,,\qquad W_2^{(\pm)}=\pm \frac{c_2}{8}\,B_{\{ l^1, l^2\}}\left(z_1,z_2;\tfrac{1}{2},\tfrac{1}{2}\right)\,.
\end{equation}

As in the previous example, these functions are the restrictions to the critical point $z_3=1$ of the off-shell tensions, which can be obtained as the solutions to the large GKZ system \eqref{gkz} of the relative cohomology problem derived in refs.~\cite{Alim:2009rf,Alim:2009bx,Li:2009dz}. For the family \eqref{D112227}, the additional charge vector is 

\begin{center}\vskip-0.3cm
\begin{tabular}{cc|ccccccccc}
& $\x_0$ & $\x_1$ & $\x_2$ & $\x_3$ & $\x_4$ & $\x_5$ & $\x_6$ & $\x_7$ & $\x_8$\\
\hline
$l^3$ & 0 & 0 & 0 & 1 & -1 & 0 & 0 & -1 & 1 
\end{tabular} .
\end{center}
This leads to the generalized hypergeometric system 
\begin{align} \label{LD12227}
\mathcal{\tilde{L}}_1 &= (\theta_1+\theta_3)(\theta_1-\theta_3) \left(7\theta_1-2 \theta_2\right)-7 z_1 \left(z_2 \left(28 \theta_1-4 \theta_2+18\right)-3 \theta_1+\theta_2-2\right)\times \nonumber\\&
\qquad \times\left(z_2\left(28 \theta_1-4 \theta_2+10\right)-3 \theta_1+\theta_2-1\right)\left(z_2 \left(28 \theta_1-4 \theta_2+2\right)-3 \theta_1+\theta_2\right)\,,\nonumber\\
\mathcal{\tilde{L}}_2 &= \theta_2(\theta_2-3\theta_1)-z_2(7\theta_1-2\theta_2-1)(7\theta_1-2\theta_2)\,,\\
\mathcal{\tilde{L}}_3 &= \theta_3(\theta_1+\theta_3)+z_3\theta_3(\theta_1-\theta_3)\,,\nonumber
\end{align}
annihilating the relative period integrals on the relative cohomology $H^3(Z^*,D_1)$ near the critical locus $y=\ln(z_3)=0$. Again this system has an alternative origin as the GKZ system associated to an F-theory compactification on a dual 4-fold described in app.~\ref{apptor}.

Alternatively, one may characterize the normal functions as solutions to an inhomogeneous Picard-Fuchs equation. From
$$
\ti{\cx L}_1=\Lbulk_1-(7\th_1-2\th_2)\th_3^2,\qquad 
\ti{\cx L}_2=\Lbulk_2,\qquad 
$$
one sees that only the first operator acquires an inhomogeneous term, which is determined by the leading part of the surface periods $\taun_a$. Acting with $(7\th_1-2\th_2)\th_3$ on the terms in \eqref{D1P12227} one obtains the inhomogeneous Picard-Fuchs equations
\bee\label{IH12227c}
\begin{aligned}
\Lbulk_1 W_1^{(\pm)} &= \mp\frac{7c_1}{16\pi^2}\sqrt{z_1}\, {}_2F_1\left(-\tfrac{3}{4},-\tfrac{5}{4},-\tfrac{1}{2},4z_2\right)\, = \pm c_1 f_1(\al_1)\ , \\
\Lbulk_1 W_2^{(\pm)} &= \pm\frac{35 c_2}{16\pi^2}\, z_1^{1/2}z_2^{3/2}\, {}_2F_1\left(\tfrac{1}{4},\tfrac{3}{4},\tfrac{5}{2},4z_2\right)\ \, = \pm c_2 f_1(\al_2)\ .
\end{aligned}
\eee
The inhomogeneous terms can be summarized as 
\bee\label{IH12227a}
f_1(\al)=-\frac{7i}{16\pi^2}\cdot \frac{\phi(\al+\phi)(6\al+\phi)}{\al^{1/2}\psi^{7/2}}\,,
\eee
where 
\bee\label{RD112227}
\al_{1/2}=\frac{1}{2}\left(-\phi\pm\sqrt{\phi^2-4}\right)\,,
\eee
denote the roots of the quadratic equation in the defining equation \eqref{C12227}.

\subsubsection*{{\it A-model expansion}}
The superpotential $W_1^{(+)}$ is associated with the curve $C_{\al_1,+}$ and similarly  $W_2^{(+)}$ with $C_{\al_2,+}$. With the normalization 
$c_1=c_2=1$ and the multi-cover formulae \eqref{MC2} and \eqref{MC22}, we obtain the integer invariants in Tab.~\ref{Tab12227}. Similarly as in the previous example, the two superpotentials are related by a $\IZ_2$ symmetry arising from the monodromy associated with an $A_1$ curve singularity \cite{Katz:1996ht,Klemm:1996kv}. On the $B$-model side, the $\IZ_2$ monodromy around the singular locus $\Delta=0$ acts on the periods as $t_1\to t_1+7 t_2,\,t_2\to -t_2\,.$ The invariants of $W_{2}$ are related to that of $W_1$ by the $\IZ_2$ quantum symmetry $q_1\to q_1q_2^7,\ q_2\to q_2^{-1}$ induced by this monodromy.

\begin{table}
\begin{tiny}
\centerline{$\frac{1}{2}\cdot n^{(1,+)}_{d_1,d_2}$}
\begin{tabular}{|c|cccccccc|}
\hline
$q_1^{1/2} \backslash q_2$&0&1&2&3&4&5&6&7\\
\hline
1& 1 & -14 & -35 & 0 & 0 & 0 & 0 & 0\\
3& -1 & 14 & -56 & -126 & -3416 & -42182 & -19481 & -396 \\
5& 5 & -126 & 1351 & -8358 & 41643 & -157990 & 87339 & -27425384\\
7& -42 & 1414 & -21455 & 195790 & -1271585 & 6722898 & -30564891 & 152513340\\
9& 429 & -18200 & 357070 & -4322640 & 37056327 & -248175368 & 1390770059 & -7006648980\\
11& -4939 & 252854 & -6077729 & 91502334 & -980198345 & 8110498760 & -55066462542 & 322702120822\\
13& 61555 & -3691114 & 104989899 & -1889415220 & 24334523486 & -241697136212 & 1953204386721 & -13402394296330\\
\hline
\end{tabular}\\[2mm]

\centerline{$\frac{1}{2}\cdot n^{(2,+)}_{d_1,d_2}$}
\hskip4cm
\begin{tabular}{|c|cccccc|}
\hline
$q_1^{1/2} \backslash q_2^{1/2}$&1&3&5&7&9&11\\
\hline
1& 0 & -35 & -14 & 1 & 0 & 0 \\
3& 0 & 0 & 28 & -396 & -19481 & -42182 \\
5& 0 & 0 & -70 & 1582 & -16212 & 179144 \\
7& 0 & 0 & 448 & -13804 & 195552 & -1907430 \\
9& 0 & 0 & -4004 & 157525 & -2892204 & 34409872\\
\hline
\end{tabular}
\end{tiny}
\caption{\label{Tab12227} Disc invariants for the on-shell superpotentials $W_a^{(+)}$ of the threefold $\IP_{1,2,2,2,7}[14]$.}
\end{table}

\subsubsection*{{\it Extremal transition and a non-compact limit }}
The above results and the normalization obtained by integration from the subsystem can be further verified by taking two different one parameter limits. At the singular locus $\Delta=0$, there is an extremal transition to the one parameter family mirror to a degree eight hypersurface in $\IP_{1,1,1,1,4}$ \cite{Berglund:1995gd}. To study this transition, we rewrite the hypersurface constraint \eqref{W12227} as 
\begin{equation}\label{W12227b}
P=\big(-\al \psi x_1^8x_2x_3x_4+x_2^7\big)+\big(x_3^7+x_4^7\big)+\big(x_5-\psi x_1x_2x_3x_4+(\al+\phi)x_1^7\big)\big(x_5-\al x_1^7\big)\ .
\end{equation}
The three summands indicated by the brackets vanish individually on the curves $C_{\al,\pm}$.
At the singular locus $\phi=\pm2$, the map to the hypersurface in $\IP_{1,1,1,1,4}$ is provided by the identifications 
$$
x_1^8x_2x_3x_4 = y_1^8,\quad  x_2^7 = y_2^8,\quad x_3^7 = y_3^8,\quad x_4^7 = y_4^8,\quad x_5\pm x_1^7 = y_5\ , 
$$
and this maps the curves $C_{\al,\pm}$ to the curves $C_{\zeta\mu}$ of ref.~\cite{Krefl:2008sj} in $\IP_{1,1,1,1,4}[8]$.\footnote{Here $\mu$ labels the two roots of the last summand in \eqref{W12227b} and $\zeta$ corresponds to a choice of the sign in \eqref{C12227}.}

From the symmetry $t_2\to -t_2$ it follows that the transition takes place at $q_2=1$, predicting the relation 
\bee
\sum_{i=0}^{7k}n^{(a,+)}_{k,i} (\IP_{1,2,2,2,7}[14])=n_k(\IP_{1,1,1,1,4}[8])\, ,
\eee
where $(k,i)$ denote the degree in $q_1$ and $q_2$, respectively. From the left hand side of the above equation one gets from the above tables
$$
-\frac{1}{2}\, n_k=\ 48,\ 65616,\ 919252560,\ ...
$$
for the invariants of $\IP_{1,1,1,1,4}[8]$. This is in agreement with the results of \cite{Krefl:2008sj,Knapp:2008uw}, up to a sign, which is convention.

On the other hand, the the $q_2^0$ term of the superpotential $W_1$ reproduces the invariants of the superpotential in the non-compact geometry $\cx O(-3)_{\IP^2}$ studied in ref.~\cite{Walcher:2007qp}, $n_{k,0}=n_k^{[3]}$ of Tab.~\ref{Tabnc} in app.~\ref{sec:AJ}. To recover this limit geometrically from eq.~\eqref{W12227b} we define
\ben
&&y_0=-\al \psi x_1^8x_2x_3x_4,\quad  y_1=x_2^7,\quad y_2=x_3^7,\quad y_3=x_4^7,\nn\\ 
&&x=\frac{x_5}{\phi}-\frac{\psi}{\phi} x_1x_2x_3x_4+\left(\frac{\al}{\phi}+1\right)x_1^7\,,\quad z=\phi x_5-\al\phi x_1^7\,,\nn
\een
to write the hypersurface constraint as 
$
P=(y_0+y_1)+(y_2+y_3)+xz\,.
$
The two roots $\al_{1/2}$ behave in the limit as $\al_{1/2}\sim-\phi^{\mp1}$. Choosing $\al=\al_1$ in \eqref{W12227b} and rescaling $x_5\to \frac{x_5}{\phi}$, one finds
$$
x=z_1^{-1/7}x_1x_2x_3x_4+x_1^7+\cx O(\phi^{-2})\,.
$$
Taking the root $x=0$ imposes a constraint on the $x_i$, and it allows us to rewrite the terms in the first two brackets as 
\bee\label{om3}
(y_0z_1^{-1/3}+y_1)+(y_2+y_3),\qquad y_0^3=y_1y_2y_3\ .
\eee
This is the equation for the Riemann surface $\Sigma$ representing the mirror of $\cx O(-3)_{\IP^2}$ \cite{Hori:2000kt,Aganagic:2001nx}. It can be verified that the factors in the holomorphic (3,0) form work out as well. After a final rescaling $y_0\to z_1^{1/3}y_0$, the integral for the domain wall interpolating between the curves $C_{\al_1,\pm}$ becomes 
\bee\label{intom3}
T^{(+,-)}_1(z_1,z_2=0)\, \sim \, \int_{y_2=-\sqrt{z_1}}^{y_2=+\sqrt{z_1}} \ln (y_1) d\ln y_2\ .
\eee
This is a 'half-cycle' on the Riemann surface, which reproduces the results for the local brane of ref.~\cite{Walcher:2009uj}.

\subsubsection*{\it Second divisor}
The same domain walls can be alternatively studied via the family of divisors
\begin{equation}\label{12227D2}
Q(\cx D_{2})=x_5^{2}+z_3z_2^{-1/2} x_1^{7}x_5\,,
\end{equation}
with the curves $C_{\alpha,\pm}$ contained in the divisor with $z_3=-\alpha\,z_2^{1/2}$. Following the same steps as in the previous example, one recovers the superpotentials \eqref{SP12227} as the integrals 
$$
W^{(\pm)}_a\,=\,\frac{1}{2\pi i}\int_{*}^{\be_a} \frac{c}{2} B_{\ti l}(u;\tfrac{1}{2})\ \frac{d\xi}{\xi}\ , \quad a=1,2 \ ,
$$
where $\be_{1/2}=\pm i (z_2^{1/2}\al_{1/2})^{1/2}$ with $\al_{1/2}$ defined in eq.~\eqref{RD112227}. The charge vector $\ti l=(-7,4,1,1,1)$ describes the subsystem defined by $\cx D_2$, and $u=-z_1z_2^3z_3^{-7}(z_2-z_3+z_3^2)^4$ is the single algebraic modulus associated with it.

\newpag \subsection{Degree 18 hypersurface in $\IP_{1,1,1,6,9}$}
The degree 18 manifold is one of the first examples, for which Ooguri--Vafa invariants for supersymmetric branes with a large volume phase have been obtained from open-string mirror symmetry \cite{Alim:2009rf}. Here we study branes near critical points of the generic type. The charge vectors of the GLSM for the A model manifold are given by:
\begin{center}
\begin{tabular}{cc|ccccccc}
& $\x_0$ & $\x_1$ & $\x_2$ & $\x_3$ & $\x_4$ & $\x_5$ & $\x_6$\\
\hline
$l^1$ & -6 & 0 & 0 & 0 & 2 & 3 & 1 \\
$l^2$ & 0 & 1 & 1 & 1 & 0 & 0 & -3 \\
\end{tabular} .
\end{center}
In homogeneous coordinates of $\IP_{1,1,1,6,9}$ the hypersurface constraint for the mirror manifold becomes 
\bee
\label{HS11169}
P=x_1^{18}+x_2^{18}+x_3^{18}+x_4^3+x_5^2+\psi\,(x_1 x_2 x_3 x_4 x_5)+\phi\,(x_1 x_2 x_3)^6\ ,
\eee
where $\psi=z_1^{-1/6}z_2^{-1/18}$ and $\phi=z_2^{-1/3}$. The Greene-Plesser orbifold group acts as $x_i\to \la_k^{g_{k,i}}\, x_i$ with $\lambda_1^{18}=1$, $\lambda_2^6=1$ and the weights
\begin{equation}
\IZ_{18}:\, g_1=(1,-1,0,0,0),\quad {\IZ_6}:\,g_2=(0,1,3,2,0)\, .
\end{equation}
In this geometry we consider the curves
\ben
\label{C11169}
C_{\al,\pm}&=&\{x_2=\eta_1x_1,\ x_5=\eta_2x_3^9-\frac{\psi}{2}x_1x_2x_3x_4,\ x_4=\psi^2\al(x_1x_2x_3)^2\}\,,\nn\\ 
&&\qquad \eta_1^{18}=\eta_2^2=-1,\quad \al^3-\frac{1}{4}\al^2+\frac{\phi}{\psi^6}=0\,,
\een
where different choices for $\eta_{1,2}$ are identified under the orbifold group as $(\eta_1,\eta_2,\al)\sim(\eta_1\la_1^2\la_2^{-1},\eta_2\la_2^3,\al)$, and we distinguish the curves $C_{\al,+}$ and $C_{\al,-}$ by the orbits of the labels $(\eta_1,\eta_2,\al)$ under this orbifold action. Specifically the orbits $C_{\al,\pm}$ contain the components $\eta_1^9=\pm i$ for fixed $\eta_2=i$ and fixed $\al$, respectively.

\subsubsection*{\it Divisor geometry and tensions}
We study the family of divisors
\begin{equation}
\label{D11169}
Q(\cx D)= x_2^{18}+z_3 x_1^{18}\,.
\end{equation}
The periods on this family are captured by the GLSM with charges
\begin{center}
\begin{tabular}{cc|ccccc}
& $\x_0$ & $\x_1 $ & $\x_2$ & $\x_3$ & $\x_4$ & $\x_5$ \\
\hline
$\ti l^1$ & -6 & 0 & 0 & 2 & 3 & 1 \\
$\ti l^2$ & 0 & 1 & 2 & 0 & 0 & -3 
\end{tabular} ,
\end{center}
where the two algebraic moduli are $u_1=z_1$ and $u_2=-\frac{z_2}{z_3}(1-z_3)^2$. The exceptional solutions
\bee\label{DP11169}
\begin{aligned}
\taun_1(u)&=\frac{c_1}{2}\,B_{\{\ti l^1,\ti l^2\}}(u_1,u_2;0,\tfrac{1}{2})=-\frac{c_1}{2\pi}\sqrt{u_2}\  {}_2F_1\left(\tfrac{1}{6},\tfrac{5}{6},-\tfrac{1}{2},432u_1\right)+\cx O(u_2^{3/2})\,,\\
\taun_2(u)&=\frac{c_2}{2}\,B_{\{\ti l^1,\ti l^2\}}(u_1,u_2;\tfrac{1}{2},\tfrac{1}{2})=\frac{2048c_2}{\pi}u_1^{3/2}\sqrt{u_2}\ {}_2F_1\left(\tfrac{5}{3},\tfrac{7}{3},\tfrac{5}{2},432u_1\right)+\cx O(u_2^{3/2})\,,
\end{aligned}
\eee
vanish at the critical point $u_2=0$. Similarly to eq.~\eqref{Int12234} we define off-shell superpotentials by
$$
\SP_a^{(\pm)}(z_1,z_2,z_3) = \frac{1}{2\pi i}\int_{\xi_0}^{\pm\sqrt{z_3}} \pi_a(u(z_1,z_2,\xi^2)) \, \frac{d\xi}{\xi} \ ,
$$
with the fixed reference point $\xi_0$. For $\xi_0=i$ the contribution of the reference point vanishes, and at the critical value $z_3=1$ we arrive at the on-shell superpotentials $\sp_a^{(\pm)}$, where the $\pm$-label is now correlated with the orbits of the curves \eqref{C11169}
\bee  \label{DW11169}
\begin{aligned}
\sp_1^{(\pm )}&=\pm \frac{c_1}{8}\sum_{n_i\geq 0}  \frac{\Gamma \left(6 n_1+1\right)\,z_1^{n_1} z_2^{n_2+\frac{1}{2}}}{ \Gamma \left(2 n_1+1\right) \Gamma \left(3 n_1+1\right) \Gamma \left( n_1-3n_2-\frac{1}{2}\right)\Gamma\left(n_2+\frac{3}{2}\right)^3}\ ,\\
\sp_2^{(\pm )}&=\pm \frac{c_2}{8}\sum_{n_i\geq 0} \frac{\Gamma \left(6 n_1+4\right)\,z_1^{n_1+\frac{1}{2}} z_2^{n_2+\frac{1}{2}}}{ \Gamma \left(2 n_1+2\right) \Gamma \left(3 n_1+\frac{5}{2}\right) \Gamma \left( n_1- 3 n_2\right)\Gamma\left(n_2+\frac{3}{2}\right)^3}\ .
\end{aligned}
\eee
They can be expressed in terms of the bulk generating function as
\begin{align}
\sp_1^{(\pm)}=\pm \frac{c_1}{8}\,B_{\{l^1, l^2\}}\left(z_1,z_2;0,\tfrac12 \right)\ , \quad
\sp_2^{(\pm)}=\pm \frac{c_2}{8}\,B_{\{l^1,l^2\}}\left(z_1,z_2;\tfrac12,\tfrac12\right)\ .
\end{align}
Again these functions can be also obtained as solutions to the large GKZ system \eqref{gkz} of the relative cohomology problem. For the family \eqref{D11169} we add the additional charge vector
\begin{center}\vskip-0.5cm
\begin{tabular}{cc|ccccccccc}
& $\x_0$ & $\x_1$ & $\x_2$ & $\x_3$ & $\x_4$ & $\x_5$ & $\x_6$ & $\x_7$ & $\x_8$\\
\hline
$l^3$ & 0 & -1 & 1 & 0 & 0 & 0 & 0 & 1 & -1 
\end{tabular} .
\end{center}
This leads to the generalized hypergeometric system 
\begin{align} \label{LD11169}
\mathcal{\tilde{L}}_1 &= \theta_1(\theta_1-3\theta_2)-12z_1 (6\theta_1+1)(6\theta_1+5)\nonumber\,, \\
\mathcal{\tilde{L}}_2 &= \theta_2(\theta_2-\theta_3)(\theta_2+\theta_3)-z_2(\theta_1-3\theta_2)(\theta_1-3\theta_2-1)(\theta_1-3\theta_2-2)\,,\\
\mathcal{\tilde{L}}_3 &= \theta_3(\theta_2+\theta_3)+z_3\theta_3(\theta_2-\theta_3)\,,\nonumber
\end{align}
annihilating the relative period integrals. There are two solutions with a minimum at the critical locus $\ln(z_3)=0$ that restrict to the on-shell superpotentials $\sp_1^{(\pm)}$ and $\sp_2^{(\pm)}$, respectively.

To characterize the on-shell superpotentials $\sp^{(\pm)}_a$ as solutions to an inhomogeneous Picard-Fuchs equation we note that
$$
\ti{\cx L}_1=\Lbulk_1,\qquad \,
\ti{\cx L}_2=\Lbulk_2-\th_2\th_3^2\,.\qquad 
$$
So only the second operator acquires an inhomogeneous term, which is determined by the leading part of the surface periods $\taun_a(u)$. Acting with $\th_2\th_3$ on the terms in \eqref{DP11169} one obtains the inhomogeneous Picard-Fuchs equations
\ben\label{IH11169}
A^{(\pm)}_1:=\ \Lbulk_2 \sp_1^{(\pm)} &=&\pm\frac{-c_1}{16 \pi^2} \sqrt{z_2}\; {}_2F_1\left(\tfrac{1}{6},\tfrac{5}{6};-\tfrac{1}{2};432z_1\right)\,,\nn\\[2mm]
A^{(\pm)}_2:=\ \Lbulk_2 \sp_2^{(\pm)} &=&\pm\frac{4096 c_2}{16 \pi^2} z_1^{3/2} \sqrt{z_2} \; {}_2F_1\left(\tfrac{5}{3},\tfrac{7}{3};\tfrac{5}{2};432z_1\right)\,.
\een
To find the geometric domain wall tensions, we note that the three roots $\al_\ell$ of the cubic equation~\eqref{C11169} can be written as 
$$
\al_\ell \,=\, \frac1{12}
\left(1+e^{\frac{2\pi i}{3}(\ell-1)}\,\Delta+e^{-\frac{2\pi i}{3}(\ell-1)}\,\frac{1}{\Delta}\right) \ , \quad \ell = 1, 2, 3 \ ,
$$  
with 
$$
\Delta\,=\,\sqrt[3]{1-864 z_1+2 \sqrt{432 z_1 \left(432 z_1-1\right)}} \ .
$$
Under a monodromy $z_1^{-1}\to e^{2\pi i}\,z_1^{-1}$ around $z_1=\infty$, $\Delta$ transforms as $\Delta\to e^\frac{2\pi i}{3}\Delta$ and the three roots are permuted according to $\al_\ell\to \al_{\ell+1}$. On the curves $C_{\al,\pm}$ the monodromy acts as the $\IZ_6$ symmetry
$$
M(z_1=\infty): \ \ \begin{pmatrix} \al_{1,\pm} \\ \al_{2,\pm} \\ \al_{3,\pm}
\end{pmatrix} \mapsto
\begin{pmatrix} \al_{2,\mp} \\ \al_{3,\pm} \\ \al_{1,\pm} \end{pmatrix} \ .
$$
It follows that the domain walls between the curves $C_{\al,\pm}$ for fixed $\al$ must be permuted under the monodromy as well. To this end note that the hypergeometric functions in eq.~\eqref{IH11169} are solutions of the same hypergeometric differential equation and in fact are related by monodromy. Indeed, for $c\equiv c_1 \equiv c_2$, the inhomogeneous pieces can be expressed in terms of the single rational expression 
\bee
f_1(\al) = 0 \ , \qquad
f_2(\al)=\frac{c}{4\pi^2}\frac{1-12 \al}{\psi^9\,\al^3 \left(1-6 \al\right){}^3} \ ,
\eee
as
\bee\label{IH11169R}
f_2(\al_1)=-2A^{(+)}_2\,,\qquad f_2(\al_2)=-A_1^{(+)}+A_2^{(+)}\,,\qquad f_2(\al_3)=A_1^{(+)}+A_2^{(+)}\,.
\eee
From the above we obtain the following linear combinations for the geometric superpotentials 
\bee
\sp_{\al_1}^{(\pm)}=2 \,\sp_2^{(\pm)}\,,\qquad
\sp_{\al_2}^{(\pm)}=\sp_1^{(\pm)}-\sp_2^{(\pm)}\,,\qquad
\sp_{\al_3}^{(\pm)}=\sp_1^{(\pm)}+\sp_2^{(\pm)}\,,
\eee
which satisfy $\Lbulk_2\, \sp_\al^{(\pm)}=\pm f(\al)$.

The inhomogeneous term~\eqref{IH11169R} becomes singular at the zeros of the open-string discriminant $$\Delta_\al=z_1(1-432z_1)\,,$$ of the cubic equation, where two roots coincide. This leads to the appearance of tensionless domain walls
\bee \label{TLess11169}
\begin{aligned}
z_1&=0  \quad &\Rightarrow\quad (\al_\ell)&=(\tfrac{1}{4},\,0,\,0) &\Rightarrow\quad  
&\begin{matrix}
     T_{\al_1,\al_1}^{(+,-)}=\sp_{\al_1}^{(+)}-\sp_{\al_1}^{(-)} =0 \\[2mm]
     T_{\al_2,\al_3}^{(\pm,\pm)}=\sp_{\al_2}^{(\pm)}-\sp_{\al_3}^{(\pm)} =0
\end{matrix}\ \ ,\\[2mm]
z_1&=\tfrac1{432}&\Rightarrow\quad (\al_\ell)&=(\tfrac{1}{6},-\tfrac{1}{12},\tfrac{1}{6})&\Rightarrow\quad
&\begin{matrix}
     T_{\al_2,\al_2}^{(+,-)}=\sp_{\al_2}^{(+)}-\sp_{\al_2}^{(-)} =0 \\[2mm]
     T_{\al_1,\al_3}^{(\pm,\pm)}=\sp_{\al_1}^{(\pm)}-\sp_{\al_3}^{(\pm)}=0
 \end{matrix} \ \ . 
\end{aligned}
\eee
\vskip2ex

\subsubsection*{{\it A-model expansion}}
In the Tab.~\ref{Tab11169} we list the integer invariants of the superpotentials $\sp_a^{(+)}$ obtained with the modified multicover formulas \eqref{MC2} and \eqref{MC22} for the normalization $c=c_1=c_2=1$.

\begin{table}
\begin{tiny}\begin{center}
$\frac{1}{2}\cdot n_{d_1,d_2}^{(1,+)}$\\[2mm]
\begin{tabular}{|c|ccccc|}
\hline
$q_1\backslash q_2^{1/2} $ & $1$ & 3 & 5 & 7 & 9  \\
\hline
0& 1 & -1 & 5 & -42 & 429 \\
1& -270 & 270 & -2430 & 27270 & -351000  \\
2& -35235 & 0 & 467775 & -7767495 & 131193270  \\
3& -1129110 & -3171960 & -56432160 & 1346568000 & -30388239450  \\
4& -19625112 & -9840669480 & 18001000575 & -268964593065 & 6132575901195  \\
5& -237548052 & -4228413761754 & 2588348258640 & 38534260978296 & -1115308309663386  \\
6& -2241975315 & -593578595396565 & 241002579933810 & -5655664165568310 & 165340822601302875  \\
\hline
\end{tabular}
\\[2mm]
$\frac{1}{8192}\cdot n_{d_1,d_2}^{(2,+)}$ \\[2mm]
\begin{tabular}{|c|ccccc|}
\hline
$q_1^{1/2}\backslash q_2^{1/2} $& $1$ & 3 & 5 & 7 & 9\\
\hline
1 & 0 & 0 & 0 & 0 & 0 \\
3 & -1 & 0 & 0 & 0 & 0 \\
5 & -54 & 108 & -270 & 1728 & -15444 \\
7 & -1215 & -24300 & 99630 & -918540 & 10783125 \\
9 & -17290 & -60310547 & -15819570 & 220135880 & -3485260710 \\
\hline
\end{tabular}
\end{center}\end{tiny}
\caption{\label{Tab11169} Disc invariants for the on-shell superpotentials $\sp_a^{(+)}$ of the threefold $\IP_{1,1,1,6,9}[18]$.}
\end{table}

In the limit $q_1\rightarrow 0$ the superpotential $\sp_1^{(+)}$ reproduces the numbers $n_k^{[3]}$ of the local Calabi-Yau geometry $\cx O(-3)_{\IP^2}$ given in Tab.~\ref{Tabnc} in sect.~\ref{sec:AJ}. Therefore in this local limit the domain wall between the curves $C_{\al_2,+}$ and $C_{\al_3,-}$, which yields the on-shell tension $T_{\al_2,\al_3}^{(+-)}\equiv\sp_{\al_2}^{(+)}-\sp_{\al_3}^{(-)}=2\,W_1^{(+)}$, becomes equivalent to the local domain wall of the local threefold $\cx O(-3)_{\IP^2}$ for the numbers $n_k^{[3]}$. The on-shell superpotentials $\sp_2^{(\pm)}$ vanish in this limit and give rise to tensionless domain walls~\eqref{TLess11169}.

\subsubsection*{{\it Non-compact limit}}
We exhibit the non-compact limit by redefining the projective coordinate of $\IP_{1,1,1,6,9}/(\IZ_{18}\times\IZ_6)$ according to
$$
y_1\,=\,x_1^6 \ , \quad y_2\,=\,x_2^6 \ , \quad y_3\,=\,x_3^6 \ , \quad x\,=\,x_5 \ , \quad z\,=\,x_5 + \psi\,x_1x_2x_3x_4 \ ,
$$
where $y_\ell \in \IC^*,\ x, z \in \IC$. In these local coordinates the Greene-Plesser orbifold group reduces to $\IZ_3$. It acts on the coordinates $y_\ell$ as $y_\ell\rightarrow \la^\ell y_\ell$, with $\la^3=1$, while the coordinates $x,z$ remain invariant. In the limit $z_1\rightarrow 0$, which is mirror symmetric to the limit $q_1\rightarrow 0$, we arrive at the local Calabi-Yau geometry
$$
0 \,=\, y_1^3+y_2^3+y_3^3 + z_2^{1/3}\,y_1y_2y_3 + x\,z + \cx O(\sqrt{z_1}) \ ,
$$
together with the associated local holomorphic three-form $\Omega$. This limit has already been studied in detail in ref.~\cite{Alim:2009rf}. The local geometry is related to the (mirror) cubic elliptic curve with the points $y_\ell =0$ removed, and it captures the local mirror of the non-compact threefolds $\cx O(-3)_{\IP^2}$ studied in app.~\ref{sec:AJ}. This explains the appearance of the disc invariants $n_k^{[3]}=n^{(1,+)}_{0,d_2}$ in Tab.~\ref{Tab11169}. 

\newpag \subsection{Degree 9 hypersurface in $\IP_{1,1,1,3,3}$}
Ooguri-Vafa invariants for supersymmetric branes with a large volume phase on this manifold have been computed in ref.~\cite{Alim:2009rf}. Here we study branes near critical points of the generic type. The charge vectors of the GLSM for the A model manifold are given by:
\begin{center}
\begin{tabular}{cc|ccccccc}
& $\x_0$ & $\x_1$ & $\x_2$ & $\x_3$ & $\x_4$ & $\x_5$ & $\x_6$\\
\hline
$l^1$ & -3 & 0 & 0 & 0 & 1 & 1 & 1 \\
$l^2$ & 0 & 1 & 1 & 1 & 0 & 0 & -3 \\
\end{tabular} .
\end{center} The hypersurface constraint for the mirror manifold, written in homogeneous coordinates of $\IP_{1,1,1,3,3}$, is
\begin{equation}
\label{HS11133}
P=x_1^{9}+x_2^{9}+x_3^{9}+x_4^3+x_5^3-\psi\,(x_1 x_2 x_3 x_4 x_5)+\phi\,(x_1 x_2 x_3)^3\,,
\end{equation}
where $\psi=z_1^{-1/3}z_2^{-1/9}$ and $\phi=z_2^{-1/3}$. 
The Greene-Plesser orbifold group acts as $x_i\to \la_k^{g_{k,i}}\, x_i$ with $\lambda_1^{9}=\lambda_2^9=1$, $\lambda_3^3=1$ and weights
\begin{equation}
\IZ_{9}:\, g_1=(1,-1,0,0,0),\quad {\IZ_9}:\,g_2=(1,0,-1,0,0),\quad {\IZ_3}:\,g_2=(0,0,0,1,-1)\, .
\end{equation}

\noi In this geometry, we study the family of divisors
\begin{equation}
\label{D11133}
Q(\cx D)= x_2^{9}+z_3 x_1^{9}\,,
\end{equation}
near the point $z_3=-1$. The Calabi-Yau threefold is an elliptic fibration over $\IP^2$ similar to the previous example and the steps of the computation of the periods of the relative cohomology group defined by the divisor $\cx D$ are straightforward. Despite these similarities, we could not identify a complete intersection representation of the type \eqref{C11169} for an appropriate curve. In the following we proceed to compute the superpotential and the disc invariants for the critical point without knowing such an explicit representation.

The surface periods defined by the family \eqref{D11133} are captured by the GLSM with charges
\begin{center}
\begin{tabular}{cc|cccccc}
& $\x_0$ & $\x_1$ & $\x_3$ & $\x_4$ & $\x_5$ & $\x_6$ \\
\hline
$\ti l^1$ & -3 & 0 & 0 & 1 & 1 & 1  \\
$\ti l^2$ & 0 & 2 & 1 & 0 & 0 & -3  
\end{tabular} ,
\end{center}
depending on the two algebraic moduli $u_1=z_1$ and $u_2=-\frac{z_2}{z_3}(1-z_3)^2$. The exceptional solutions 
\bee\label{DP11133}
\begin{aligned}
\taun_1&=\frac{c_1}{2}\ B_{\{\ti l^1,\ti l^2\}}(u_1,u_2;0,\tfrac{1}{2})
=\frac{-c_1}{2\pi}\, \sqrt{u_2}\ _2F_1(\tfrac{1}{3},\tfrac{2}{3},-\tfrac{1}{2},27z_1)+\cx O(u_2^{3/2})\,,\\
\taun_2&=\frac{c_2}{2}\ B_{\{\ti l^1,\ti l^2\}}(u_1,u_2;\tfrac{1}{2},\tfrac{1}{2})=\frac{105c_2}{2\pi}\, \sqrt{u_2}\ z_1^{3/2}\, _2F_1(\tfrac{11}{6},\tfrac{13}{6},\tfrac{5}{2},27z_1)+\cx O(u_2^{3/2})\,,
\end{aligned}
\eee
vanish at the point $z_3-1=0=u_2.$
The sign of the root $\sqrt{z_3}$ distinguishes the two sheets of the coordinate change $x_1^2= \tilde{x}_1$ similarly as in eq.~\eqref{Int12234}. Integrating along similar contours as in that case, we obtain the superpotentials
\bee
\begin{aligned}
W_1^{(\pm )}&=\pm \frac{c_1}{8} \sum_{n_i\geq 0} \frac{\Gamma \left(3 n_1+1\right)\,z_1^{n_1} z_2^{n_2+\frac{1}{2}}}{\Gamma \left(n_1+1\right)^2 \Gamma \left(n_1-3n_2-\frac{1}{2}\right) \Gamma \left(n_2+\frac{3}{2}\right)^3}\ ,\\
W_2^{(\pm )}&=\pm \frac{c_2}{8} \sum_{n_i\geq 0} \frac{\Gamma \left(3 n_1+\frac{5}{2}\right)\,z_1^{n_1+\frac{1}{2}} z_2^{n_2+\frac{1}{2}}}{\Gamma\left(n_1+\frac{3}{2}\right)^2 \Gamma \left(n_1-3 n_2\right) \Gamma \left(n_2+\frac{3}{2}\right)^3}\ ,
\end{aligned}
\eee
or equivalently, expressed in terms of the bulk generating function 
\begin{align}
W_1^{(\pm)}=\pm \frac{c_1}{8}\ B_{\{l^1,l^2\}}\left(z_1,z_2;0,\tfrac{1}{2}\right)\,,\qquad
W_2^{(\pm )}=\pm \frac{c_2}{8}\ B_{\{l^1,l^2\}}\left(z_1,z_2;\tfrac{1}{2},\tfrac{1}{2}\right)\,.
\end{align}

These functions are solutions to the large GKZ system \eqref{gkz} of the relative cohomology problem. For the family \eqref{D11133} the additional extended charge vector is
\begin{center}
\begin{tabular}{cc|ccccccccc}
& $\x_0$ & $\x_1$ & $\x_2$ & $\x_3$ & $\x_4$ & $\x_5$ & $\x_6$ & $\x_7$ & $\x_8$\\
\hline
$l^3$ & 0 & -1 & 1 & 0 & 0 & 0 & 0 & 1 & -1 
\end{tabular} ,
\end{center}
which, together with the charge vectors of the threefold, gives rise to the differential operators according to eq.~\eqref{gkz}
\bee \label{LD11133}
\begin{aligned}
\mathcal{\tilde{L}}_1 &= \theta_1(\theta_1-3\theta_2)- 3z_1 (3\theta_1+1)(3\theta_1+2)\ ,\\
\mathcal{\tilde{L}}_2 &= (\theta_2-\theta_3)(\theta_2+\theta_3)\theta_2-z_2(\theta_1-3\theta_2)(\theta_1-3\theta_2-1)(\theta_1-3\theta_2-2)\ , \\
\mathcal{\tilde{L}}_3 &= \theta_3(\theta_2+\theta_3)+z_3\,\theta_3(\theta_2-\theta_3)\ .
\end{aligned}
\eee
The solutions to these operators are the relative period integrals. In particular there are two solutions with a minimum at the critical locus $\ln(z_3)=0$, which restrict to the on-shell superpotentials $W_1^{(\pm)}$ and $W_2^{(\pm)}$, respectively.

To characterize the critical superpotentials $W^{(\pm)}_a$ as solutions to an inhomogeneous Picard-Fuchs equation, we observe
$$
\ti{\cx L}_1=\Lbulk_1,\qquad \,
\ti{\cx L}_2=\Lbulk_2-\th_2\th_3^2\,.\qquad 
$$
So only the second operator acquires an inhomogeneous term, which is determined by the leading part of the surface periods $\taun_a$. Acting with $\th_2\th_3$ on the leading coefficients of \eqref{DP11133} one obtains the inhomogeneous Picard-Fuchs equations 
\ben\label{IH11133c}
\Lbulk_2 W_1^{(\pm)} &=&\mp \frac{c_1}{16\pi^2} \sqrt{z_2}\; {}_2F_1\left(\tfrac{1}{3},\tfrac{2}{3};-\tfrac{1}{2};27z_1\right)\,,\nn\\[2mm]
\Lbulk_2 W_2^{(\pm)} &=&\pm\frac{105 c_2}{16\pi^2} z_1^{3/2}\sqrt{z_2}\; {}_2F_1\left(\tfrac{11}{6},\tfrac{13}{6};\tfrac{5}{2};27z_1\right)\,.
\een
The inhomogeneous terms are again solutions to the same hypergeometric equation and related by monodromy. The differential operator is obtained by specializing the Picard-Fuchs operator of the surface $\cx L_1^{\cx D}=\ti\th_1(\ti\th_1-3\ti\th_2)-3u_1(3\ti\th_1+1)(3\ti\th_1+2)$ to the critical point $u_2=0$: 
$$
\cx L^{inh}\, f_2(z_1,z_2)=0\,,\qquad \cx L^{inh}=\cx L_1^{\cx D}|_{\zh_{crit}}=(1-z)z\frac{d^2}{dz^2}+\left(-\frac{1}{2}-2z\right)\frac{d}{dz}-\frac{2}{9}\  ,
$$
with $\ti \th_a=u_a\tfrac{d}{du_a}$, $z=27 z_1$. The specialization to the leading term in the limit $u_2=0$ is achieved by setting $\th_2=\tfrac{1}{2}$. Similarly as in the other examples one can verify that the hypergeometric functions \eqref{IH11133c} can be written in closed form.

In Tab.~\ref{Tab11133} we list the integer invariants obtained with the modified multicover formula \eqref{MC2}, \eqref{MC22} for the normalization $c_1=c_2=1$. Similarly as in the previous examples, the hypersurface degenerates to the non-compact threefold \eqref{om3} in the limit $z_1\to 0$ \cite{Alim:2009rf}. This explains the appearance of the invariants $n^{[3]}$ (c.f. Tab.~\ref{Tabnc}) in the superpotential $W_1^{(+)}$, which are listed in the first row of the first table in Tab.~\ref{Tab11133}. 

\begin{table}[htb]
\begin{tiny}\begin{center}
$ \frac{1}{2}\cdot n_{d_1,d_2}^{(1,+)}$\\[2mm]
\begin{tabular}{|c|cccccc|}
\hline
$q_1\,\backslash\,q_2^{1/2}$ & 1 & 3 & 5 & 7 & 9 & 11  \\[0.5mm]
\hline
0 & 1 & -1 & 5 & -42 & 429 & -4939 \\
1 & -27 & 27 & -243 & 2727 & -35100 & 487647 \\
2 & -243 & 0 & 4131 & -71442 & 1230795 & -21333942 \\
3 & -1347 & -2295 & -33804 & 979800 & -24220836 & 544584789 \\
4 & -6021 & -231876 & 532575 & -10061955 & 319551804 & -9298367514 \\
5 & -22356 & -7276878 & 5101407 & 73610289 & -3196953927 & 117194205483 \\
\hline
\end{tabular}
\\[2mm]
$ \frac{1}{2}\cdot n_{d_1,d_2}^{(2,+)}$
\\[2mm]
\begin{tabular}{|c|ccccc|}
\hline
$q_1^{1/2}\,\backslash\,q_2^{1/2}$ & 1 & 3 & 5 & 7 & 9\\
\hline
1 & 0 & 0 & 0 & 0 & 0 \\
3 & -105 & 0 & 0 & 0 & 0 \\
5 & -567 & 1134 & -2835 & 18144 & -162162 \\
7 & -2916 & -18954 & 81648 & -826686 & 10133100 \\
9 & -11904 & -1421850 & -498555 & 13289664 & -255008817 \\
\hline
\end{tabular}
\end{center}
\end{tiny}
\caption{\label{Tab11133} {Disc invariants for the on-shell superpotentials $W_a^{(+)}$ of the threefold $\IP_{1,1,1,3,3}[9]$.}}
\end{table}

\newpag \subsection{Degree 12 hypersurface in $\IP_{1,1,2,2,6}$}
Ooguri--Vafa invariants for supersymmetric branes with a large volume phase on this manifold have been computed in ref.~\cite{Jockers:2009mn}. Here we study branes near critical points of the generic type. The critical value of the superpotential for these branes was computed already in ref.~\cite{Walcher:2009uj}. This gives a check on the off-shell superpotential obtained from the GKZ system for the relative periods by restriction to the critical point. The charges of the GLSM for the A-model manifold are given by:
\begin{center}
\begin{tabular}{cc|ccccccc}
& $\x_0$ & $\x_1$ & $\x_2$ & $\x_3$ & $\x_4$ & $\x_5$ & $\x_6$\\
\hline
$l^1$ & -6 & 0 & 0 & 1 & 1 & 3 & 1  \\
$l^2$ & 0 & 1 & 1 & 0 & 0 & 0 & -2  
\end{tabular} .
\end{center}
The hypersurface constraint  for the mirror manifold in homogeneous coordinates of $\IP_{1,1,2,2,6}$ is
\begin{equation}
P = x_1^{12}+x_2^{12}+x_3^6+x_4^6+x_5^2+\psi\,x_1 x_2 x_3 x_4 x_5+\phi\,(x_1 x_2)^6\,,
\end{equation}
where $\psi =z_1^{-1/6}z_2^{-1/12}$ and $\phi =z_2^{-1/2}$. The Greene-Plesser orbifold group acts as $x_i\to \la_k^{g_{k,i}}\, x_i$ with generators $\lambda_1^{6}=\lambda_2^6=\lambda_3^2=1$ and the weights
\begin{equation}
\IZ_{6}:\, g_1=(1,0,-1,0,0)\,,\quad {\IZ_6}:\, g_2=(1,0,0,-1,0)\,,\quad {\IZ_2}:\,g_3=(1,0,0,0,-1)\, .
\end{equation}
In this geometry we consider the same curves as in ref.~\cite{Walcher:2009uj},
\ben
\label{C11226}
&&C_{\alpha,\eta}=\{x_3=\eta_1 x_1^2\;,\;x_4=\eta_2 x_2^2\;,\;x_5=\alpha x_1x_2x_3x_4\}\,,\nn\\&&\hskip1.5cm \eta_1^6=\eta_2^6=-1\,,\qquad \alpha^2+\psi\alpha+\frac{\phi}{(\eta_1\eta_2)^2}=0\,,
\een
which under the orbifold group are identified as $(\eta_1,\eta_2,\alpha)\sim(\eta_1\lambda_1^3\lambda_2^2,\eta_2 \lambda_2,\alpha)$. The 36 choices for $\eta_1$ and $\eta_2$ form 3 orbits of length 12. Together with the two choices for $\alpha$ there are 6 different curves, that we choose to label by $(\alpha,\eta)$, where $\eta=(\eta_1\eta_2)^{-2}$ and $\eta^3=1$.

\subsubsection*{\it Divisor geometry and tensions}
We study the family of divisors
\begin{equation}\label{D111226}
Q(\cx D)=x_5^2 - z_3z_1^{-1/6}z_2^{-1/12} x_1x_2x_3x_4x_5\,,
\end{equation}
which contains the curves $C_{\alpha,\eta}$ at the critical points $z_3=\alpha z_1^{1/6}z_2^{1/12}$. Note that the chosen open coordinate $z_3$ arises naturally in the associated non-compact fourfold defined by the additional charge vector
\begin{center}\vskip-0.5cm
\begin{tabular}{cc|ccccccccc}
& $\x_0$ & $\x_1$ & $\x_2$ & $\x_3$ & $\x_4$ & $\x_5$ & $\x_6$ & $\x_7$ & $\x_8$\\
\hline
$l^3$ & -1 & 0 & 0 & 0 & 0 & 1 & 0 & -1 & 1 
\end{tabular} .
\end{center}

Derivatives of the relative periods with respect to $z_3$ are related to the surface periods of the intersection $Q(\cx D)=0=P$. The relevant surface is captured by the GLSM with the charges
\begin{center}
\begin{tabular}{cc|cccccc}
& $\x_0$ & $\x_1$ & $\x_2$ & $\x_3$ & $\x_4$ & $\x_5$ \\
\hline
$\ti l^1$ & -3 & 0 & 0 & 1 & 1 & 1  \\ 
$\ti l^2$ & 0 & 1 & 1 & 0 & 0 & -2 
\end{tabular} .
\end{center}
The algebraic moduli of the surface are related to those of the threefold by $u_1=-\frac{z_1}{z_3^3(1+z_3)^3}$ and $u_2=z_2$. 

In the moduli of the intersection surface the critical points $z_3=\alpha z_1^{1/6}z_2^{1/12}\equiv \ti\al$ are given in terms of the condition $u_1=u_2$ or equivalently in terms of $-\frac{z_1}{z_2}=z_3^3(1+z_3)^3$. Then the characteristic equation for the curves $C_{\al,\eta}$ becomes
\begin{equation} \label{Crit11226}
   \eta y + \ti\al(1+\ti\al)=0 \ , \qquad y \equiv \left(\frac{z_1}{z_2}\right)^{1/3} \ ,
\end{equation}
with $\eta^3=1$, and the critical points are given by
\begin{equation}
\ti\al_\pm(\eta) = z_1^{1/6}z_2^{1/12}\alpha_{\pm}(\eta)=\frac{1}{2}\left(-1\pm\sqrt{1-4\eta y}\right) \ .
\end{equation}
Hence the critical points $\ti\al(\eta)$ are in one-to-one correspondence to the labels $(\al,\eta)$ of the curves $C_{\alpha,\eta}$.

The solutions of this subsystem can be generated with the Frobenius method from the generating function,
\begin{equation}
B_{\{\ti l^1,\ti l^2\}}(u_1,u_2;\rho_1,\rho_2)=\sum_{n_i\in \IZ+\rho_i}\frac{\Gamma(1+3n_1)\,u_1^{n_1} u_2^{n_2}}{\Gamma(1+n_1)^2\,\Gamma(1+n_1-2n_2)\,\Gamma(1+n_2)^2}\,.
\end{equation}
The linear combination 
\begin{equation}
\taun(u_1,u_2) =\frac{c}{2\pi i}\ (\partial_{\rho_1}B_{\{\ti l^1,\ti l^2\}}-\partial_{\rho_2}B_{\{\ti l^1,\ti l^2\}})\big|_{\rho_i=0}:=c(t_1-t_2)\,,
\end{equation}
vanishes at $u_1=u_2$ or equivalently at the critical points $z_3= \ti\al$ of eq.~\eqref{Crit11226}. Note that $t_1$ and $t_2$ are the volumes of two generators of  $H_2(K3,\IZ)$, and the zero of the period arises from the coincidence of their volumes. 

In order to derive the superpotentials we need to integrate the surface periods $\taun(u)$. Note that for the divisor family \eqref{D111226} the induced holomorphic two form of the embedding surface differs from the canonically normalized holomorphic two form of the corresponding isogenic K3 surface by a moduli dependent pre-factor. As a consequence the relation \eqref{surfper} must also be modified by a moduli-dependent measure factor \cite{Jockers:2009mn}
$$
 2\pi i\,\theta_{z_3}\cx W(z_1,z_2,z_3)=\frac{1}{1+z_3}\taun(u_1,u_2) \ ,
$$ 
where now both the superpotential $\cx W$ and the surface period $\taun(u_1,u_2)$ are canonically normalized.  
Thus integrating the surface period $\taun(u_1,u_2)$ together with the measure factor according to 
\begin{equation}
W^{(\alpha_\pm,\eta)}=\frac{1}{2\pi i}\int^{\ti\al_\pm(\eta)}_{*} \frac{1}{1+z_3}\,\taun(z_3) \frac{dz_3}{z_3}\,,
\end{equation}
we find the on-shell superpotentials for the curves $C_{(\al,\eta)}$
\begin{align}
\label{T11226}
W^{(\alpha_\pm,\eta)}(y,z_2)=\mp\ \frac{c}{4\pi^2} \left(\frac{3}{2}S_0\,\log(-\eta y)^2+(S_1-S_2)\,\log(-\eta y)+\frac{5\pi^2}{2}+S_{\alpha_+,\eta}(y,z_2)\right)\,.
\end{align}
Here $S_0,S_1$ and $S_2$ are the power-series\footnote{$S_0$ is the fundamental closed string period and $S_a$, $a=1,2$, the series part of the single logarithmic closed string periods $(2\pi i) t_a = \log(z_a)+S_a$, which determine the closed string mirror map. However there is no double logarithimic closed string period that has the same classical terms as eq.~\eqref{T11226}.}
$$
\begin{aligned}
S_0 &=\sum_{n_i\geq 0}\frac{\Gamma(1+6n_1)}{\Gamma(1+3n_1)\Gamma(1+n_1)^2\,\Gamma(1+n_1-2n_2)\Gamma(1+n_2)^2}\,y^{3n_1}\,z_2^{n_1+n_2}  \\
	&= 1 + 120 y^3 z_2 + 83160 y^6 z_2^2 + 166320 y^6 z_2^3 + 81681600 y^9 z_2^3+\dots \ , \\
S_1 &=\sum_{n_i\geq 0}\frac{\Gamma(1+6n_1)\left(6\psi(1+n_1)-3\psi(1+3n_1)-2\psi(1+n_1)-\psi(1+n_1-2n_2) \right)}{\Gamma(1+3n_1)\Gamma(1+n_1)^2\,\Gamma(1+n_1-2n_2)\Gamma(1+n_2)^2}\,y^{3n_1}\, z_2^{n_1+n_2} \nn\\ 
	&= -z_2+744 y^3 z_2+-\frac{3 z_2^2}{2}+120 y^3 z_2^2+562932 y^6 z_2^2+\dots \ , \\
S_2&= \sum_{n_i\geq 0}\frac{2\Gamma(1+6n_1)\left(\psi(1+n_1-2n_2)-\psi(1+n_2)\right)}{\Gamma(1+3n_1)\Gamma(1+n_1)^2\,\Gamma(1+n_1-2n_2)\Gamma(1+n_2)^2}\,y^{3n_1}\, z_2^{n_1+n_2} \\ 
	&= 2 z_2+240 y^3 z_2+ 3 z_2^2-240 y^3 z_2^2+249480 y^6 z_2^2 +\dots\ ,
\end{aligned}
$$
while the instanton part reads
\begin{equation}
S_{\alpha_+,\eta}=   \left(6 (\eta y) +z_2\right)+ \left(\frac{9 (\eta y)^2}{2}+\frac{15 z_2^2}{4}\right)+ \left(\frac{20 y^3}{3}+81 (\eta y)^2 z_2+\frac{191
 z_2^3}{18}\right)+\dots \ .
\end{equation}

In ref.~\cite{Walcher:2009uj} the on-shell superpotentials~\eqref{T11226} were obtained as the solutions to inhomogeneous Picard-Fuchs equations. To calculate these inhomogeneous terms, we rewrite the bulk operators
\bee
\begin{aligned}
\mathcal{L}^{bulk}_1&=\theta_1^2(\theta_1-2\theta_2)-8z_1(6\theta_1+1)(6\theta_1+3)(6\theta_1+5)\ ,\\
\mathcal{L}^{bulk}_2&=\theta_2^2-z_2(\theta_1-2\theta_2)(\theta_1-2\theta_2-1)\ ,
\end{aligned}
\eee
in terms of the coordinates $y$ and $z_2$, we act with them upon the superpotentials~\eqref{T11226}, and we find for the inhomogeneous terms
$$
\begin{aligned}
\mathcal{L}_1^{bulk}\, W^{(\alpha_\pm,\eta)} &= \pm\frac{c}{4\pi^2} \left(\frac{2}{3} \eta y + 4 (\eta y)^2 + 20 (\eta y)^3 
+\dots \right)= \pm \frac{c}{6\pi^2} \frac{\eta y}{(1-4\eta y)^{3/2}} \  , \\
\mathcal{L}_2^{bulk}\,W^{(\alpha_\pm,\eta)} &= \pm\frac{c}{4\pi^2}\left(  \frac{1}{3} + \frac{2}{3} \eta y + 2 (\eta y)^2 + \frac{20}{3} (\eta y)^3 
+\dots \right)  = \pm\frac{c}{12\pi^2} \frac{1}{(1-4\eta y)^{1/2}}\,.
\end{aligned}
$$

\subsubsection*{{\it A-model expansion}}
\begin{table}\begin{tiny}\begin{center}
\begin{tabular}{|c|cccccccccc|}
\hline
$q_y \backslash q_{z_2} $& 0 & 1 & 2 & 3 & 4 & 5 & 6 & 7 & 8 & 9 \\
\hline
0&0& 0& 0& 0& 0& 0& 0& 0& 0& 0\\ 
1&6& 6& 0& 0& 0& 0& 0& 0& 0&0\\ 
2&3& 90& 3& 0& 0& 0& 0& 0& 0& 0\\ 
3&6& 388& 388& 6& 0& 0& 0& 0& 0& 0\\ 
4&12& -258& 2934& -258& 12& 0& 0& 0& 0& 0\\ 
5&30& -540& 11016& 11016& -540& 30& 0& 0& 0& 0\\ 
6&75& -1388& 67602& 348774& 67602& -1388& 75& 0& 0& 0\\ 
7&210& -3960& -44496& 731952& 731952& -44496& -3960& 210& 0& 0\\ 
8&600& -12042& -75036&3177414& 20289960& 3177414& -75036& -12042& 600& 0\\ 
9&1800& -38236& -136672& 20383740& 399653208& 399653208& 20383740& -136672& -38236& 1800\\
\hline
\end{tabular}
\end{center}\end{tiny}
\caption{\label{Tab11226} Symmetric disc invariants for the on-shell superpotentials $W^{(\alpha_+,1)}$ of the threefold $\IP_{1,1,2,2,6}[12]$.}
\end{table}

For completeness we quote in Tab.~\ref{Tab11226} the disc instantons for the on-shell superpotentials in terms of $q_y=\frac{1}{3}(z_1-z_2)+\dots$ and $q_{z_2}=z_2+\dots$ for $c=1$. These numbers have already been computed in ref.~\cite{Walcher:2009uj} by deriving the inhomogeneous Picard-Fuchs equations. As in ref.~\cite{Walcher:2009uj} we have added a rational multiple of a closed-string period with leading behavior $\Pi=t_1t_2+\dots$ to get invariants $n_{d_1,d_2}$ symmetric under the $\IZ_2$ symmetry
$
t_1\to t_1+t_2,\ t_2\to -t_2;\  q_y \to q_yq_2,\ q_2\to q_2^{-1}\ .
$
This is the Weyl symmetry of a non-perturbative $SU(2)$ gauge symmetry appearing in the type II compactification at the transition point \cite{Katz:1996ht,Klemm:1996kv}. The domain wall is a singlet under this global symmetry as can be seen from the defining equation \eqref{C11226}.

\newpag \subsection{Degree 8 hypersurface in $\IP_{1,1,2,2,2}$}

The charge vectors of the GLSM for the A-model manifold are given by:
\begin{center}
\begin{tabular}{cc|ccccccc}
& $\x_0$ & $\x_1$ & $\x_2$ & $\x_3$ & $\x_4$ & $\x_5$ & $\x_6$ \\
\hline
$l^1$ & -4 & 0 & 0 & 1 & 1 & 1 & 1  \\
$l^2$ & 0 & 1 & 1 & 0 & 0 & 0 & -2  
\end{tabular}
\end{center}
The hypersurface constraint  for the mirror manifold in homogeneous coordinates of $\IP_{1,1,2,2,2}$ is
\ben
P = x_1^{8}+x_2^{8}+x_3^4+x_4^4+x_5^4+\psi\,x_1 x_2 x_3 x_4 x_5 +\phi\,(x_1 x_2)^4
\een
where $\psi = z_1^{-1/4}z_2^{-1/8}$ and $\phi = z_2^{-1/2}$. The Greene-Plesser orbifold group acts as $x_i\to \la_k^{g_{k,i}}\, x_i$ with generators $\lambda_1^{8}=1$, $\lambda_2^4=\lambda_3^4=1$ and weights
\begin{equation}
\IZ_{8}:\, g_1=(1,-1,0,0,0),\quad {\IZ_4}:\,g_2=(1,0,-1,0,0),\quad {\IZ_4}:\,g_2=(1,0,0,-1,0)\, .
\end{equation}

In this geometry we consider the curves
\ben \label{C11222}
C_{\alpha}&=&\{x_3=\eta_1 x_1^2\,,\,x_4=\eta_2 x_2^2\,,\, \eta_1\eta_2 x_5=\alpha x_1x_2\}\,,\nn\\
&&\quad \eta_1^4=\eta_2^4=-1\,, \alpha^4+\psi\,\alpha+\phi=0\,.
\een
where $\eta_1^4=\eta_2^4=-1$. Under the orbifold action the curves are identified as $(\eta_1,\eta_2,\alpha)\sim(\eta_1\lambda_1^2\lambda_2^3\lambda_3^{2},\eta_2 \lambda_1^{-2}\lambda_3,\alpha)$. The curves are labeled by the four roots $\alpha$, while under the $\mathbb{Z}_8\times \mathbb{Z}_4^2$ orbifold action the 16 distinct choices for the phases $\eta_1$ and $\eta_2$ are identified. Thus we find four distinct orbits of curves $C_\al$. 

\subsubsection*{\it Divisor geometry and tensions}
To compute DW tensions for these curves we study the family of divisors
\begin{equation}
 Q(\cx D)=x_5^4 - z_3\,z_1^{-1/4}z_2^{-1/8}\, x_1x_2x_3x_4x_5\ .
\end{equation}
The curves $C_\al$ are included in $\cx D$ for the critical values $z_3 = z_1^{1/4} z_2^{1/8}\,\al^3 \equiv \tilde \al$, where the new label $\tilde\al$ obeys the fourth order equation
\bee\label{tal11222}
\tilde\al ( 1+ \tilde\al)^3 + y \,=\,0 \ , \qquad y\,\equiv\,\frac{z_1}{z_2} \ .
\eee
Note that the roots $\tilde\al$ of this fourth order equation are in one-to-one correspondence with the curves $C_\alpha$. 

The chosen open-string coordinate $z_3$ is the natural coordinate on the non-compact fourfold defined by adding

\begin{center}\vskip-0.5cm
\begin{tabular}{cc|ccccccccc}
& $\x_0$ & $\x_1$ & $\x_2$ & $\x_3$ & $\x_4$ & $\x_5$ & $\x_6$ & $\x_7$ & $\x_8$\\
\hline
$l^3$ & -1 & 0 & 0 & 0 & 0 & 1 & 0 & 1 & -1 
\end{tabular}
\end{center}
to the GLSM for the A-model manifold. Periods on the intersection $Q(\cx D)=P=0$ are captured by a GLSM with charges
\begin{center}
\begin{tabular}{cc|cccccc}
& $\x_0$ & $\x_1$ & $\x_2$ & $\x_3$ & $\x_4$ & $\x_5$ \\
\hline
$\ti l^1$ & -3 & 0 & 0 & 1 & 1 & 1 \\
$\ti l^2$ & 0 & 1 & 1 & 0 & 0 & -2  
\end{tabular},
\end{center}
where the algebraic moduli are $u_1=-\frac{z_1}{z_3(1+z_3)^3}$ and $u_2=z_2$. In these coordinates the critical points $z_3=\tilde\al$ arise at $u_1=u_2$. This condition corresponds to the fourth order equation~\eqref{tal11222} for the label $\tilde\al$.

The solutions of this subsystem can be generated with the Frobenius method from the generating function
\ben
B_{\{\ti l^1,\ti l^2\}}(u_1,u_2;\rho_1,\rho_2)=\sum_{n_i\in \IZ+\rho_i}\frac{\Gamma(1+3n_1)\,u_1^{n_1} u_2^{n_2}}{\Gamma(1+n_1)^2\,\Gamma(1+n_1-2n_2)\,\Gamma(1+n_2)^2}\,.
\een
The linear combination
\ben \label{subper11222}
\tau =  \frac{c}{2\pi i}  (\partial_{\rho_1}B_{\{\ti l^1,\ti l^2\}}-\partial_{\rho_2}B_{\{\ti l^1,\ti l^2\}})\big|_{\rho_i=0}:=c(t_1-t_2)\,,
\een
vanishes at the critical locus $u_1=u_2$. Again $t_1$ and $t_2$ measure the volumes of two generators of $H_2(K3,\IZ)$ and at criticality the zero of the period arises because their volumes coincide. The four critical points $\tilde\al_k, k=0,1,2,3$, which are given in terms of the fourth order equation~\eqref{tal11222}, enjoy in terms of $y=\frac{z_1}{z_2}$ the expansion
\bee \label{Roots11222}
\begin{aligned}
\tilde\al_0(y)&=-y \left(1+3 y+15 y^2+91 y^3+612 y^4+4389 y^5+32890 y^6 +\dots \right)\ ,\\
\tilde\al_\ell(y)&= -1+\nu_\ell  y^{1/3}+\frac{1}{3} \nu^2_\ell y^{2/3}+\frac{\nu^3_\ell y}{3}+\frac{35}{81} \nu^4_\ell y^{4/3}+\frac{154}{243} \nu^5_\ell y^{5/3}+\nu^6_\ell y^2+\dots \ ,
\end{aligned}
\eee
with $\nu_\ell=e^{\frac{2\pi i}{3}(\ell-1)}, \ell=1,2,3$. Similarly to the related example $\IP_{1,1,2,2,6}$, there appears an additional measure factor for the integration of subsystem period to the superpotential, namely 
$$
2\pi i\,\theta_{z_3}\cx W(z_1,z_2,z_3)=\frac{1}{1+z_3}\taun(u_1,u_2) \ .
$$
Hence, integrating the discussed subsystem period~\eqref{subper11222} with the additional measure factor, we obtain for the critical point $\tilde\al_0$ the on-shell superpotential
\bee \label{T11222}
W^{(\alpha_0)}(y,z_2)=-\frac{c}{4\pi^2}\left(\frac{1}{2}\,S_0\,\log(-y)^2+(S_1-S_2)\,\log(-y)+S_{\alpha_0}(y,z_2)\right)\ .
\eee
Here $S_0,S_1$ and $S_2$ are the power-series\footnote{$S_0$ is the fundamental closed string period and $S_a$, $a=1,2$, the series part of the single logarithmic closed string periods $(2\pi i) t_a = \log(z_a)+S_a$, which determine the closed string mirror map. However there is no double logarithimic closed-string period that has the same classical terms as eq.~\eqref{T11222}.}
\bee
\begin{aligned}
S_0 &= 1 + 24 y z_2 + 2520 y^2 z_2^2 + 5040 y^2 z_2^3 + 369600 y^3 z_2^3 + 2217600 y^3 z_2^4+\dots\,,\\
S_1 &= -z_2 + 104 y z_2 - \frac{3}{2} z_2^2 + 24 y z_2^2 + 12276 y^2 z_2^2 - \frac{10}{3}z_2^3 + 12 y z_2^3 - \frac{35}{4} z_2^4+\dots \,,\\
S_2 &= 2 z_2 + 48 y z_2 + 3 z_2^2 - 48 y z_2^2 + 7560 y^2 z_2^2 + \frac{20}{3} z_2^3 - 24 y z_2^3 + \frac{35}{2} z_2^4 +\dots\, .
\end{aligned}
\eee
For the instanton part $S_{\al_0}$ we get
\bee
S_{\alpha_0}(y,z_2)=\left(4 y+3 z_2\right)+\left(7 y^2-64 y z_2+\frac{45 z_2^2}{4}\right)+\left(\frac{220 y^3}{9}+210 y^2 z_2+528 y z_2^2+\frac{191
 z_2^3}{6}\right) + \dots \,.
\eee
Finally, we note that in terms of the GLSM charges suitable for the coordinates $y, z_2$
\begin{center}
\begin{tabular}{cc|ccccccc}
& $\x_0$ & $\x_1$ & $\x_2$ & $\x_3$ & $\x_4$ & $\x_5$ & $\x_6$ \\
\hline
$h^1$ & -4 & -1 & -1 & 1 & 1 & 1 & 3  \\
$h^2$ & 0 & 1 & 1 & 0 & 0 & 0 & -2  
\end{tabular}
\end{center}
we can express the superpotential $W^{(\al_0)}$ as
\bee
W^{(\alpha_0)}(y,z_2)\,=\,-\frac{c}{8\pi^2}\,\partial^2_{\rho_1}\,B_{\{h^1,h^2\}}(y,z_2,\rho_1,\rho_2)|_{\rho_i=0} \ .
\eee
Integrating the subsystem period with the additional measure factor to the other roots $\tilde\al_\ell(y)$ one finds similar expansions for the on-shell superpotentials $W^{(\al_\ell)}(y,z_2)$ associated to these roots.

To characterize the superpotential $W^{(\alpha_0)}$ by an inhomogeneous Picard-Fuchs equation we calculate the inhomogeneous pieces with the following bulk operators
\begin{align}
\mathcal{L}^{bulk}_1&=\theta_1^2(\theta_1-2\theta_2)-4z_1(4\theta_1+1)(4\theta_1+2)(4\theta_1+3)\,,\\
\mathcal{L}^{bulk}_2&=\theta_2^2-z_2(\theta_1-2\theta_2)(\theta_1-2\theta_2-1)\,,
\end{align}
and we obtain 
\bee
\begin{aligned}
\mathcal{L}_1^{bulk}\, W^{(\alpha_0)} 
&=-\frac{3\,c}{4\pi^2}\ \th_y\, _3F_2\left(\tfrac{1}{4},\tfrac{2}{4},\tfrac{3}{4};\tfrac{1}{3},\tfrac{2}{3};\tfrac{256\,y}{27}\right)
= 3\, \theta_y f(\al_0) \  , \\
\mathcal{L}_2^{bulk}\,W^{(\alpha_0)} 
&=-\frac{c}{4\pi^2}\ _3F_2\left(\tfrac{1}{4},\tfrac{2}{4},\tfrac{3}{4};\tfrac{1}{3},\tfrac{2}{3};\tfrac{256\,y}{27}\right)
= f(\al_0) \ ,
\end{aligned}
\eee
where the label $\al_0$ refers to the root of the quartic equation in \eqref{C11222} associated to the corresponding root $\tilde\al_0$ in eq.~\eqref{Roots11222}.
As in previous examples we can also express the inhomogeneous terms as functions in the coefficients of the defining equations, i.e.  
\bee\label{IH11222R}
  f(\al) \,=\,
  \frac{c}{4\pi^2}\, \frac{z_2^{1/8}\al}{4y^{1/4}+3z_2^{1/8}\al}
  \,=\,
  -\frac{c}{4\pi^2} \cdot \frac{1}{4\tilde{\alpha}+1} \ .
\eee
The open string discriminant is $\Delta_\al= y(1-\tfrac{256y}{27})$, with the three roots $\ti\al_\ell$, $\ell>0$ colliding for $y=0$ at $\ti\al_\ell=-1$, see eq.~\eqref{Roots11222}, while at $y=\tfrac{27}{256}$ one has $\ti \al_0=-\tfrac{1}{4}=\ti \al_1$. The inhomogeneous term~\eqref{IH11222R} become singular at the second zero, indicating a tensionless domainwall between the curves associated with $\ti\al_{0,1}$.

For the other on-shell superpotentials $W^{(\alpha_\ell)}(y,z_2)$, we find the same inhomogeneous terms
\bee
 \mathcal{L}_1^{bulk}\, W^{(\alpha_\ell)}\,=\,3\,\theta_y f(\al_\ell) \ , \quad
 \mathcal{L}_2^{bulk}\, W^{(\alpha_\ell)}\,=\, f(\al_\ell) \ , \quad \ell=1,2,3 \ ,
\eee
where again the roots $\al_\ell$ are associated to the corresponding roots $\tilde\al_\ell$. 

\subsubsection*{{\it A-model expansion}}
Using the standard multicover formula 
$$
\frac{W^{(\alpha_0)}(z(q))}{\omega_0(z(q))}=\frac{1}{(2\pi i)^2}\sum_{k}\sum_{d_{1,2}\geq 0} n^{(\alpha_1)}_{d_1,d_2}\, \frac{q_1^{k d_1}\,q_2^{k d_2}}{k^2}
$$
we obtain for $c=1$ the integer invariants in Tab.~\ref{Tab11222}. Here $q_y=z_1-z_2+\dots$ and $q_{2}=z_2+\dots$. Again we have added a rational multiple of a closed-string period with leading behavior $\Pi=\frac{3}{2}t_1t_2+\dots$ to get invariants $n_{d_1,d_2}$ symmetric under the $\IZ_2$ symmetry
$
t_1\to t_1+t_2,\ t_2\to -t_2;\ q_y \to q_yq_2,\ q_2\to q_2^{-1}\ .
$
This is the Weyl symmetry of a non-perturbative $SU(2)$ gauge symmetry appearing in the type II compactification at the transition point \cite{Katz:1996ht,Klemm:1996kv}. The domain wall is a singlet under this global symmetry as can be seen from the defining equation \eqref{C11222}.

\begin{table}
\hskip-1cm
\begin{tiny}
\begin{tabular}{|c|ccccccccc|}
\hline
$\hskip-0.15cm q_y\backslash q_2\hskip-0.15cm $& 0 & 1 & 2 & 3 & 4 & 5 & 6 & 7 & 8 \\
\hline
0&0&0& 0& 0& 0& 0& 0& 0& 0\\ 
1&4& 188& 188& 4& 0& 0& 0& 0& 0\\
2&6& -68& 5194& 19024& 5194& -68& 6& 0& 0 \\ 
3&24& -292& -3232& 259524& 3569704& 3569704& 259524& -3232& -292\\ 
4&112& -1660& -10996& -4092& 13712184&
   555071696& 1455120104& 555071696& 
  13712184\\ 
5&620& -10768& -42752& 383424& -256440& 
  695568492& 74900481736& 418921719720& 418921719720\\ 
6&3732& -75468& -140150& 4170468& 6794752& -464516720& 
  32227348614& 9235136625472& 97930146122188\\ 
7&24164& -556600& 5648& 37548816& 
  24834800& -2671560544& -62352854944& 991475402468& 
  1066545645786456\\ 
8&164320& -4256460& 7444296& 
  318651284& -286806192& -20467318044& -282718652536& \
-7115509903004& -64593220192464\\ 
9&1162260& -33442800& 114057840& 
  2622725460& -7347237536& -170307380384& -1384203066912& \
-28014543398208& -915396773309428\\ 
\hline
\end{tabular}
\end{tiny}
\caption{\label{Tab11222}Symmetric disc invariants for the on-shell superpotentials $\sp^{(\alpha_0)}$ of the threefold $\IP_{1,1,2,2,2}[8]$.}
\end{table}

For the candidate superpotentials $W^{(\al_\ell)},\, \ell=1,2,3$, which have an expansion in fractional powers $q_y^{d_y/3}$ with $d_y\in \IZ$, we did not find integral invariants with the multi-cover formula used in the other examples and in ref.~\cite{Walcher:2009uj}. It appears that only the numbers $n_{d_y,d_2}\cdot Z^{d_y}$, with $Z$ a small power of 3, are integral. The solution to this problem might require a shift of the open string mirror map or a refinement of the multi-cover formula.

\subsubsection*{{\it Extremal transition}}
At the singular locus $\phi^2=1$ $(z_2=\frac{1}{4})$ there is an extremal transition to the mirror of the one-parameter model $\IP^5[2,4]$ \cite{Candelas:1993dm}. The large complex structure parameters $z^{(1)}$ of the one-parameter model and the two-parameter model are related by
\ben
z^{(1)}=\frac{1}{2}\,z_1
\een
To restrict the superpotential found in the two-parameter model to that of the one-parameter model we have to add as in ref.~\cite{Walcher:2009uj} an additional linear combination of bulk periods
\ben
\tilde{W}^{(\alpha_0)}(y,z_2)=W^{(\alpha_0)}(y,z_2)+3 F_1(y,z_2)+\frac{3}{2}F_2(y,z_2)
\een
where $(2\pi i)^2\,F_1=\partial_{\rho_1}\partial_{\rho_2}B_{\{l^1,l^2\}}|_{\rho_i=0}$ and $(2\pi i)^2\,F_2=\partial^2_{\rho_1}B_{\{l^1,l^2\}}|_{\rho_i=0}$. We then obtain
\ben
W^{(1,\alpha_0)}(z^{(1)})=\tilde{W}^{\alpha_0}(8z^{(1)},\tfrac{1}{4})
\een
Using the Picard-Fuchs operator of the one-parameter model
\ben
\mathcal{L}^{(1)}=\theta^4-8z^{(1)}(4\theta+1)(4\theta+2)(4\theta+3)(2\theta+1)
\een
with $\theta=z^{(1)}\partial_{z^{(1)}}$ one obtains the inhomogeneous term
\ben
\mathcal{L}^{(1)}W^{(1,\alpha_0)}=\frac{224z^{(1)}}{(2\pi i)^2} \left(   1+272 z^{(1)}+\frac{285120 (z^{(1)})^2}{7}+4925440 (z^{(1)})^3+\dots \right)
\een
For the integer invariants we expect the following relation
\ben
\sum_{l=0}^{3k} n_{k,l}^{(\alpha_0)}(\IP_{1,1,2,2,2}[8])=n_k^{(\alpha_0)}(\IP^5[2,4])\,.
\een
However such a relation only emerges after the addition of an additional bulk period, again as in \cite{Walcher:2009uj}
\ben
\tilde{W}^{(1,\alpha_0)}(z^{(1)})=W^{(1,\alpha_0)}(z^{(1)})-\frac{3}{2}F^{(1)}(z^{(1)})\,,
\een
where $(2\pi i)^2\,F^{(1)}=\partial_{\rho}^2B_{\{l^{(1)}\}}|_{\rho=0}$ with $l^{(1)}=(-4,-2\,|\, 1,1,1,1,1,1)$. The invariants of $\tilde{W}^{(1,\alpha_0)}$ are given by
\ben
n_k^{(\alpha_0)}=384, 29288, 7651456, 2592654592, 989035688064,\dots
\een
It would be interesting to also get a better understanding of the restriction of the superpotentials $W^{(\al_\ell)}, \ell=1,2,3$, to the one-parameter model.

\newpag\subsection{Degree 18 hypersurface in $\IP_{1,2,3,3,9}$}
This is a three parameter Calabi--Yau manifold with the charge vectors of the GLSM given by \cite{Hosono:1995bm}: 

\begin{center}
\vskip-0.4cm\begin{tabular}{cc|cccccccc}
& $\x_0$ & $\x_1$ & $\x_2$ & $\x_3$ & $\x_4$ & $\x_5$ & $\x_6$ & $\x_7$\\
\hline
$l^1$ & -6 & -1 & 0 & 1 & 1 & 3 & 2 & 0 \\
$l^2$ & 0 & 1 & 0 & 0 & 0 & 0 & -2 & 1 \\
$l^3$ & 0 & 0 & 1 & 0 & 0 & 0 & 1 & -2
\end{tabular}
\end{center}
The hypersurface constraint is
\begin{equation} \label{P12339}
P = x_1^{18}+x_2^{9}+x_3^{6}+x_4^6+x_5^{2}+\psi\,x_1 x_2 x_3 x_4 x_5 +\phi\,x_1^{12}x_2^3 +\chi\, x_1^6 x_2^6\,,
\end{equation}
where $\psi=z_1^{-1/6}z_2^{-2/9}z_3^{-1/9}$, $\phi=z_2^{-2/3}z_3^{-1/3}$ and $\chi=z_2^{-1/3}z_3^{-2/3}$. The orbifold group acts as $x_i\to \la_k^{g_{k,i}}\, x_i$ with the weights
\begin{equation}
\IZ_9:\, g_1=(1,-1,0,0,0)\,,\ \ \IZ_6:\, g_2=(1,0,-1,0,0)\,, \ \ \IZ_6:\, g_3=(1,0,0,-1,0)\,,
\end{equation}
with $1=\la_1^9=\la_{2,3}^6$. 
In this geometry we consider the set of curves
\ben
\label{C12339}
&&C_{\pm}=\{x_3^3=\pm i x_4^3\;,\;x_5=\pm i x_1^9\;,\;x_2=0\}\,,
\een
with the identifications $(+,+)\sim(-,-)$ and $(+,-)\sim(-,+)$ for the possible choices of sign under the orbifold group.
The divisor
\begin{equation}
Q(\cx D)=x_3^{6}+z_4 x_4^{6}
\end{equation}
leads by the now familiar steps to a GLSM for a K3 manifold with charges 
\begin{center}
\begin{tabular}{cc|ccccccc}
& $\x_0$ & $\x_1$ & $\x_2$ & $\x_4$ & $\x_5$ & $\x_6$ & $\x_7$   \\
\hline
$\ti l^1$ & -6 & -1 & 0 & 2 & 3 & 2 & 0   \\
$\ti l^2$ & 0 & 1 & 0 & 0 & 0 & -2 & 1 \\ 
$\ti l^2$ & 0 & 0 & 1 & 0 & 0 & 1 & -2  
\end{tabular}
\end{center}
where the moduli of the surface are related to that of the Calabi-Yau threefold by $u_1=-\frac{z_1}{z_4}(1-z_4)^2$, $u_2=z_2$ and $u_3=z_3$. The GLSM is again at a special codimension one locus in the moduli space, with the coefficient of the monomial $x_5x_1^9$ set to zero. The solution
\begin{align}\label{D1tau12339}
\taun(u) &=\frac{c}{2}\ B_{\{\ti l^1,\ti l^2,\ti l^3\}}(u_1,u_2,u_3;\tfrac{1}{2},0,0)=\frac{4c}{\pi} \sqrt{u_1}+\cx O(u_1^{3/2})\,,
\end{align}
vanishes at the critical locus $u_1=0$ and integrates to the superpotential
\begin{align}
W^{(\pm )}&=\mp \frac{c}{8}\ B_{\{ l^1, l^2, l^3\}}\left(z_1,z_2,z_3;\tfrac{1}{2},0,0\right)\,.
\end{align}
Using the multicover formula
\bee\label{MC3}
\frac{W^{(\pm)}(z(q))}{\omega_0(z(q))}=\frac{1}{(2\pi i)^2}\sum_{k\, odd}\sum_{d_1\, odd \atop d_{2,3}\geq 0} n^{(\pm)}_{d_1,d_2,d_3}\, \frac{q_1^{k d_1/2}\,q_2^{k d_2}\,q_3^{k d_3}}{k^2}
\eee
we obtain, for $c=1$, the integer invariants in Tab.~\ref{Tab12339}.
\begin{table}
\begin{tiny}
$q_3^0$
\begin{tabular}{|c|ccccccc|}
\hline
$q_1^{1/2} \backslash q_2$&0&1&2&3&4&5&6\\ 
\hline 
1& 1& 1& 0& 0& 0& 0& 0\\ 
3& -27& -10& -10& -27& 0& 0& 0\\ 
5& 2840& -1629& 2034& 2034& -1629& 2840& 0\\ 
7& -450807& 523790& -501714& 37970& 37970& -501714& 523790\\ 
9& 87114366& -143646335& 151709190& -82679940& 42724232& 42724232& -82679940\\ 
11& -18907171063& 39698748864& -48496621950& 38005868880& -25022027880& 6124612608& 6124612608\\ 
\hline
\end{tabular}
\vskip0.2cm \ \\
$q_3^1$
\begin{tabular}{|c|ccccccc|}
\hline
$q_1^{1/2} \backslash q_2$&0&1&2&3&4&5&6\\ 
\hline 
1& 0& 1& 0& 0& 0& 0& 0\\ 
3& 0& -10& 876& -10& 0& 0& 0\\ 
5& 0& -1629& -2520& 595890& -2520& -1629& 0\\ 
7& 0& 523790& -3041532& 702090& 393040296& 702090& -3041532\\ 
9& 0& -143646335& 913643880& -2889725838& 1131043400& 248949858594& 1131043400\\ 
11& 0& 39698748864& -261938878740& 899363170080& -2195675791704& 998105927940& 153662218213536\\ 
\hline
\end{tabular}
\vskip0.2cm \ \\
$q_3^2$
\begin{tabular}{|c|ccccccc|}
\hline
$q_1^{1/2} \backslash q_2$&0&1&2&3&4&5&6\\ 
\hline 
1& 0& 0& 0& 0& 0& 0& 0\\ 
3& 0& 0& -10& -10& 0& 0& 0\\ 
5& 0& 0& 2034& 595890& 595890& 2034& 0\\ 
7& 0& 0& -501714& 702090& 1648025820& 1648025820& 702090\\ 
9& 0& 0& 151709190& -2889725838& 691571574& 2721112372690& 2721112372690\\ 
11& 0& 0& -48496621950& 899363170080& -7230517669764& 2911708467972& \\ 
\hline
\end{tabular}
\vskip0.2cm \ \\
$q_3^3$
\begin{tabular}{|c|ccccccc|}
\hline
$q_1^{1/2} \backslash q_2$&0&1&2&3&4&5&6\\ 
\hline 
1& 0& 0& 0& 0& 0& 0& 0\\ 
3& 0& 0& 0& -27& 0& 0& 0\\ 
5& 0& 0& 0& 2034& -2520& 2034& 0\\ 
7& 0& 0& 0& 37970& 393040296& 1648025820& 393040296\\ 
9& 0& 0& 0& -82679940& 1131043400& 2721112372690& 8512061067684\\ 
11& 0& 0& 0& 38005868880& -2195675791704& & \\ 
\hline
\end{tabular}
\end{tiny}
\caption{\label{Tab12339}Disc invariants $\frac{1}{16}\cdot n_{d_1,d_2,d_3}$ for the on-shell superpotential $W^{(+)}$ of the threefold $\IP_{1,2,3,3,9}[18]$.}
\end{table}
\vskip0.5cm

The closed-string type II compactification has a non-perturbative enhanced gauge symmetry with gauge group $G=SU(3)$ at the special values $t_2=t_3=0$ of the closed-string moduli. The monodromy around the special locus acts as 
$$
m_1:\ t_1\to t_1+2t_2,\ t_2\to -t_2,\ t_3\to t_2+t_3,\qquad
m_2:\ t_1\to t_1,\ t_2\to t_2+t_3,\ t_3\to -t_3,\qquad
$$
and generates the Weyl group of $SU(3)$. The superpotential $W^{(\pm)}$ is a singlet under this group while the individual BPS states counted by the disc invariants are exchanged under the group action as $m_1:\ n_{d_1,d_2,d_3}\to n_{d_1,2d_1-d_2+d_3,d_3}$ and $m_2:\ n_{d_1,d_2,d_3}\to n_{d_1,d_2,d_2-d_3}$

The off-shell superpotentials are solutions of the following extended hypergeometric system
\begin{align}
\mathcal{  L}_1 &= (\theta_2-\theta_1)(\theta_2-2\theta_3)-z_2(2\theta_1-2\theta_2+\theta_3-1)(2\theta_1-2\theta_2+\theta_3)\nn\,,\\
\mathcal{  L}_2 &= \theta_3(2\theta_1-2\theta_2+\theta_3)-z_3(\theta_2-2\theta_3-1)(\theta_2-2\theta_3)\nn\,,\\
\mathcal{  L}_3 &= \theta_3(\theta_2-\theta_1)-z_2 z_3 (2\theta_1-2\theta_2+\theta_3)(\theta_2-2\theta_3)\,,\\
\mathcal{  L}_4 &= (\theta_1+\partial_y)(\theta_1-\partial_y)(2\theta_1-2\theta_2+\theta_3)\nn\\&\quad-24z_1(6\theta_1+1)(6\theta_1+5)((4z_2-1)\theta_1+(3z_2 z_3-4z_2+1)\theta_2+(2z_2-6z_2z_3)\theta_3)\nn\,,\\
\mathcal{  L}_5 &= (\theta_1+\partial_y)(\theta_1-\partial_y)(\theta_2-2\theta_3)-8z_1 z_2 (6\theta_1+5)(6\theta_1+3)(6\theta_1+1)\nn\,,\\
\mathcal{  L}_6 &= \partial_y(\theta_1+\partial_y)+e^y\partial_y(\theta_1-\partial_y)\,,\nn
\end{align}
where $y=\log(z_4)$. 

To compute the inhomogeneous terms we note that the above differential operators are related to that of the Calabi--Yau threefold derived in \cite{Hosono:1995bm} as
\begin{align}
\mathcal{L}_a&=\mathcal{L}_a^{bulk}\,,\qquad a=1,2,3\,, \nn\\
\mathcal{L}_4 &=\mathcal{L}_4^{bulk} -(2\th_1-2\th_2+\th_3)\th_4^2\,,\\
\mathcal{L}_5 &=\mathcal{L}_5^{bulk} -(\th_2-2\th_3)\th_4^2\,.\nn
\end{align}
to obtain from \eqref{D1tau12339}
\bee
\mathcal{L}_4^{bulk} W^{(\pm )} = \mp \frac{c}{\pi^2} \sqrt{z_1}\,,\qquad 
\mathcal{L}_a^{bulk} W^{(\pm )} = 0\,, a=1,2,3,5\,.
\eee
Note that $\sqrt{z_1}=\psi^{-3}\phi$ is a rational function in terms of $\psi$ and $\phi$ appearing in the hypersurface equation~\eqref{P12339}. The appearance of the square root is related to the non-trivial Greene-Plesser orbifold action on the defining equations \eqref{C12339} for the curves $C_\pm$.

As in the previous examples one may study the relation of the above brane geometry to (two and) one parameter configurations in a certain limit in the moduli. For $z_2=z_3=0$ the geometry approximates the non-compact Calabi--Yau of degree six discussed in App.~\ref{sec:AJ}, explaining the relation $n_{k,0,0}=n^{[6]}_k$ between the invariants in Tab.~\ref{Tab12339} and Tab.~\ref{Tabnc}.

At the point $t_2=t_3=0$ of $SU(3)$ gauge enhancement there is a transition to the one modulus Calabi-Yau $\IP_{1,1,1,1,2,3}[3, 6]$ \cite{Katz:1996ht}, leading to the prediction 
$$
\sum_{i,j}n_{k,i,j} (\IP_{1,2,3,3,9}[18])=n_k(\IP_{1,1,1,1,2,3}[6,3])\, ,
$$
where the first numbers are 
\bee
\frac{1}{16}\, n_k =3,\ 735,\ 1791060,\ 6117294147,\ 25579918417320
.
\eee
The superpotential of the one parameter model is the solution of the inhomogeneous Picard-Fuchs equation
$$
\Lbulk W=\frac{3}{(2\pi i)^2}\sqrt{z_1}\,, \qquad \Lbulk=\th_1^4-36 z_1 (3 \th_1+1) (3 \th_1+2) (6 \th_1+5) (6 \th_1+1)\ .
$$

\newpag\subsection{Degree 12 hypersurface in $\IP_{1,2,3,3,3}$}
This example is very similar to the hypersurface in $\IP_{1,2,3,3,9}$ studied above. The charge vectors of the GLSM given by \cite{Hosono:1995bm}: 

\begin{center}
\vskip-0.4cm\begin{tabular}{cc|cccccccc}
& $\x_0$ & $\x_1$ & $\x_2$ & $\x_3$ & $\x_4$ & $\x_5$ & $\x_6$ & $\x_7$\\
\hline
$l^1$ & -4 & -1 & 0 & 1 & 1 & 1 & 2 & 0 \\
$l^2$ & 0 & 1 & 0 & 0 & 0 & 0 & -2 & 1 \\
$l^3$ & 0 & 0 & 1 & 0 & 0 & 0 & 1 & -2
\end{tabular}
\end{center}
The hypersurface constraint is
\begin{equation}
P = x_1^{12}+x_2^{6}+x_3^{4}+x_4^4+x_5^{4}+\psi\,x_1 x_2 x_3 x_4 x_5 +\phi\,x_1^{8}x_2^2 +\chi\, x_1^4 x_2^4\,,
\end{equation}
where $\psi=z_1^{-1/4}z_2^{-1/3}z_3^{-1/6}$, $\phi=z_2^{-2/3}z_3^{-1/3}$ and $\chi=z_2^{-1/3}z_3^{-2/3}$. The orbifold group acts as $x_i\to \la_k^{g_{k,i}}\, x_i$ with the weights
\begin{equation}
\IZ_6:\, g_1=(1,-1,0,0,0)\,,\ \ \IZ_4:\, g_2=(1,0,-1,0,0)\,, \ \ \IZ_4:\, g_3=(1,0,0,-1,0)\,,
\end{equation}
with $1=\la_1^6=\la_{2,3}^4$.
In this geometry we consider the set of curves
\ben
\label{C12333}
&&C_{\pm}=\{x_3^2=\pm i x_4^2\;,\;x_5^2=\pm i x_1^6\;,\;x_2=0\}\,,
\een
with the identifications $(+,+)\sim(-,-)$ and $(+,-)\sim(-,+)$ for the possible choices of sign under the orbifold group.
The divisor
\begin{equation}
Q(\cx D)=x_3^{4}+z_4 x_4^{4}
\end{equation}
leads by the now familiar steps to a GLSM for a K3 manifold with charges 
\begin{center}
\begin{tabular}{cc|ccccccc}
& $\x_0$ & $\x_1$ & $\x_2$ & $\x_4$ & $\x_5$ & $\x_6$ & $\x_7$   \\
\hline
$\ti l^1$ & -4 & -1 & 0 & 2 & 1 & 2 & 0   \\
$\ti l^2$ & 0 & 1 & 0 & 0 & 0 & -2 & 1 \\ 
$\ti l^2$ & 0 & 0 & 1 & 0 & 0 & 1 & -2  
\end{tabular}
\end{center}
where the moduli of the surface are related to that of the Calabi-Yau threefold by $u_1=-\frac{z_1}{z_4}(1-z_4)^2$, $u_2=z_2$ and $u_3=z_3$. The GLSM is again at a special co-dimension one locus in the moduli space. The solution
\begin{align}\label{D1tau12333}
\taun &=\frac{c}{2} B_{\{\ti l^1,\ti l^2,\ti l^3\}}(u_1,u_2,u_3;\tfrac{1}{2},0,0)=\frac{2c}{\pi} \sqrt{u_1}+\cx O(u_1^{3/2})\,,
\end{align}
vanishes at the critical locus $u_1=0$ and integrates to the superpotential
\begin{align}
W^{(\pm )}&=\mp \frac{c}{8}\,B_{\{ l^1, l^2, l^3\}}\left(z_1,z_2,z_3;\tfrac{1}{2},0,0\right)\,.
\end{align}
Using the multicover formula \eqref{MC3} we obtain, for $c=1$ the integer invariants in Tab.~\ref{Tab12333}.

\begin{table}
\begin{tiny}
$q_3^0$
\begin{tabular}{|c|cccccc|}
\hline
$q_1^{1/2} \backslash q_2$&0&1&2&3&4&5\\ 
\hline 
1 & 1 & 0 & 0 & 0 & 0 & 0 \\
3 & -3 & 0 & 0 & 0 & 0 & 0 \\
5 & 40 & 0 & 0 & 0 & 0 & 0 \\
7 & -847 & 0 & 0 & 0 & 0 & 0 \\
9 & 21942 & 0 & 0 & 0 & 0 & 0 \\
11 & -640431 & 0 & 0 & 0 & 0 & 0 \\
\hline
\end{tabular}
\vskip0.2cm \ \\
$q_3^1$
\begin{tabular}{|c|cccccc|}
\hline
$q_1^{1/2}  \backslash q_2$&0&1&2&3&4&5\\ 
\hline 
1 & 1 & 1 & 0 & 0 & 0 & 0 \\
3 & -2 & -2 & 0 & 0 & 0 & 0 \\
5 & -45 & -45 & 0 & 0 & 0 & 0 \\
7 & 1750 & 1750 & 0 & 0 & 0 & 0 \\
9 & -61551 & -61551 & 0 & 0 & 0 & 0 \\
11 & 2233440 & 2233440 & 0 & 0 & 0 & 0 \\
\hline
\end{tabular}
\vskip0.2cm \ \\
$q_3^2$
\begin{tabular}{|c|cccccc|}
\hline
$q_1^{1/2}  \backslash q_2$&0&1&2&3&4&5\\ 
\hline 
1 & 0 & 0 & 0 & 0 & 0 & 0 \\
3 & -2 & 108 & -2 & 0 & 0 & 0 \\
5 & 50 & -56 & 50 & 0 & 0 & 0 \\
7 & -1962 & -11196 & -1962 & 0 & 0 & 0 \\
9 & 86630 & 439560 & 86630 & 0 & 0 & 0 \\
11 & -3842790 & -16939860 & -3842790 & 0 & 0 & 0 \\
\hline
\end{tabular}
\vskip0.2cm \ \\
$q_3^3$
\begin{tabular}{|c|cccccc|}
\hline
$q_1^{1/2}  \backslash q_2$&0&1&2&3&4&5\\ 
\hline 
1 & 0 & 0 & 0 & 0 & 0 & 0 \\
3 & -3 & -2 & -2 & -3 & 0 & 0 \\
5 & 50 & 11090 & 11090 & 50 & 0 & 0 \\
7 &  506 & 1634 & 1634 & 506 & 0 & 0 \\
9 & -67884 & -1577166 & -1577166 & -67884 & 0 & 0 \\
11 & 4125840 & 66691520 & 66691520 & 4125840 & 0 & 0 \\ 
\hline
\end{tabular}
\vskip0.2cm \ \\
$q_3^4$
\begin{tabular}{|c|cccccc|}
\hline
$q_1^{1/2}  \backslash q_2$&0&1&2&3&4&5\\ 
\hline 
1 & 0 & 0 & 0 & 0 & 0 & 0 \\
3 & 0 & 0 & 0 & 0 & 0 & 0 \\
5 & -45 & -56 & 11090 & -56 & -45 & 0 \\
7 & 506 & 1127464 & 4423692 & 1127464 & 506 & 0 \\
9 & 28776 & 517288 & 46134 & 517288 & 28776 & 0 \\
11 & -3030696 & -185400024 & -566257044 & -185400024 & -3030696 & 0 \\ 
\hline
\end{tabular}
\vskip0.2cm \ \\
$q_3^5$
\begin{tabular}{|c|cccccc|}
\hline
$q_1^{1/2}  \backslash q_2$&0&1&2&3&4&5\\ 
\hline 
1 & 0 & 0 & 0 & 0 & 0 & 0 \\
3 & 0 & 0 & 0 & 0 & 0 & 0 \\
5 & 40 & -45 & 50 & 50 & -45 & 40 \\
7 & -1962 & 1634 & 4423692 & 4423692 & 1634 & -1962 \\
9 & 28776 & 111025794 & 1085027250 & 1085027250 & 111025794 & 28776 \\
11 & 1030368 & 74577268 & 129171092 & 129171092 & 74577268 &  1030368 \\
\hline
\end{tabular}
\end{tiny}
\caption{\label{Tab12333}Disc invariants $\frac{1}{8}\cdot n_{d_1,d_2,d_3}$ for the on-shell superpotential $W^{(+)}$ of the threefold $\IP_{1,2,3,3,3}[12]$.}
\end{table}
\vskip0.5cm

The closed-string type II compactification has a non-perturbative enhanced gauge symmetry with gauge group $G=SU(3)$ at $t_2=t_3=0$. The monodromy around this special locus acts as 
$$
m_1:\ t_1\to t_1,\ t_2\to -t_2,\ t_3\to t_2+t_3,\qquad
m_2:\ t_1\to t_1+t_3,\ t_2\to t_2+t_3,\ t_3\to -t_3,\qquad
$$
and generates the Weyl group of $SU(3)$. The superpotential $W^{(\pm)}$ is a singlet under this group while the individual BPS states counted by the disc invariants are exchanged under the group action as $m_1:\ n_{d_1,d_2,d_3}\to n_{d_1,-d_2+d_3,d_3}$ and $m_2:\ n_{d_1,d_2,d_3}\to n_{d_1,d_2,d_1+d_2-d_3}$.

The off-shell superpotentials are solutions of the following extended hypergeometric system 
\begin{align}
\mathcal{  L}_1 &= (\theta_2-\theta_1)(\theta_2-2\theta_3)-z_2(2\theta_1-2\theta_2+\theta_3-1)(2\theta_1-2\theta_2+\theta_3)\nn\,,\\
\mathcal{  L}_2 &= \theta_3(2\theta_1-2\theta_2+\theta_3)-z_3(\theta_2-2\theta_3-1)(\theta_2-2\theta_3)\nn\,,\\
\mathcal{  L}_3 &= \theta_3(\theta_2-\theta_1)-z_2 z_3 (2\theta_1-2\theta_2+\theta_3)(\theta_2-2\theta_3)\,,\\
\mathcal{  L}_4 &= (\theta_1+\partial_y)(\theta_1-\partial_y)(2\theta_1-2\theta_2+\theta_3)\nn\\&\quad-8z_1(4\theta_1+3)(4\theta_1+1)((4z_2-1)\theta_1+(3z_2 z_3-4z_2+1)\theta_2-(6z_2z_3-2z_2)\theta_3)\nn\,,\\
\mathcal{  L}_5 &= (\theta_1+\partial_y)(\theta_1-\partial_y)(\theta_2-2\theta_3)-4z_1 z_2 (4\theta_1+3)(4\theta_1+2)(4\theta_1+1)\nn\,,\\
\mathcal{  L}_6 &= \partial_y(\theta_1+\partial_y)+e^y\partial_y(\theta_1-\partial_y)\,,\nn
\end{align}
where $y=\log(z_4)$. 

To compute the inhomogeneous terms we note that the above differential operators are related to that of the Calabi--Yau threefold as
\begin{align}
\mathcal{L}_a&=\mathcal{L}_a^{bulk}\,,\quad a=1,2,3\,, \nn\\
\mathcal{L}_4 &=\mathcal{L}_4^{bulk} -(2\th_1-2\th_2+\th_3)\th_4^2\,,\\
\mathcal{L}_5 &=\mathcal{L}_5^{bulk} -(\th_2-2\th_3)\th_4^2\,.\nn
\end{align}
Then we obtain from eq.~\eqref{D1tau12333}
\bee
\mathcal{L}_4^{bulk} W^{(\pm )} = \mp \frac{c}{2\pi^2}\cdot \sqrt{z_1}\,,\qquad 
\mathcal{L}_a^{bulk} W^{(\pm )} = 0\,, a=1,2,3,5\, ,
\eee
where, similarly as in the previous example, $\sqrt{z_1} = \psi^{-2}\phi$ is a rational function in $\psi$ and $\phi$, and the appearance of the square root is related to the non-trivial action of the Greene-Plesser orbifold group on the defining equations for the curves $C_\pm$.
 
In the limit $z_2=z_3=0$ we can again make contact with a non-compact Calabi-Yau. Here it is the degree four hypersurface discussed in App.~\ref{sec:AJ}. This explains the relation $n_{k,0,0}=n^{[4]}_k$ between the invariants in Tab.~\ref{Tab12333} and Tab.~\ref{Tabnc}.

At the point $t_2=t_3=0$ of $SU(3)$ gauge enhancement there is a transition to the one modulus Calabi--Yau $\IP_{1,1,1,1,1,2}[3, 4]$ \cite{Katz:1996ht}, leading to the prediction 
$\sum_{i,j}n_{k,i,j} (\IP_{1,2,3,3,3}[12])=n_k(\IP_{1,1,1,1,1,2}[3,4])\, ,$ with the first invariants being 
\begin{small}\bee
\frac{1}{8}\,n_k=3,\ 87,\ 33252,\ 16628907,\ 10149908544,\ 6979959014559,5196581251886028\,.
\eee
\end{small}
The superpotential is a solution of the inhomogeneous Picard-Fuchs equation of the one modulus problem
$$
\Lbulk W=-\frac{3}{8\pi^2}\sqrt{z_1}\,, \qquad \Lbulk=\th_1^4-12 z_1 (3 \th_1+1) (3 \th_1+2) (4 \th_1+1) (4 \th_1+3)\ .
$$

\section{Conclusions and outlook}\label{sec:con}
In this work we studied off-shell brane superpotentials for four-dimensional type II/F-theory compactifications depending on several open-closed deformations as well as their specialization to the on-shell values in the open-string direction. Mathematically the two potentials are respectively related to the integral period integrals on the (relative) cohomology group defined by the family of branes \cite{Lerche:2002ck,Lerche:2002yw,Jockers:2008pe,Alim:2009rf,Alim:2009bx,Aganagic:2009jq,Li:2009dz}, which depend on both open and closed deformations,  and the so-called normal functions, depending only on closed-string moduli \cite{Walcher:2009uj,Morrison:2007bm}. Both objects can be studied Hodge theoretically by computing the variation of Hodge structure on the relevant (co-)homology fibers over the open-closed-string deformation space $\cx M$. Ultimately, this determines the superpotential as a particular solution of a system of generalized GKZ type differential equations determined by the integral (relative) homology class of the brane. 

The D-brane superpotentials computed in this way are relevant in different contexts. From the phenomenological point of view, the superpotential determines the vacuum structure of four-dimensional F-theory compactifications. The complicated structure of the superpotential for this class of compactifications, described by infinite generalized hypergeometric series, should be contrasted with the simple structure of F-theory superpotentials in other classes of compactifications, as e.g. in refs.~\cite{Douglas:2006es,Blumenhagen:2006ci}. These hypergeometric series have sometimes a dual interpretation as D-instanton corrections and heterotic world-sheet corrections \cite{Jockers:2009ti}, and the rich structure of non-perturbative corrections to the brane superpotential should lead to interesting hierarchies of masses and couplings in the low-energy effective theory. 

As shown in ref.~\cite{Jockers:2009ti}, the solutions to the generalized GKZ system representing the F-theory superpotential do not only capture the superpotentials of dual Calabi-Yau threefold compactifications, but more generally of type II and heterotic compactifications on generalized Calabi-Yau manifolds of complex dimension three.\footnote{The first examples of dual compactifications of this type were given in ref.~\cite{Dasgupta:1999ss}. See also refs.~\cite{Haack:2000di,Haack:2001jz} for related works and examples.} This offers a powerful tool to study more generally the vacuum structure of  phenomenologically interesting F-theory/type~II/heterotic compactifications. It would be interesting to apply the Hodge theoretic approach described in this paper to examples of phenomenologically motivated F-theory scenarios, as described e.g. in refs.~\cite{Beasley:2008dc,Heckman:2008es,Marsano:2008jq}.\footnote{See also ref.~\cite{Weigand:2010wm}, for a recent review on this subject, and further references therein.} In the search for vacua, the step of passing from relative periods depending on open and closed-string deformations to normal functions depending only on closed-string moduli provides a natural split in the minimization process, which should be helpful in a regime of small string coupling. On the other hand, this distinction between closed and open-string moduli disappears away from this decoupling limit, for finite string coupling, where the two types of fields mix in a way determined by a certain degeneration of the F-theory fourfold described in \cite{Alim:2009bx,Jockers:2009ti}. 

A complementary aspect of the $B$-type superpotentials considered in this paper is the prediction of $A$ model disc invariants by open-closed mirror symmetry. For almost flat open-string directions (characterized by a generalized large complex structure point in the open-closed deformation space \cite{Alim:2009bx,Li:2009dz}), already the off-shell superpotential computed by the relative period integral has an $A$ model expansion in closed- and open-string parameters, leading to predictions for  integral Ooguri-Vafa invariants \cite{Alim:2009bx,Jockers:2009mn,Li:2009dz,Jockers:2009ti}. In the present work we instead concentrated on the critical points of the type studied in refs.~\cite{Walcher:2006rs,Morrison:2007bm,Pandharipande:2008,Krefl:2008sj,Knapp:2008uw,Krefl:2009md}, where the $A$ model expansion emerges only after integrating out the open-string directions. The on-shell computations of refs.~\cite{Walcher:2006rs,Morrison:2007bm,Pandharipande:2008} are conceptually well understood and provided the first examples of open-string mirror symmetry in compact Calabi--Yau. Our main motivation to study the type of critical points accessible also in the on-shell formalism was to gain a better understanding of the minimization in the open-string direction, which relates the on-shell computation to the off-shell framework of refs.~\cite{Lerche:2002ck,Lerche:2002yw,Jockers:2008pe,Alim:2009rf,Alim:2009bx}. On the $B$ model side, the relation is provided by the connection between integral relative period integrals and normal functions described in sect.~\ref{sec:RelCoh}. An important datum in this correspondence is the period vector on the surface, that is the brane 4-cycle. It classifies the D-brane vacua by the vanishing condition \eqref{critcon2} and determines the inhomogeneous term in the Picard-Fuchs equation for the normal function. 

In the relative cohomology approach of refs.~\cite{Lerche:2002ck,Lerche:2002yw,Jockers:2008pe,Alim:2009rf,Alim:2009bx}, the open-string deformations are off-shell yet one avoids working in string field theory by perturbing the unobstructed F-theory moduli space associated with the family of surfaces $\cx D$ by a probe brane representing an element in $H_2(\cx D)$. This leads to well-defined {\it finite dimensional} off-shell deformation spaces associated with a particular parametrization by 'light' fields in the superpotential. The parametrization of off-shell deformations  is adapted to the topological string and leads to a definition of off-shell mirror maps and off-shell invariants consistent with expectations. By the general arguments of sect.~\ref{sec:22}, different parametrizations are bound to fit together in an consistent way, as is explicitly demonstrated in some of the examples, where we parametrized the off-shell superpotentials by different choices of open-string deformation parameters. This means starting from a given supersymmetric configuration, we compare different off-shell deformation directions in the infinite-dimensional open-closed deformation space, and we find that the obtained on-shell tensions are independent from the chosen off-shell directions.\footnote{See ref.~\cite{Aganagic:2009jq} for an earlier example of this kind.} This is a gratifying result as the on-shell domain wall tensions should not depend on the details of integrating out the heavy modes.

The relative cohomology approach to open-closed deformations has successfully passed other non-trivial checks \cite{Aganagic:2009jq,Baumgartl:2010ad}. In leading order the computed off-shell superpotentials are compatible with derivations of effective superpotentials using open-string worldsheet and matrix factorization techniques \cite{Hori:2004ja,Ashok:2004xq,Aspinwall:2004bs,Baumgartl:2010ad,Jockers:2007ng,Knapp:2008tv,Carqueville:2009ay}. Beyond leading order, however, the discussed off-shell  superpotentials predict in the context of type II theories higher order open-closed CFT correlators, which (at present) are difficult to compute by other means.

There are many other open questions that need further exploration. For examples with a single open-string deformation a detailed analysis of the Hodge structure of the K3 surface, equivalent to the subsystem defined by the Hodge structure on the surface $\cx D$, might be rewarding. In this work we explained how the analyzed supersymmetric domain wall tensions arise at enhancement points of the Picard lattice in the K3 moduli space. The leading term of the K3 periods near these specially symmetric points is a rational function in the deformations $z$ and the roots $\al$ of the defining equation. As argued in sect.~\ref{sec:X12234}, the global symmetry seems to be related to the discrete symmetry in the $A$-type brane in the mirror $A$-model configuration. It would be interesting to study in detail the structure of Picard lattice enhancement loci in order to systematically explore the web of $\cx N=1$ domain wall tensions in Calabi-Yau threefolds. Such an analysis potentially sheds light on the global structure of $\cx N=1$ superpotentials (see e.g. refs.~\cite{Uranga:2008nh}), and should be related to the wall-crossing phenomena described in refs.~\cite{Gaiotto:2008cd,Kontsevich:2009xt,Cecotti:2009uf}.

In this work we have focused on a single open-string deformation. Then the subsystem of the extended hypergeometric GKZ system, which governs the open deformations, describes the periods of an isogenic K3 surface. The presented techniques are directly applicable also to examples with several open deformations \cite{Toappear}. Then the subsystem geometry is not anymore governed by K3 periods but instead by the periods of a complex surface of a higher geometric genus. Exploring such examples is technically more challenging but new phenomena and interesting structures, like non-commutativity in the open-string sector, are likely to emerge. A related question in this context is the contribution from D-instanton corrections, which are also computed by GKZ system for the F-theory compactification \cite{Jockers:2009ti}. It would be very interesting to connect these results to the recent progress in computing D-brane instantons by different methods \cite{Rocek:2005ij,RoblesLlana:2006ez,RoblesLlana:2006is,Alexandrov:2008gh,Billo:2010mg}.

\vskip2cm
\noindent {\bf Acknowledgements:} 
We would like to thank 
Ilka Brunner,
Albrecht Klemm and especially 
David Morrison 
and Johannes Walcher 
for discussions and correspondence.
M.A. is supported by the Hausdorff Center for Mathematics and DFG fellowship
AL 1407/1-1.
The work of M.H. and P.M. is supported by the program
``Origin and Structure of the Universe'' of the German Excellence Initiative.
The work of H.J. is supported by the Stanford Institute of Theoretical Physics
and the NSF Grant 0244728 and also by the Kavli Institute for Theoretical Physics
and the NSF Grant PHY05-51164. The work of A.M. is supported by the 
Studienstiftung des deutschen Volkes. The work of M.S. is supported by a EURYI award of the European Science Foundation.

\appendix
\section{Appendix}\label{sec:app}
\subsection{Toric  hypersurfaces for type II and F-theory compactifications}\label{apptor}
In the framework of \cite{Batyrev:1994hm} a mirror pair $(X^*,X)$ of Calabi-Yau threefolds is 
given as a pair of hypersurfaces defined in two toric ambient spaces $(V^*,V)$ as follows. The toric varieties $(V^*,V)$ are associated to the fans $(\Sigma(\Delta^*),\Sigma(\Delta))$ obtained from the set of cones over the faces of two dual reflexive polyhedra $(\Delta^*,\Delta)$. The polyhedron $\Delta^*$ is the convex hull of $p$ integral points $\nus_i \in \mathbbm{Z}^5 \in \mathbbm{R}^5$ lying in a hyperplane of distance one to the origin and $\Delta$ is the dual polyhedron with integral points $\nu_i$. The mirror 3-folds $X$ are defined in $V$ as the zero locus of the hypersurface constraint
$$
P = \sum_{i=0}^{p-1}  a_i \prod_{k=1}^4 X_k^{\nu^*_{i,k}}\, .
$$
Here the $X_k, \, k=1,\dots,4$ are coordinates on an open torus $(\mathbbm{C}^*)^4 \subset V$ and $a_i$ are complex parameters which determine the complex structure of $X$. Alternatively, one may write the hypersurface in homogeneous coordinates $x_j$ on the toric ambient space as 
\begin{equation}\label{defW} 
P = \sum_{i=0}^{p-1}  a_i \prod_{\nu_j\in \Delta} x_j^{<\nu_j,\nu^*_i>+1} \,  .
\end{equation}
Keeping only the coordinates $x_i$ associated with the vertices of $\Delta$ in the product on the right hand side, one obtains the simplified expression used in the text, e.g. eq.\eqref{W12234} in the first example. 

The integral points $\nus_i$ of $\Delta^\star$ fulfill a set of linear relations $\sum_{i=0}^{p-1} l^a_i \nus_i = 0$ specified by $M= h^{(1,1)}(X^*)=h^{(2,1)} (X)$ vectors $l^a, \, a=1,\dots,M$ with integral entries, given e.g. in \eqref{ls12234} for the first example. The vectors $l^a$ represent the $U(1)$ charges of the two-dimensional fields in the GLSM associated to $X$ \cite{Witten:1993yc}. The first index $i=0$ refers to the single interior point of $\Delta$, which corresponds to the distinguished field of negative charge that multiplies the hypersurface constraint in the two dimensional superpotential.

The open-string sector for the compactification with  $B$-type branes on $X$ is captured by the family of hypersurfaces $\cx D$ defined as the complete intersections 
$
\cx D: \ \{P(X)=0\}\cap \{Q(\cx D)=0\} 
$
in $V$ \cite{Aganagic:2000gs,Lerche:2002yw,Alim:2009bx,Aganagic:2009jq}. Locally, one may write $Q(D)$ as
\begin{equation}\label{defQ}
Q(\cxH) = \sum_{i=p}^{p+p'-1}  a_i X_k^{\nus_i,k}\ ,
\end{equation}
where the right hand side defines $p'$ additional (not necessarily integral) vertices $\nus_i$ associated with the monomials in $Q(\cx D)$. The coefficients $a_i,\ i\geq p$\ are complex parameters of a family of embeddings $\cx D\hookrightarrow X$ for a fixed set of monomials in $Q(\cx D)$ and determine a point on the fiber $\hat{\cx M}$ of the deformation space
\begin{equation}\label{ocdia}
\hat{\cx M}(\zh) \xrightarrow{\ \ \phantom{\pi} \ \ } \cx M \xrightarrow{\ \ \pi\ \ } \cx M_{CS}(z) \ .
\end{equation}

The combined data for the closed and the open-string sector can be expressed in terms of extended vertices $\nub_i$, which makes contact to the F-theory compactification on a fourfold dual to the brane geometry \cite{Alim:2009rf,Aganagic:2009jq,Li:2009dz,Jockers:2009ti}. To this end, consider the set of extended vertices
$$
\nub_i=\begin{cases}(\nus_i,0)& i=0,...,p-1\\(\nus_i,1)&i=p,...,p+p'-1 \end{cases}\,,
$$
which determine the (ordered) monomials in \eqref{defW} and \eqref{defQ}. Define the set $L=\{\lb\}$ as the set of solutions to the equations
\begin{equation}\label{defextmori}
\sum_{i=0}^{p+p'-1} \lb_i\nub_i=0,\qquad \sum_{i=0}^{p-1} \lb_i=0=\sum_{i=p}^{p+p'-1} \lb_i\,.
\end{equation}

At this point, the $\nub_i$ for $i\geq p$ are defined only up to an overall shift $\nu_i\to \nub_i+(\mu,0)$ for a constant four-vector $\mu$ (corresponding to multiplication of $Q$ by an overall factor), but this shift is not relevant in \eqref{defextmori} because of the last condition.
For the generators of $L$ one may choose the charge vectors of the closed-string GLSM extended by $p'$ zeros to the left and in addition $p'-1$ vectors describing relations between the monomials in $Q(\cx D)$ and those in $P$. 
\begin{equation}
\lb^a=\begin{cases}(\lb^a,0,...,0)& i=1,...,M\\(...)&a=M+1,...,M+p'-1 \end{cases}\,.
\end{equation}
From these vectors one obtains the parameters
\bee z_a=(-)^{\lb^a_0} \prod_i a_i^{\lb^a_i},\qquad  a=1,...,h^{2,1}(X)+p'-1\,,\eee invariant under the torus action. For $a\leq h^{2,1}(X)$, these are the coordinates on the base $\mcs$ and the $z_a$ for $a>h^{2,1}(X)$ describe open-string deformations. If the vertices $\nub_i$ satisfy appropriate extra conditions discussed in \cite{Alim:2009rf,Alim:2009bx,Li:2009dz}, the $z_a$ define local coordinates near an open-string generalization of a large complex structure point $\cx P$ in $\cx M$, where the superpotential has an $A$ model expansion in Ooguri--Vafa invariants . 

The extended vertices $\nub_i$ for the brane geometry on the threefold $X$ define an extended polyhedron $\ux \Delta^\star$ of one dimension higher, which can be associated to mirror pair of non-compact Calabi--Yau fourfolds $(X_4^{*,\sharp},X_4^\sharp)$ \cite{Lerche:2002ck,Lerche:2002yw,Alim:2009rf,Li:2009dz}. M/F-theory compactification on $X_4^\sharp$ gives a dual compactification without branes but with flux, related to the brane compactification on the threefold $X$ by open-closed duality \cite{Mayr:2001xk,Alim:2009bx,Aganagic:2009jq}. Under duality, the RR brane (and flux) superpotential on the threefold $X$ maps to the leading order term of the GVW superpotential on $X_4^\sharp$ in the expansion \eqref{GVW} in $g_s$, that is 
\bee\label{GVWl}
W_{GVW}(X^\sharp_4)=\sum_{\Si} N_\Si(G)\, \ux \Pi_\Si(z,\zh) \ .
\eee 
This open-closed duality at $g_s=0$ extends to a full string duality between the brane compactification on $X$ and F-theory compactification on a compact fourfold $X_4$ \cite{Alim:2009bx,Jockers:2009ti}. The details of the compactification capture the coupling of the brane to the global geometry and affect only the higher order terms in $g_s$, but not the disc invariants. 

We hence restrict to report the polyhedra for the non-compact 4-folds $X_4^\sharp$ below, which determine the leading order superpotential by eq.~\eqref{GVWl}. In the following table we collect the (extended) points $\nub_i$ for the brane geometry in the examples and the dual vertices $\nu_i$ defining the homogeneous coordinates used in the text via eq.~\eqref{defW}. The points $\nus_i$ for the threefold $X$ are given by the subset of the $\nub_i$ with vanishing last entry, $\nub_i=(\nus_i,0)$.
\newpage

\def\-{\hphantom{-}}
\begin{small}
$$
\vbox{\offinterlineskip\halign{
#&~$#$\hfil
&~$#$\hfil&\hfil$#$\hfil &\qquad $#$\hfil&\hfil$#$\hfil\cr
&\quad&&\bar{\Delta}^\star(X_4^\sharp)\ \supset\  \Delta^\star(X)&&\Delta(X^\star)\cr
\noalign{\hrule}
&&\phantom{XXX} &\cr
&\IP_{1,2,2,3,4}[12]\quad&\nub_1=& (\- 2,\- 2,\- 3,\- 4;\- 0) &\nu_1=& (\- 1,\- 1,\- 1,\- 1)\cr
&&\nub_2=& (-1,\- 0,\- 0, \- 0;\- 0) &\nu_2=& ( - 5,\- 1,\- 1,\- 1)\cr
&&\nub_3=& (\- 0,  -1,\- 0,\- 0;\- 0) &\nu_3=& (\- 1, - 5,\- 1,\- 1)\cr
&&\nub_4=& (\- 0,\- 0, - 1,\- 0;\- 0) &\nu_4=& (\- 1,\- 1, - 3,\- 1)\cr
&&\nub_5=& (\- 0,\- 0,\- 0, - 1;\- 0) &\nu_5=& (\- 1,\- 1,\- 1, - 2)\cr
&&\nub_6=& (\- 1,\- 1,\- 1,\- 2;\- 0) \cr
&&\phantom{XXX} &\cr
&X_4^\sharp(\cx D_1)&\nub_7=& (\nus_2;1),\ \nub_8=(\nus_3;1) \cr
&X_4^\sharp(\cx D_2)&\nub_7=& (\nus_4;1),\ \nub_8=(\nus_6;1) \cr
&&\phantom{XXX} &\cr
&&\phantom{XXX} &\cr
&\IP_{1,2,2,2,7}[14]\quad&\nub_1=& (\- 2,\- 2,\- 2,\- 7;\- 0) &\nu_1=& (\- 1,\- 1,\- 1,\- 1)\cr
&&\nub_2=& (-1,\- 0,\- 0, \- 0;\- 0) &\nu_2=& ( - 6,\- 1,\- 1,\- 1)\cr
&&\nub_3=& (\- 0,  -1,\- 0,\- 0;\- 0) &\nu_3=& (\- 1, - 6,\- 1,\- 1)\cr
&&\nub_4=& (\- 0,\- 0, - 1,\- 0;\- 0) &\nu_4=& (\- 1,\- 1, - 6,\- 1)\cr
&&\nub_5=& (\- 0,\- 0,\- 0, - 1;\- 0) &\nu_5=& (\- 1,\- 1,\- 1, - 1)\cr
&&\nub_6=& (\- 1,\- 1,\- 1,\- 3;\- 0) \cr
&&\phantom{XXX} &\cr
&X_4^\sharp(\cx D_1)&\nub_7=& (\nus_3;1),\ \nub_8=(\nus_4;1) \cr
&X_4^\sharp(\cx D_2)&\nub_7=& (\nus_5;1),\ \nub_8=(\nus_6;1) \cr
&&\phantom{XXX} &\cr
&&\phantom{XXX} &\cr
&\IP_{1,1,1,6,9}[18]\quad&\nub_1=& (\- 1,\- 1,\- 6,\- 9;\- 0) &\nu_1=& (\- 1,\- 1,\- 1, \- 1)\cr
&&\nub_2=& (-1, \- 0,\- 0,\- 0;\- 0) &\nu_2=& ( - 17,\- 1,\- 1,\- 1)\cr
&&\nub_3=& (\- 0,-1,\- 0,\- 0;\- 0) &\nu_3=& (\- 1, -17,\- 1,\- 1)\cr
&&\nub_4=& (\- 0,\- 0, -1,\- 0;\- 0) &\nu_4=& (\- 1,\- 1, - 2,\- 1)\cr
&&\nub_5=& (\- 0,\- 0,\- 0, - 1;\- 0) &\nu_5=& (\- 1,\- 1,\- 1, -1)\cr
&&\nub_6=& (\- 0,\- 0,\- 2,\- 3;\- 0) \cr
&&\phantom{XXX} &\cr
&X_4^\sharp(D)&\nub_7=& (\nus_1;1),\ \nub_8=(\nus_2;1) \cr
&&\phantom{XXX} &\cr
&&\phantom{XXX} &\cr
&\IP_{1,1,1,3,3}[9]\quad&\nub_1=& ( - 1,\- 0,\- 1,\- 1;\- 0) &\nu_1=& ( - 6,\- 3,\- 1,\- 1)\cr
&&\nub_2=& (\- 0, - 1,\- 1,\- 1;\- 0) &\nu_2=& (\- 3, - 6,\- 1,\- 1)\cr
&&\nub_3=& (\- 1,\- 1,\- 1,\- 1;\- 0) &\nu_3=& (\- 3,\- 3,\- 1,\- 1)\cr
&&\nub_4=& (\- 0,\- 0, - 1,\- 0;\- 0) &\nu_4=& (\- 0,\- 0, -2,\- 1)\cr
&&\nub_5=& (\- 0,\- 0,\- 0, - 1;\- 0) &\nu_5=& (\- 0,\- 0,\- 1,-2)\cr
&&\nub_6=& (\- 0,\- 0,\- 1,\- 1;\- 0) \cr
&&\phantom{XXX} &\cr
&X_4^\sharp(D)&\nub_7=& (\nus_1;1),\ \nub_8=(\nus_2;1) \cr
&&\phantom{XXX} &\cr
&&\phantom{XXX} &\cr
&\IP_{1,1,2,2,6}[12]\quad&\nub_1=& (\- 1,\- 2,\- 2,\- 6;\- 0) &\nu_1=& ( \- 1,\- 1,\- 1,\- 1)\cr
&&\nub_2=& (-1, \- 0,\- 0,\- 0;\- 0) &\nu_2=& (- 11,\- 1,\- 1,\- 1)\cr
&&\nub_3=& (\- 0,-1,\- 0,\- 0;\- 0) &\nu_3=& (\- 1,- 5,\- 1,\- 1)\cr
&&\nub_4=& (\- 0,\- 0, -1,\- 0;\- 0) &\nu_4=& (\- 1,\- 1, - 5,\- 1)\cr
&&\nub_5=& (\- 0,\- 0,\- 0, - 1;\- 0) &\nu_5=& (\- 1,\- 1,\- 1, -1)\cr
&&\nub_6=& (\- 0,\- 1,\- 1,\- 3;\- 0) \cr
&&\phantom{XXX} &\cr
&X_4^\sharp(\cx D)&\nub_7=& (\nus_5;1),\ \nub_8=(0,0,0,0;1) \cr
&&\phantom{XXX} &\cr
&&\phantom{XXX} &\cr
&\IP_{1,1,2,2,2}[8]\quad&\nub_1=& (\- 1,\- 2,\- 2,\- 2;\- 0) &\nu_1=& ( \- 1,\- 1,\- 1,\- 1)\cr
&&\nub_2=& (-1, \- 0,\- 0,\- 0;\- 0) &\nu_2=& (- 7,\- 1,\- 1,\- 1)\cr
&&\nub_3=& (\- 0,-1,\- 0,\- 0;\- 0) &\nu_3=& (\- 1,- 3,\- 1,\- 1)\cr
&&\nub_4=& (\- 0,\- 0, -1,\- 0;\- 0) &\nu_4=& (\- 1,\- 1, - 3,\- 1)\cr
&&\nub_5=& (\- 0,\- 0,\- 0, - 1;\- 0) &\nu_5=& (\- 1,\- 1,\- 1, -3)\cr
&&\nub_6=& (\- 0,\- 1,\- 1,\- 1;\- 0) \cr
&&\phantom{XXX} &\cr
&X_4^\sharp(\cx D)&\nub_7=& (\nus_5;1),\ \nub_8=(0,0,0,0; 1) \cr
&&\phantom{XXX} &\cr
&&\phantom{XXX} &\cr
}}
$$
\end{small}
\begin{table}[!h]
\begin{small}
$$
\vbox{\offinterlineskip\halign{
#&~$#$\hfil
&~$#$\hfil&\hfil$#$\hfil &\qquad $#$\hfil&\hfil$#$\hfil\cr
&\quad&&\bar{\Delta}^\star(X_4^\sharp)\ \supset\  \Delta^\star(X)&&\Delta(X^\star)\cr
\noalign{\hrule}
&&\phantom{XXX} &\cr
&\IP_{1,2,3,3,9}[18]\quad&\nub_1=& (\- 2,\- 3,\- 3,\- 9;\- 0) &\nu_1=& ( \- 1,\- 1,\- 1,\- 1)\cr
&&\nub_2=& (-1, \- 0,\- 0,\- 0;\- 0) &\nu_2=& (- 8,\- 1,\- 1,\- 1)\cr
&&\nub_3=& (\- 0,-1,\- 0,\- 0;\- 0) &\nu_3=& (\- 1,- 5,\- 1,\- 1)\cr
&&\nub_4=& (\- 0,\- 0, -1,\- 0;\- 0) &\nu_4=& (\- 1,\- 1, - 5,\- 1)\cr
&&\nub_5=& (\- 0,\- 0,\- 0, - 1;\- 0) &\nu_5=& (\- 1,\- 1,\- 1, -1)\cr
&&\nub_6=& (\- 1,\- 2,\- 2,\- 6;\- 0) \cr
&&\nub_7=& (\- 0,\- 1,\- 1,\- 3;\- 0) \cr
&&\phantom{XXX} &\cr
&X_4^\sharp(\cx D)&\nub_8=& (\nus_3;1),\ \nub_9=(\nus_4;1) \cr
&&\phantom{XXX} &\cr
&&\phantom{XXX} &\cr
&\IP_{1,2,3,3,3}[12]\quad&\nub_1=& (\- 2,\- 3,\- 3,\- 3;\- 0) &\nu_1=& ( \- 1,\- 1,\- 1,\- 1)\cr
&&\nub_2=& (-1, \- 0,\- 0,\- 0;\- 0) &\nu_2=& (- 5,\- 1,\- 1,\- 1)\cr
&&\nub_3=& (\- 0,-1,\- 0,\- 0;\- 0) &\nu_3=& (\- 1,- 3,\- 1,\- 1)\cr
&&\nub_4=& (\- 0,\- 0, -1,\- 0;\- 0) &\nu_4=& (\- 1,\- 1, - 3,\- 1)\cr
&&\nub_5=& (\- 0,\- 0,\- 0, - 1;\- 0) &\nu_5=& (\- 1,\- 1,\- 1, -3)\cr
&&\nub_6=& (\- 1,\- 2,\- 2,\- 2;\- 0) \cr
&&\nub_7=& (\- 0,\- 1,\- 1,\- 1;\- 0) \cr
&&\phantom{XXX} &\cr
&X_4^\sharp(\cx D)&\nub_8=& (\nus_3;1),\ \nub_9=(\nus_4;1) \cr
&&\phantom{XXX} &\cr
&&\phantom{XXX} &\cr
}}
$$
\end{small}
\vskip-1.2cm
\caption{Vertices of the toric polyhedra for the threefolds for type II compactification and the non-compact limit of the F-theory fourfolds.}
\end{table}

\subsection{From three-chains to Abel-Jacobi maps on the elliptic curve}\label{sec:AJ}
In some of the examples considered in sect.~3, the domain wall tensions can be directly related to the Abel-Jacobi map on an elliptic curve in a certain limit in the moduli. This gives a check on the normalization obtained from the geometric surface periods. To this end, we consider the non-compact Calabi-Yau manifolds $X^\flat$ of ref.~\cite{Lerche:1996ni}
\ben\label{nccys}
\IP^4_{3,2,1,1,-1}[6]:\quad P= y_1^2+y_2^3+y_3^6+y_4^6+\frac{1}{y_5^6}+\hat\psi \,y_1y_2y_3y_4y_5\ ,
\quad z=\hat\psi^{-6}\ ,\nn\\
\IP^4_{2,1,1,1,-1}[4]:\quad P= y_1^2+y_2^4+y_3^4+y_4^4+\frac{1}{y_5^4}+\hat\psi \,y_1y_2y_3y_4y_5\ ,
\quad z=\hat\psi^{-4}\ ,\\
\IP^4_{1,1,1,1,-1}[3]:\quad P= y_1^3+y_2^3+y_3^3+y_4^3+\frac{1}{y_5^3}+\hat\psi \,y_1y_2y_3y_4y_5\ ,
\quad z=\hat\psi^{-3}\ .\nn
\een
The closed-string periods on the non-compact threefolds are solutions of the Picard-Fuchs operators 
\bee\label{ncpf}
\cx L^{[n]}=\cx L^{[n]}_E(-z)\cdot \th\,,
\eee
where $\cx L^{[n]}_E(z)$ denote the Picard-Fuchs operators for the representations of the elliptic curve~$E$
\ben\label{Eop}
\IP_{3,2,1}[6]:&\qquad \cx L^{[6]}_E(z) &= \th^2-12z(6\th+5)(6\th+1)\,,\nn\\
\IP_{2,1,1}[4]:&\qquad \cx L^{[4]}_E(z) &= \th^2-4z(4\th+3)(4\th+1)\,,\\
\IP_{1,1,1}[3]:&\qquad \cx L^{[3]}_E(z) &= \th^2-3z(3\th+1)(3\th+2)\,,\nn
\een
with $\th = z\frac{d}{dz}$. The equation for the elliptic curve is given by the restriction to $(y_4y_5)^n=-1$ in \eqref{nccys}.\footnote{Keeping the convention eq.~\eqref{Defz}, the algebraic modulus of the Calabi-Yau manifold and the curve differ by a minus sign, as indicated in eq.~\eqref{ncpf} and below.} Eq.~\eqref{ncpf} implies the relation $2\pi i\,\th\Pi_\ell(z)=\pi_\ell(-z)$ between the periods $\Pi_\ell(z)$ of the non-compact threefold and the periods $\pi(z)$ on the elliptic curve.

A similar relation 
\bee \label{Trelii}
2\pi i\,\th T(z)=\tau(-z)\,,
\eee
holds for the chain integrals between the domain wall tension $T$ of the non-compact threefold and the line integral $\tau$ of the associated elliptic curve $E$. They fulfill the inhomogeneous differential equation
\bee \label{PFinhom}
\cx L^{[n]} T(z)\,=\, -\frac{c^{[n]}}{16\pi^2}\sqrt{z}  \ , \qquad \cx L^{[n]}_E(z) \tau(z) \,=\, -\frac{c^{[n]}}{8\pi}\sqrt{z} \ ,
\eee
in terms of the constants $c^{[n]}$
\bee \label{cnorm}
c^{[6]}=16 \ , \qquad c^{[4]}=8 \ , \qquad c^{[3]}=2 \ ,
\eee
which determine the normalization of the of the domain wall tension $T$. Then the domain wall tensions $T$, which are now solutions to the normalized inhomogeneous Picard-Fuchs equations~\eqref{PFinhom}, contains the quantum instanton contribution $T_{\rm inst}$, which starts as
$$
T_{\rm inst}(z)\,=\,-\frac{1}{2\pi^2}\left(c^{[n]} \sqrt{z} + \ldots \right) \ ,
$$  
and yields for the three geometries \eqref{nccys} the normalized disc invariants in Tab.~\ref{Tabnc}.\footnote{Here we list the integral disc instanton numbers $n_k^{[n]}$. These invariants are related to the real invariants $n_{k,{\rm real}}^{[n]}$ in ref.~\cite{Walcher:2007qp} by a factor $2$, {\it i.e.} $n_k^{[n]}=2 \cdot n_{k,{\rm real}}^{[n]}$.}

\begin{table}
\leftskip0.5cm
\begin{tiny}
\def\tbsp{\ \hskip1.5cm\ }
\begin{tabular}{|r|c|c|c|}
\hline
& & & \\[-2mm]
$k$&\tbsp$n_k^{[6]}$\tbsp&\tbsp$n_k^{[4]}$\tbsp&\tbsp$n_k^{[3]}$\tbsp\\[0.6mm]
\hline
1&16 & 8 & 2 \\
3&-432 & -24 & -2 \\
5& 45\,440 & 320 & 10 \\
7& -7\,212\,912 & -6\,776 & -84 \\
9& 1\,393\,829\,856 & 175\,536 & 858 \\
11& -302\,514\,737\,008 & -5\,123\,448 & -9\,878 \\
13& 70\,891\,369\,116\,256 & 161\,777\,200 & 123\,110 \\
15& -17\,542\,233\,743\,427\,360 & -5\,401\,143\,120 & -1\,622\,890 \\
17& 4\,520\,954\,871\,206\,554\,016 & 187\,981\,969\,232 & 22\,308\,658 \\
19&-1\,202\,427\,473\,254\,100\,406\,128 & -6\,756\,734\,860\,408 & -316\,775\,410 \\
21& 327\,947\,495\,234\,600\,477\,004\,048 & 249\,179\,670\,525\,576 & 4\,616\,037\,426 \\
23& -91\,298\,034\,448\,725\,882\,319\,078\,384 & -9\,384\,048\,140\,182\,200 & -68\,700\,458\,258\\
\hline
\end{tabular}
\end{tiny}
\caption{\label{Tabnc} Disc invariants for the on-shell superpotentials $W_{\rm inst}=\tfrac{1}{2} T_{\rm inst}$ for the non-compact hypersurfaces $X^\flat$ of degree $d=6,4,3$.}
\end{table}
The normalization constants $c^{[n]}$ are determined by requiring integrality of the monodromy matrices with respect to the singularities of the moduli space of the extended period vector. The extended period vector consists of the bulk periods $\Pi$ and the domain wall tension $T$. Alternatively, the constants $c^{[n]}$ can be determined by directly evaluating the line integral $\tau$ on the curve $E$ and by exploiting its relation to the 3-chain integral $T$ according to eq.~\eqref{Trelii}. In the following we exemplify the two approaches for the non-compact sextic threefold \eqref{nccys} to determine the normalization constant $c^{[6]}$. The other two normalization constants $c^{[4]}$ and $c^{[3]}$ are obtained analogously.

The moduli space of the non-compact sextic threefold \eqref{nccys} exhibits three singularities $z=0$, $z=-\frac{1}{432}$, and $z=\infty$, which correspond to a large radius, a conifold, and a orbifold point of the moduli space. In the vicinity of the large radius point $|z|<\frac{1}{432}$ a complete set of solutions to the Picard-Fuchs operator $\cx L^{[6]}$ is given by
\bee
\begin{aligned}
\tilde\Pi_0(z)=&1 \ ,\\
\tilde\Pi_1(z)=& \log z + \sum_{k=1}^{+\infty}\frac{(6k)!}{k!\,(2k)!\,(3k)!}\cdot\frac{(-z)^k}{k} \ ,\\
\tilde\Pi_2(z)=&\frac{1}{2}(\log z)^2+\sum_{k=1}^{+\infty}\frac{(6k)!}{k!\,(2k)!\,(3k)!}\cdot\frac{(-z)^k}{k}
\cdot\left( \log z - \frac{1}{k}+6 \Psi(6k+1)\right. \\
&\qquad\qquad\qquad\qquad\qquad\qquad\quad\ \left. -\Psi(k+1)-2\Psi(2k+1)-3\Psi(3k+1)\vphantom{\frac{1}{k}}\right) \, ,
\end{aligned}
\eee
in terms of the Polygamma function $\Psi$. Together with the solution $\tilde{T}$ to the inhomogeneous Picard-Fuchs equation ${\cx L}^{[6]}\tilde{T}(z) \sim \sqrt{z}$
\bee
\tilde{T}(z)\,=\, \frac{\pi}{32}\sqrt{z} \ 
\sum_{k=0}^{+\infty}\frac{\Ga(6k+4)}{\Ga(3k+\frac{5}{2})\Ga(2k+2)\Ga(k+\frac{3}{2})(k+\frac{1}{2})}(-z)^k \ , 
\eee
they form the extended period vector $\tilde\Pi = \left(\tilde\Pi_0,\tilde\Pi_1,\tilde\Pi_2,\tilde{T}\right)$. For this vector we determine the large radius monodromy matrix $\tilde M_{\rm LR}$. Furthermore, by analytically continuation with the help of Barnes integrals to the other singular points in the moduli space we also infer the conifold and orbifold monodromy matrices $\tilde M_{\rm con}$ and $\tilde M_{\rm orb}$. Next we perform a change of basis to the integral extended period vector $\Pi=\left(\Pi_0,\Pi_1,\Pi_2,T\right)$ by demanding integrality of all the monodromy matrices. For the bulk sector these steps can be found in detail in ref.~\cite{Lerche:1996ni}. In addition to integrality of the monodromy matrices we require that the domain wall tension $T$ vanishes at $z=\infty$. The latter condition arises because the domain wall tension $T$ interpolates between two supersymmetric vacua that coincide at the orbifold point. After these steps we finally arrive at the integral periods
\bee \label{ExtVecInt}
\begin{aligned}
\Pi_0(z)&=\tilde \Pi_0(z) \,=\, 1 \ , \\
\Pi_1(z)&=\frac{1}{2\pi i} \tilde \Pi_1(z) \,=\, t(z) \ , \\
\Pi_2(z)&=\frac{1}{(2\pi i)^2} \tilde \Pi_2(z)-\frac{1}{4\pi i}\tilde\Pi_1(z)-\frac{5}{12}\tilde \Pi_0(z) \,=\, 
 \frac{1}{2} t(z)^2-\frac{1}{2} t(z) -\frac{5}{12}+ \Pi_{\rm inst}(z) \ , \\
T(z)&=\frac{32}{(2\pi i)^2} \tilde{T}(z) - \frac{1}{4\pi i} \tilde \Pi_1(z) + \frac{1}{4} \tilde \Pi_0(z) \,=\,
 - \frac{1}{2} t(z) + \frac{1}{4} + T_{\rm inst}(z) \ .
\end{aligned}
\eee  
Here we also exhibit the classical terms in terms of the flat coordinate $t$ and the instanton contributions $\Pi_{\rm inst}$ and $T_{\rm inst}$. In particular the normalized domain wall tension yields the normalized instanton contribution
$$
T_{\rm inst}(z)\,=\, - \frac{16}{2\pi^2}\left(\sqrt{z} -\frac{512}{9}z^{3/2}+\frac{229\,376}{25}z^{5/2}-\ldots \right) \ ,
$$
and hence the normalization constant $c^{[6]} = 16$ in eq.~\eqref{cnorm}. The integral monodromy matrices in the integral basis \eqref{ExtVecInt} are then given by
$$
\def\-{\phantom{-}}
M_{\rm LR}\,=\,
\begin{pmatrix}
 1&\-0&\-0&\-0\\ 1&\-1&\-0&\-0\\ 0&\-1&\-1&\-0\\ 0&-1&\-0&-1
\end{pmatrix}\,,\quad
M_{\rm con}\,=\,
\begin{pmatrix}
 1&\-0&\-0&\-0\\ 1&\-1&\-1&\-0\\ 0&\-0&\-1&\-0\\ 0&\-0&\-0&\-1
\end{pmatrix}\,,\quad
M_{\rm orb}\,=\,
\begin{pmatrix}
 1&\-0&\-0&\-0\\ 0&\-0&-1&\-0\\ 0&\-1&\-1&\-0\\ 0&-1&\-0&-1
\end{pmatrix}\,,
$$    
with $M_{\rm con}M_{\rm orb} = M_{\rm LR}$. 

As an independent calculation to determine normalized domain wall tensions we now directly reduce the three-chain integrals of the domain wall tensions between the curves $C^\flat_{\varepsilon,\pm}$ of eq.~\eqref{LimC12234} to line integrals on the elliptic curve $E$. In order to evaluate the chain integrals we first change to the inhomogeneous coordinate $\al, y, z_1, z_2$, which are suitable to evaluate the chain integrals \cite{Lerche:1996ni}
$$
y_1=y-\frac12\hat\psi \al z_1^3 z_2 \ , \quad y_2=-\al z_1^2 \ , \quad y_3 = - i z_1 \ , \quad y_4 = -i z_2 \ , \quad y_5=1 \ .
$$
In terms of these coordinates the hypersurface equation~\eqref{nccys} of the non-compact sextic Calabi-Yau threefold $X^\flat$ becomes
$$
X^\flat: \quad y^2 = z_1^6\left(1+\al^3+\frac14 (\hat\psi\al)^2 z_2^2 \right) + (z_2^6-1) \ , 
$$
while the holomorphic three form reads
$$
\Omega(\hat\psi) = \frac{6}{(2\pi i)^3} \cdot 
\frac{\hat\psi\,z_1^2\,d\al \,dz_1\, dz_2}{2\sqrt{z_1^6\left(1+\al^3+\frac14 (\hat\psi\al)^2 z_2^2 \right) + (z_2^6-1)}} \ .
$$
We can think of this geometry as a complex surface given in terms of the coordinates $z_1$ and $z_2$ fibered over a $\IP^1$ base parametrized by the affine coordinate $\al$. Furthermore, in these coordinates the curves $C^\flat_{\varepsilon,\pmk}$ are given by\footnote{For ease of notation we have chosen here the explicit root $\eta = i$ for $\eta^6 = -1$ in eq.~\eqref{LimC12234}.} 
$$
C^\flat_{\varepsilon,\pmk}\,=\,
\left\{ z_1 = i\,z_2 \,,\  \al z_2 =- i \pmk  \sqrt{\varepsilon\,i\,\hat\psi}  \,, \ \varepsilon=\frac{i}{2}\hat\psi \al z_1^4 + y \right\} \ , 
\quad \varepsilon=\pm i \quad \pmk=\pm i \,.
$$  
The goal is now to evaluate the domain wall tensions
$$
T_\ell(\hat\psi) = \int_{\Ga_\ell} \Omega(\hat\psi) \ , 
$$
where we consider the two three chains $\Ga_1$ and $\Ga_2$ bounded by
$$
\partial\Ga_1 = C_{+i,+i} - C_{-i,-i} \ , \qquad   \partial\Ga_2 = C_{-i,+i} - C_{-i,-i} \ . 
$$  
As we will see in the calculation the domain wall tensions for the remaining combinations of curves do not yield independent results. The steps to reduce the three dimensional integral to a line integral over the $\IP^1$ base are worked out and explained in detail in ref.~\cite{Lerche:1996ni}. Therefore for completeness we merely sketch the necessary steps here.

Instead of calculating the domain wall tension, it is easier to derive the line integral $\tau$ of eq.~\eqref{Trelii}. With $z=\hat\psi^{-6}$ we get
$$
\begin{aligned}  
\tau_\ell(-z) &= 2 \pi i\, \theta\, T_\ell(\hat\psi(z)) \,=\,  -\frac{2 \pi i}{6} \, \hat\psi \,\frac{d T_\ell(\hat\psi)}{d\hat\psi} \cr
&= 
 -\frac{\hat \psi}{(2\pi i)^2}  \int_{\Ga_\ell} d\al\,dz_2\,dz_1 \, \frac{d}{dz_2} 
 \frac{1}{2\sqrt{z_1^6\left(1+\al^3+\frac14 (\hat\psi\al)^2 z_2^2 \right) + (z_2^6-1)}} \ .
\end{aligned}    
$$  
The simplification occurs because for the integrand the derivative with respect to $\hat\psi$ is equivalent to the derivative with respect to $z_2$. Then the integral over $z_2$ becomes trivial.\footnote{Similarly as for the examples discussed in ref.~\cite{Krefl:2008sj}, there is no contribution from the derivative $\tfrac{d}{d\hat\psi}$ acting on the three chain $\Ga_\ell$.} We now evaluate the integral over the coordinate $z_1$ along a closed contour encircling the six branch points of the square root. Next we integrate the coordinate $z_2$ along the interval from $z_2=1$ to $z_2=-1$ to arrive at \cite{Lerche:1996ni}
\bee \label{Xflatline}
\int \Omega=-\frac{1}{2\pi i} \left( 
\int\left. \frac{\hat\psi\,z_2\,d \al}{2\sqrt{1+\al^3+\frac{1}{4} (\hat\psi z_2)^2 \al^2}} \right|_{z_2=1}
\!\!\!\!\!\!\!\!-
\int\left. \frac{\hat\psi\,z_2\,d \al}{2\sqrt{1+\al^3+\frac{1}{4} (\hat\psi z_2)^2 \al^2}} \right|_{z_2=-1}\right) \ .
\eee
Note that the performed integration is equivalent to the integration over a homology 2-sphere, as the contour in the $z_1$ coordinate can be shrunk to a point at the endpoints $z_2=\pm 1$ of the interval. 

If we now carry out the remaining integral~\eqref{Xflatline} over $\al$ along a closed contour encircling the two branch points with leading behavior $\sim \hat\psi^{-1}$ for large $\hat\psi$, we integrate over a one cycle of the elliptic curve $E$ and obtain the fundamental period of the elliptic curve
$$
\pi_0(-z)\,=\, {}_2F_1(\tfrac{1}{6},\tfrac{5}{6};1;-432z)\,=\,1-60z+13\,860z^2-\ldots \ .
$$
If, instead, we reduce the three chains $\Ga_1$ and $\Ga_2$ to line integrals over $\al$ in eq.~\eqref{Xflatline}, we need to evaluate the integrals
\bee \label{lineint}
\begin{aligned}
\tau_1(-z) \,&=\, - \frac{1}{2\pi i} \left( 
\int_{i\sqrt{\hat\psi}}^{-\sqrt{\hat\psi}} \frac{\hat\psi\,d \al}{2\sqrt{1+\al^3+\frac{1}{4} \hat\psi^2 \al^2}}+
\int_{-i\sqrt{\hat\psi}}^{\sqrt{\hat\psi}} \frac{\hat\psi\,d \al}{2\sqrt{1+\al^3+\frac{1}{4} \hat\psi^2 \al^2}} \right) \cr
\tau_2(-z) \,&=\, - \frac{1}{2\pi i} \left( 
\int_{\sqrt{\hat\psi}}^{-\sqrt{\hat\psi}} \frac{\hat\psi\,d \al}{2\sqrt{1+\al^3+\frac{1}{4} \hat\psi^2 \al^2}}+
\int_{-\sqrt{\hat\psi}}^{\sqrt{\hat\psi}} \frac{\hat\psi\,d \al}{2\sqrt{1+\al^3+\frac{1}{4} \hat\psi^2 \al^2}} \right)
\end{aligned}
\eee
Here the integration boundaries for $\al$ are determined by requiring that the coordinates $(z_2=\pm 1, \al)$ associated to the endpoints of the line integral correspond to a point on the appropriate curve $C^\flat_{\varepsilon,\pmk}$.

While the line integral \eqref{lineint} trivially vanishes for $\Ga_2$, namely $\tau_2(-z)=0$, we evaluate the integral over $\Ga_1$ and arrive at
\bee
\tau(-z)\,\equiv\,\tau_1(-z)\,=\,\frac{16\sqrt{z}}{2\pi i}\ {}_3F_2(\tfrac{2}{3},\tfrac{4}{3},1;\tfrac{3}{2},\tfrac{3}{2};-432z)
-\frac{1}{2}\pi_0(-z) \ .
\eee  
The resulting domain wall tension $\tau(-z) = 2\pi i \,\theta T(z)$ is in agreement with the result in eq.~\eqref{ExtVecInt} and in eq.~\eqref{PFinhom} together with the normalization $c^{[6]}=16$ of eq.~\eqref{cnorm}. 

\clearpage




\end{document}